\documentclass[aps,prb,twocolumn,superscriptaddress,floatfix,amsmath,amssymb]{revtex4-2}
\usepackage{amsmath}
\usepackage{amssymb}
\usepackage{braket}
\usepackage{graphicx}
\usepackage[table]{xcolor}
\usepackage{verbatim}
\usepackage{float}
\usepackage{color}
\usepackage{siunitx}
\usepackage{gensymb}
\usepackage{bm}
\usepackage{xcolor}
\usepackage{multirow}
\usepackage{footnote}
\usepackage{bbm}

\usepackage{array}
\newcolumntype{P}[1]{>{\centering\arraybackslash}p{#1}}
\newcolumntype{Q}[1]{>{\raggedleft\arraybackslash}p{#1}}
\newcolumntype{R}[1]{>{\raggedright\arraybackslash}p{#1}}

\newcommand{\rr}{\mathbf{r}}
\newcommand{\qq}{\mathbf{q}}
\newcommand{\kk}{\mathbf{k}}
\newcommand{\KK}{\mathbf{K}}
\newcommand{\QQ}{\mathbf{Q}}
\newcommand{\GG}{\mathbf{G}}
\newcommand{\gG}{\mathbf{g}}

\begin{document}
\title{Multifaceted moir{\'e} superlattice physics in twisted WSe$_2$ bilayers}

\author{S. J. Magorrian}
\affiliation{National Graphene Institute, University of Manchester, Booth St E, Manchester, M13 9PL, United Kingdom}
\affiliation{Department of Physics, University of Warwick, Coventry, CV4 7AL, United Kingdom}
\author{V. V. Enaldiev}
\affiliation{National Graphene Institute, University of Manchester, Booth St E, Manchester, M13 9PL, United Kingdom}
\affiliation{Department of Physics \& Astronomy, University of Manchester, Oxford Road, Manchester, M13 9PL, United Kingdom}
\affiliation{Kotel'nikov Institute of Radio-engineering and Electronics of the Russian Academy of Sciences, 11-7 Mokhovaya St, Moscow, 125009 Russia}
\author{V. Z\'{o}lyomi}
\affiliation{Hartree Centre, STFC Daresbury Laboratory,
Daresbury, WA4 4AD, United Kingdom}
\author{F. Ferreira}
\affiliation{National Graphene Institute, University of Manchester, Booth St E, Manchester, M13 9PL, United Kingdom}
\affiliation{Department of Physics \& Astronomy, University of Manchester, Oxford Road, Manchester, M13 9PL, United Kingdom}
\author{V. I. Fal'ko}
\email{vladimir.falko@manchester.ac.uk}
\affiliation{National Graphene Institute, University of Manchester, Booth St E, Manchester, M13 9PL, United Kingdom}
\affiliation{Department of Physics \& Astronomy, University of Manchester, Oxford Road, Manchester, M13 9PL, United Kingdom}
\affiliation{Henry Royce Institute for Advanced Materials, University of Manchester, Manchester, M13 9PL, United Kingdom}
\author{D.\ A.\ Ruiz-Tijerina}
\email{d.ruiz-tijerina@fisica.unam.mx}
\affiliation{Secretar\'ia Acad\'emica, Instituto de F\'isica, Universidad Nacional Aut\'onoma de M\'exico, Ciudad de M\'exico, C.P. 04510, M\'exico}


\begin{abstract}
    Lattice reconstruction in twisted transition-metal dichalcogenide (TMD) bilayers gives rise to piezo- and ferroelectric moir\'e potentials for electrons and holes, as well as a modulation of the hybridisation across the bilayer. Here, we develop hybrid $\mathbf{k}\cdot \mathbf{p}$ tight-binding models to describe electrons and holes in the relevant valleys of twisted TMD homobilayers with parallel (P) and anti-parallel (AP) orientations of the monolayer unit cells. We apply these models to describe moir\'e superlattice effects in twisted WSe${}_2$ bilayers, in conjunction with microscopic \emph{ab initio} calculations, and considering the influence of encapsulation, pressure and an electric displacement field. Our analysis takes into account mesoscale lattice relaxation, interlayer hybridisation, piezopotentials, and a weak ferroelectric charge transfer between the layers, and describes a multitude of possibilities offered by this system, depending on the choices of P or AP orientation, twist angle magnitude, and electron/hole valley.
\end{abstract}

\maketitle

\section{Introduction}

Moir\'e superlattices -- emergent structures with long-range stacking periodicity -- are a generic feature of van der Waals (vdW) heterostructures\cite{McGilly2020}. The presence of a small misalignment angle $\theta$ or lattice mismatch $\delta$ between their constituent layers amplifies the atomic periodicity as $a_M=a/\sqrt{\theta^2+\delta^2}$, with $a$ the monolayer lattice constant. Moir\'e superlattices induce a plethora of physical effects, such as long-range interlayer hybridisation, leading to flat minibands with strongly correlated electronic states\cite{cao2018unconventional,cao2018correlated,Yankowitz2019,Zhang2020,Wang2020,Pan2020,MoralesDuran2021,MakMott2020,PasupathyMott2020} and minibands for excitons in transition metal dichalcogenide (TMD) bilayers\cite{alexeev2019resonantly,jin2019} at twist angles $\theta \ll 10^\circ$, for which the moir\'e periodicity exceeds the exciton Bohr radius, thus affecting the system's optoelectronic properties\cite{rivera2018interlayer,tran2019,seyler2019,Scuri_PRL,sung2020broken}. Moreover, piezoelectric effects caused by lattice reconstruction in TMD bilayers \cite{Weston2020,rosenberger2020,McGilly2020} create periodic traps for charge carriers \cite{Edelberg2020,shabani2020} and excitons \cite{Enaldiev2021}, whereas interlayer charge transfer \cite{Li2017,Tong_2020} induces ferroelectric polarisation in these structures \cite{woods2020,stern2020,yasuda2020}.   

\begin{figure}
    \centering
    \includegraphics[width =\columnwidth]{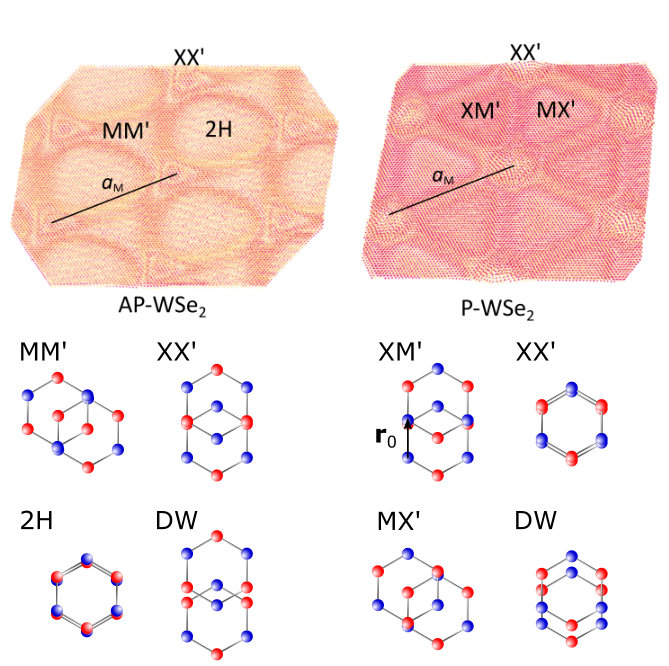}
    \caption{Examples of reconstructed twisted WSe$_2$ bilayers with antiparallel (AP) and parallel (P) orientations of layers' unit cells. In-plane lattice reconstruction promotes the growth of hexagonal domains with 2H-stacking for AP-bilayers and triangular domains with XM$'$/XM$'$ stackings for P-bilayers, whereas out-of-plane relaxation leads to bulging of energetically unfavourable XX$'$ stacking areas and domain walls. In the top panels, $a_M$ indicates the moir\'e superlattice period. In our stacking notation, A$_{\rm b}$A${}_{\rm t}'$ indicates which atoms in the bottom and top layers are vertically aligned (A${}_{\rm t/b}=$ M for metal and X for chalcogen), using the more familiar notation 2H for MX${}'$ stacking in AP structures. The bottom panels illustrate the different stacking configurations, showing the in-plane stacking vector $\mathbf{r}_0$. }
    \label{Fig:Intro}
\end{figure}

For marginal twist angles, moir\'e superlattices in twisted TMD homobilayers undergo strong lattice reconstruction, resulting in the formation of energetically preferential domains separated by networks of dislocation-like domain walls \cite{Weston2020,McGilly2020,Enaldiev_arXiv,carr2018relaxation,PRLNaik}. Due to the inversion asymmetry of TMD monolayers, the emerging domain structures differ for homobilayers with parallel (P, $\theta=\theta_P$) and anti-parallel (AP, $\theta=\pi + \theta_{AP}$) orientations of their unit cells (Fig. \ref{Fig:Intro}). Whereas for P-bilayers the reconstructed moir\'e pattern consists of alternating triangular domains with MX$'$/XM$'$-type stacking (here, MX$'$ indicates that the bottom-layer metallic atoms are vertically aligned with the top-layer chalcogen atoms, as in bulk 3R structures), domains in AP-bilayers are hexagonal and feature 2H-type stacking  \cite{Weston2020,rosenberger2020,Enaldiev_arXiv,PRLNaik}. In-plane lattice reconstruction is accompanied by interlayer distance modulation across the supercell, which is of especial importance for the hybridisation of the top valence band states at the $\Gamma$-valley, formed by $d_{z^2}$ and $p_z$ orbitals of metals and chalcogens, respectively. A theoretical analysis of the electron properties in twisted TMD homobilayers must take into account the competition between various comparable factors, such as the piezoelectric potential, variation of the local band structure throughout the moir\'e superlattice with local stacking and interlayer distance, and interlayer (ferroelectric) charge transfer, relevant for P-bilayers \cite{Li2017,Tong_2020}.

 Here, we develop a unified approach for the theoretical description of electronic properties in twisted TMD homobilayers, taking into account interlayer hybridisation, lattice reconstruction, piezoeffects and interlayer charge transfer, and demonstrate a great variety of emergent features when applying it to twisted WSe$_2$ bilayers. The proposed theory is based on the multiscale analysis \cite{Enaldiev_arXiv} of atomic reconstruction in twisted TMD bilayers, combining elasticity theory with density functional theory (DFT) modelling of the interlayer adhesion energy, and the derivation of hybrid $\mathbf{k}\cdot\mathbf{p}$ tight-binding interpolation models for the hybridisation of relevant conduction- or valence-band states based on DFT band structures. This allows us to trace the evolution of the potential energy landscapes for electrons and holes in the range of twist angles $0^{\circ}<\theta_{P,AP}\leq 4^{\circ}$. In addition, we analyse the effects of external perturbations, such as homogeneous strain, out-of-plane electric fields and pressure, on the energy and momentum of ground-state excitons in 3R- and 2H-stacked WSe$_2$ bilayers, within the framework of DFT.

Our findings for \textbf{AP-WSe$_2$ bilayers} are as follows:

\begin{itemize}
    \item The $\Gamma$-point valence band maximum is modulated across the moir\'e supercell by a combination of piezopotential and strong interlayer hybridisation. The large effective mass at the $\Gamma$ point promotes the formation of strongly localised quantum dot (QD) states for holes at superlattice regions with 2H stacking for twist angles $1^\circ<\theta_{AP}\lesssim 4^\circ$.
    
    \item The $K$-point valence band edge variation throughout the supercell is dominated by the piezopotential. At marginal twist angles $\theta_{AP}\lesssim 1^{\circ}$, piezopotential wells form QDs that localise $K$-point holes at corners of the domain wall structure with local XX$'$ stacking (see Fig. \ref{Fig:Intro}). For $1^\circ<\theta_{AP}<2^\circ$, these QD states mix to form narrow minibands for holes, realising an SU${}_4$ Hubbard model\cite{FermionLadder} on a mesoscale triangular lattice. 
    
    \item The $K$-point conduction band edge modulation is also dominated by the piezopotential. For $\theta_{AP}\lesssim 1^\circ$, MM$'$ corners host QDs for electrons, giving rise to narrow bands for $1^\circ<\theta_{AP}<2^\circ$, and again realising an SU${}_4$ Hubbard model on a mesoscale triangular lattice.
    
    \item  The $Q$-point conduction band energy landscape is dominated by the piezopotential for $\theta_{AP}\leq 1^{\circ}$, forming QDs for electrons in MM$'$ corners. The resulting QD states have a total spin- and valley degeneracy factor of 12, realising a large-$N$ SU${}_N$ Hubbard model. For $\theta_{AP}\geq 3^{\circ}$, the conduction band edge shifts to 2H regions.

\end{itemize}

Our findings for \textbf{P-WSe$_2$ bilayers} are as follows:
\begin{itemize}
    \item We quantify the stacking dependence of the ferroelectric interlayer charge transfer, and calculate the variation of areal density of electric dipole moments across the moir\'e supercell.
    
    \item The $\Gamma$-point valence band energy is highest at MX$'$ and XM$'$ sites, raised by the piezo- and ferroelectric potentials, forming a honeycomb lattice of quantum boxes for holes. For $1^\circ<\theta_P\lesssim 4^\circ$, hole states in these quantum boxes hybridise, mostly through interlayer tunnelling, producing narrow minibands with Dirac-like features, realising a narrow-band version of ``mesoscale graphene''.
    
    \item The $K$-point valence band edge behaves differently for $\theta_P\lesssim 1^\circ$ and $1<\theta_P<2^\circ$. In the former case, the band maximum appears at MX$'$ and XM$'$ regions of the superlattice, where the combined piezo- and ferroelectric potential energy is highest. In the latter case, interlayer hybridisation dominates, shifting the band maxima to XX$'$ corners, forming a mesoscale triangular QD lattice.
    
    \item The $K$-point conduction band edge modulation is dominated by the piezo- and ferroelectric potentials, defining quantum boxes for electrons at MX$'$ and XM$'$ regions across the whole studied range of misalignment angles. In contrast to $K$-point holes, interlayer tunnelling of $K$-point electrons is suppressed, so that when overlapping minibands form for $\theta>1^\circ$, they are based on separate triangular QD arrays in the top and bottom layers.

    \item The $Q$-point conduction band edge is affected by variations of the resonant interlayer coupling and piezopotential across the moir\'e superlattice. For $\theta_P\leq 2^{\circ}$, the $Q$-point conduction band edge appears at one-dimensional channels along two out of every three domain walls in each moir\'e supercell, as a consequence of the low symmetry of the $Q$-point states.
    
\end{itemize}

We note that the analysis presented here, corresponding to suspended WSe$_2$ bilayers, shows that the $\Gamma$-point valence band edge is systematically below that of the $K$-point. However, their relative energies may depend on the sample encapsulation, as these two edges are formed by orbitals with different symmetry: The $\Gamma$-point states, formed by selenium $d_z$ and tungsten $p_z$ orbitals\cite{kormanyos2015k}, are likely to interact more strongly with the environment [e.g., hexagonal boron nitride (hBN) substrate or encapsulation] than the $K$-point states, which consist of tungsten $d_{(x\pm i y)^2}$ orbitals. Therefore, for encapsulated bilayers the energy shift of the $\Gamma$ point valence band edge will be also determined by its relative order with respect to energies of orbitals of the encapsulating material. 

The main body of this paper is organised as follows: in Section \ref{sec:adhesion} we overview the model for adhesion energy and lattice reconstruction\cite{Enaldiev_arXiv}, including the analysis of piezoelectric potentials\cite{APLarxiv}. In Section \ref{sec3}, we  employ \emph{ab initio} DFT to analyse interlayer charge transfer and to develop interpolation formulae for its analytical description in P-bilayers. In Section \ref{sec:hybrydisation_Hams} we construct minimal effective Hamiltonians describing interlayer hybridisation between relevant band edge states at the  $\Gamma$- and $K$-points. In Sections \ref{sec:Twisted} and \ref{sec:twisted_P} we combine these Hamiltonians with lattice reconstruction, piezo- and ferroelectric potential contributions to study the conduction and valence band edge modulation across the moir\'e supercell in twisted WSe$_2$ bilayers, and compute the corresponding moir\'e minibands in Section \ref{sec:Kminibandstogether}. Finally, in Section \ref{sec:tuning_aligned_bilayers} we discuss how the band edge position across the Brillouin zone of WSe${}_2$ bilayers can be modified by external electric fields, strain, and encapsulation using, \emph{e.g.}, hBN.

\section{Interlayer adhesion energy}
\label{sec:adhesion}
\begin{figure*}
	\includegraphics[width = \linewidth]{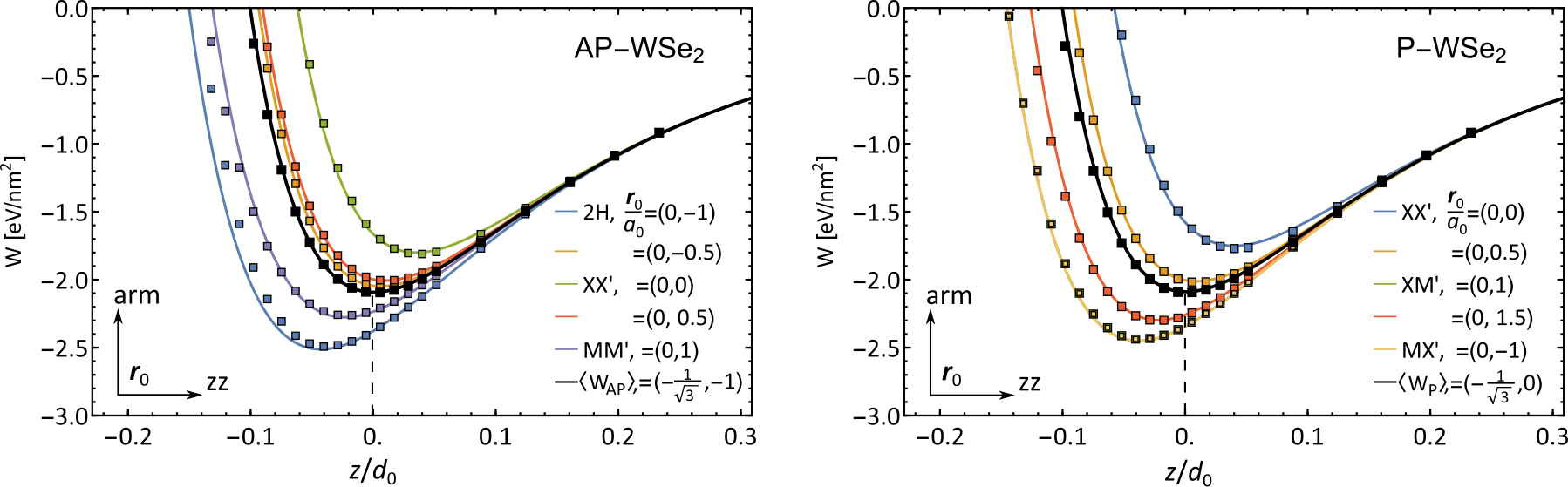}
	\caption{
		\label{vdWen}
		Adhesion energy density of AP- (left) and P- (right) WSe$_2$ bilayers with various in-plane offsets $\rr_0$ between the layers.  $\bm{r}_0=a_0(0,0)$ corresponds to XX$'$ stacking, and $a_0=3.282/\sqrt{3}$\,\AA is the in-plane metal-chalcogen distance. Dots represent data calculated in Ref.\ \cite{Enaldiev_arXiv} using the optB88 vdW-DFT functional, whereas lines are their fit by Eq. \eqref{Eq:Adhesion_energy}. Interlayer distances are counted from the configuration-averaged adhesion energy minimum $\langle W_{P/AP} \rangle_{\bm{r}_0}$. For AP-bilayers the most stable configuration is 2H, being an elementary building block in the bulk 2H-TMD crystals. For P-bilayers, the MX$'$ and XM$'$ configurations are mirror twins of each other and have the same $d$-dependence corresponding to the most energetically favourable configuration. 
	}
\end{figure*}

We begin with an analysis of the lattice structure of twisted WSe$_2$ bilayers, based on a previously established model for the adhesion energy $W_{P/AP}$ between two aligned monolayers \cite{Enaldiev_arXiv} (see Fig. \ref{vdWen}):
\begin{align}\label{Eq:Adhesion_energy}
    W_{P/AP}&(\bm{r}_0, z) = \overline{W}+\gamma z^2 + A_1e^{-\sqrt{G^2+\rho^{-2}}z}f_s(\rr_0)\\
    & \qquad\qquad+ A_2e^{-G z}f_{s/a}(\rr_0),\nonumber\\
    f_s(\bm{r}_0)&=2\cos\left(2\pi \frac{x_0}{a}\right)\cos\left(\frac{2\pi}{\sqrt{3}} \frac{y_0}{a}\right) + \cos\left(\frac{4\pi}{\sqrt{3}} \frac{y_0}{a}\right), \nonumber\\
    f_a(\bm{r}_0)&=2\sin\left(2\pi \frac{x_0}{a}\right)\cos\left(\frac{2\pi}{\sqrt{3}} \frac{y_0}{a}\right) -  \sin\left(\frac{4\pi}{\sqrt{3}} \frac{y_0}{a}\right).\nonumber
\end{align}
Here, $\bm{r}_0=(x_0,y_0)$ is a lateral offset between layers characterising different stacking configurations ($\bm{r}_0=(0,0)$ for XX$'$ stacking corresponding to overlaying of chalcogens in two layers) in a Cartesian reference frame with $x$- and $y$-axes along zigzag and armchair directions, respectively, and $a=3.282\,{\rm \AA}$ is the lattice constant. The interlayer distances $d=z+d_0$ in Eq.\ (\ref{Eq:Adhesion_energy}) are counted from an optimal interlayer distance $d_0$ obtained from the configuration-averaged adhesion energy profile $\langle W_{P/AP} \rangle_{\bm{r}_0}=-\sum_{n=1,2,3}C_n/(d_0+z)^n\approx \overline{W} + \gamma z^2$, where after the second approximate equality we leave only the lowest terms in a Taylor series over $z$ ($\gamma=190$\,eV$\cdot$nm$^{-4}$). In Eq.\ \eqref{Eq:Adhesion_energy}, $G=4\pi/a\sqrt{3}$ is the magnitude of the basis reciprocal vectors of monolayer WSe$_2$, $\GG_{1,2} = G(\pm \tfrac{\sqrt{3}}{2},\, \tfrac{1}{2})$. The values of the fitting parameters $C_{1,2,3}$, $A_{1,2}$ and $\rho$ are listed in Table \ref{tab_Fit}.

Linearisation of the exponentials in Eq.\ (\ref{Eq:Adhesion_energy}), followed by minimisation with respect to $z$, gives an expression for the optimal interlayer distance variation with stacking configuration\cite{Enaldiev_arXiv}:
\begin{multline}\label{Eq:interlayer_distance}
    z_{P/AP}(\bm{r}_0) =  \\
    =\frac{1}{2\gamma} \left[A_1\sqrt{G^2+\rho^{-2}}f_{s}(\bm{r}_0)+A_2Gf_{s/a}(\bm{r}_0)\right].
\end{multline}
Equation (\ref{Eq:interlayer_distance}) leads to a slightly larger interlayer distance ($\approx 6.66$\AA) for 2H-stacked WSe$_2$ bilayers than that extracted from experiments with bulk samples ($\approx 6.48$\AA\, \cite{schutte1987crystal}). This is because vdW-DFT calculations overestimate $d_0=6.89$\AA\, used as a reference interlayer distance, while $\gamma$, which determines the amplitude of the optimal distance variation, is computed more accurately\cite{klimevs2011van}. This is confirmed by its comparison with the frequency of the layer breathing mode measured using Raman scattering (see SI in Ref. \cite{Enaldiev_arXiv}). Below, to compensate for the discrepancy between the calculated and measured interlayer distances for 2H bilayers, we will use a shifted value $d_0=6.71$\AA\ for the reference interlayer distance in DFT calculations of the band structures.

{\setlength{\tabcolsep}{3.2pt}
\renewcommand{\arraystretch}{1.2}
\begin{table}[!h]
	\caption{Interpolation parameters for the adhesion energy density, Eq.\ (\ref{Eq:Adhesion_energy}), of WSe$_2$ bilayers. Equivalent to the results presented in Ref.\ \onlinecite{Enaldiev_arXiv}.\label{tab_Fit}}
	\begin{tabular}{cccccc}
		\hline
		\hline
		$C_1$, & $C_2$, & $C_3$,  & \mbox{A$_{1}$}, & \mbox{A$_{2}$}, & $\rho$\\ 
		\mbox{eV$\cdot$nm$^2$} &\mbox{eV$\cdot$nm$^6$} &  \mbox{eV$\cdot$nm$^{10}$} &  eV/nm$^2$ &  eV/nm$^2$ & nm \\
		\hline 
		0.1488 & 0.2478 & -0.0395 & $0.1428$ & 0.0275 &  0.0497 \\ 
		\hline
		\hline
	\end{tabular}
\end{table}
}
 
In moir\'e superlattices of twisted WSe$_2$ bilayers, the existence of an energetically favourable local stacking configuration promotes lattice reconstruction. As shown in Ref.\ \cite{Enaldiev_arXiv}, the magnitude of the twist angle distinguishes between strong ($\theta_{P/AP}<\theta_{P/AP}^*$) and weak ($\theta_{P/AP} \gtrsim \theta_{P/AP}^*$) reconstruction regimes, where $\theta_P^*=2.5^{\circ}$ and $\theta_{AP}^*=1^{\circ}$. The former regime is characterised by the expansion of the regions with the lowest energy into domains separated by domain wall networks, whereas for the latter  domains do not form, leaving a smooth variation of of the interlayer atomic registry across the supercell. Qualitatively, strong reconstruction happens when the energy gain from developing the lowest energy domains outweighs the elastic energy cost of domain wall formation. Since the gain grows as the square of the superlattice period ($\propto a_M^2$) and the cost linearly ($\propto a_M$), the moir\'e superlattice experiences a commensurate-incommensurate transition only at sufficiently large periods, \emph{i.e.}, below some critical angle, as described above.

\begin{figure}
	\includegraphics[width = 1.0\columnwidth]{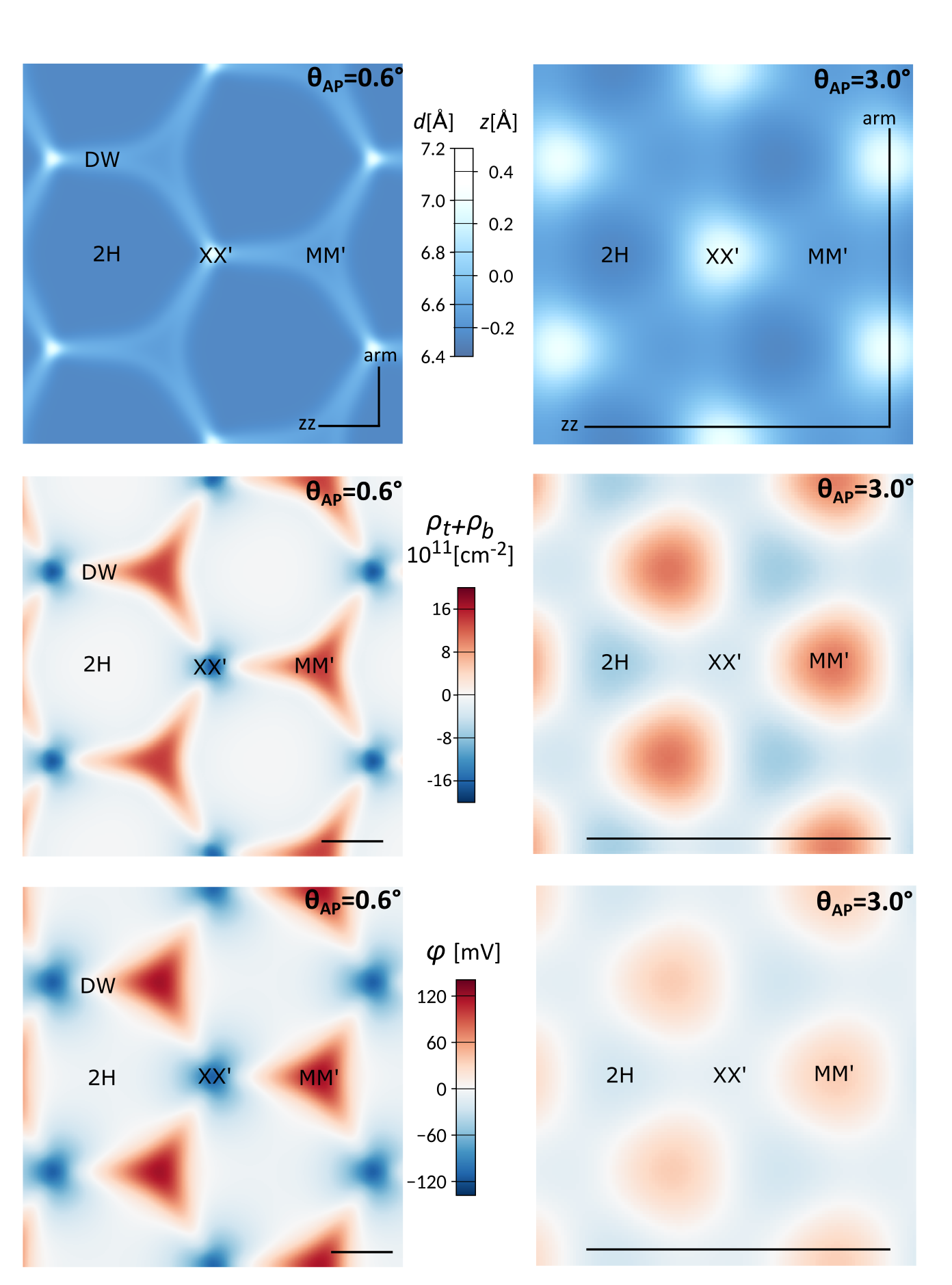}
	\caption{
		\label{Fig:AP_relaxed_bands} 
		Maps of interlayer distance $d(\bm{r})=d_0+z_{AP}(\bm{r}_0(\bm{r}))$ (top), total piezocharge density in the top- and bottom layers $\rho_t+\rho_b$ (middle), and piezoelectric potential $\varphi_t=\varphi_b\equiv\varphi$ (bottom), in the reconstructed superlattices of twisted AP-WSe$_2$ bilayers at two different twist angles $\theta_{AP}=0.6^\circ$ (left panels) and $\theta_{AP}=3^\circ$ (right panels). We used $d_0=6.71$\AA\, as the reference distance and local lateral offset $\bm{r}_0(\bm{r})=\theta_{P/AP}\hat{z}\times \bm{r} + \bm{u}^t-\bm{u}^b$ for producing the interlayer distance maps. For calculation of piezocharge densities and piezopotential we took into account screening coming from polarisation of filled bands and encapsulation in hBN (see Appendix \ref{app_electrostatics}). The piezocharge magnitudes remain the same for smaller twist angles. The scale bar in all panels is 10\,nm. On the top panels we show armchair (arm) and zigzag (zz) crystallographic axes in each layer; all maps are shown at the same crystallographic orientation.
	}
\end{figure}

For twisted AP-bilayers, the reconstructed moir\'e superlattice consists of 2H-stacking domains, each analogous to the layer alignment found in bulk crystals. These domains are separated by a network of domain walls, each of which is a full screw dislocation. The other high symmetry registries, XX$'$ and MM$'$, occupy sites of the domain wall network (Fig. \ref{Fig:AP_relaxed_bands}). 

In the moir\'e supercells of P-WSe$_2$ bilayers there are two registries, MX$'$ and XM$'$, representing the same energetically favourable layer alignment, analogous to 3R stacking in bulk crystals. This allows an easier transition into the commensurate phase for P-bilayers (see Fig. \ref{Fig:P_relaxed}), as in this case triangular MX$'$ and XM$'$ domains are separated by less energetically expensive partial-screw-dislocation-like domain walls  \cite{Enaldiev_arXiv}. 

In-plane strain $u_{ij}^{t/b}$, caused by reconstruction, induces piezoelectric charges
\begin{equation}\label{Eq:piezo_chargs}
    \rho^{t/b}=e_{11}^{t/b}\left[2\partial_xu_{xy}^{(t/b)}+\partial_y(u_{xx}^{(t/b)}-u_{yy}^{(t/b)})\right],
\end{equation}
in the top and bottom layers of P- and AP-WSe$_2$ structures, due to the lack of inversion symmetry of the individual layers (the piezocoefficient for WSe$_2$ monolayers is $\left|e_{11}^{t/b}\right|=2.03\times 10^{-10}$\,C/m \cite{rostami2018piezoelectricity}). For AP-bilayers, both layers have equal piezocharge densities as a result of a sign compensation between the piezocoefficients ($e_{11}^t = -e_{11}^b$) and the strain tensors ($u_{ij}^t=-u_{ij}^b$) of opposite layers. The latter is due to the tendency of the monolayers to deform toward each other. In Fig.\ \ref{Fig:AP_relaxed_bands} we show the distribution of  piezocharges, piezopotentials and interlayer distances in the two reconstruction regimes for AP-WSe$_2$ bilayers. At small twist angles $\theta_{AP}<\theta^*_{AP}$, the piezocharge density extrema 
appear at the corners of hexagonal 2H domains, with opposite signs in XX$'$ and MM$'$ areas. For large twist angles $\theta_{AP}\gtrsim \theta^*_{AP}$, the piezocharge and potential modulation amplitudes decay significantly. 

Because of the negligible energy cost of bending
deformations of WSe$_2$ monolayers \cite{Enaldiev_arXiv} as compared
with in-plane strain and adhesion energy variation, interlayer distance modulation, expressed by Eq.\ (\ref{Eq:interlayer_distance}) with local lateral shift $\bm{r}_0(\bm{r})=\theta_{P/AP}\hat{z}\times \bm{r} + \bm{u}^t-\bm{u}^b$, occurs in both the strong and weak reconstruction regimes, as shown in the top panels of Figs.\ \ref{Fig:AP_relaxed_bands} and \ref{Fig:P_relaxed}.

The spatial variation of the interlayer distance in twisted P- and AP- bilayers can be expressed as a Fourier series over moir\'e superlattice reciprocal vectors $\bm{g}_j$ [see Eq.\ \eqref{eq:moirebragg} below]:
\begin{equation}\label{Eq:z_local_P}
    z_{P/AP}\left(\bm{r}_0(\bm{r})\right) = z_0 + \sum_j\left[ z_j^s\cos\left(\bm{g}_j\cdot\bm{r}\right)+z_j^a\sin\left(\bm{g}_j\cdot\bm{r}\right)\right],
\end{equation}
where $$z_j^{s}+i z_j^{a}=\frac{2}{S_{\rm sc}}\int_{\rm sc} d^2\bm{r}z_{P/AP}\left(\bm{r}_0(\bm{r})\right)e^{i\bm{g}_j\bm{r}}$$
are Fourier coefficients, with $z_0\equiv z_0^s$ and $z_j^a=0$ for P-bilayers, and $S_{\rm sc}$ is the supercell area. For not too small angles ($\theta_{P/AP}\gtrsim 1^{\circ}$) the summation in Eq.\ \eqref{Eq:z_local_P} involves a few stars of the moir\'e harmonics (see Appendix \ref{sec:minibands}), simplifying the calculation of miniband structures presented in Section \ref{sec:Kminibandstogether}.

\section{Interlayer charge transfer from density functional theory modelling of P-WS\MakeLowercase{e}$_2$ bilayers}\label{sec3}

\begin{figure}
    \centering
    \includegraphics[width = 0.9\columnwidth]{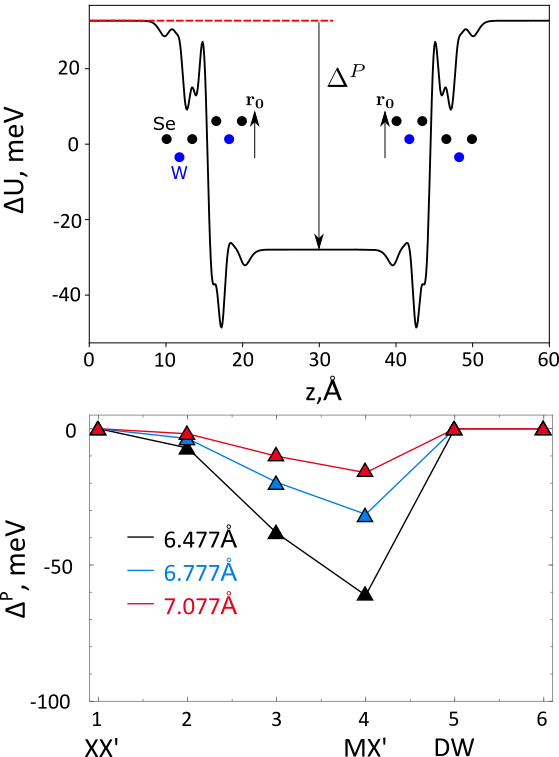}
    \caption{Top panel: Difference between the plane-averaged local potential for XM$'$-stacked bilayer WSe$_2$ with $d=6.477$~\AA~(calculated using a supercell containing two images of the bilayer $\sigma_h$ reflected relative to each other as shown in the schematic) and that from the sum of isolated monolayers. The net charge transfer between the layers gives a potential difference across the bilayer, with the majority of the potential drop taking place between the layers. Bottom panel: Dependence of the difference between the DFT-calculated vacuum potentials (triangles) $\Delta^P$ (indicated in the top panel) on interlayer distance and stacking configuration, fitted (lines) according to Eq.\ (\ref{Eq:potential_jump}).}
    \label{fig:pot_diff}
\end{figure} 

Using the interlayer adhesion model described in Ref.\ \onlinecite{Enaldiev_arXiv} and set out above, we determined the stacking patterns and interlayer distances realised in twisted WSe$_2$ bilayers. To find the resulting band energies, we constructed the model Hamiltonians presented below, fully parametrised using DFT band structure calculations for aligned bilayers with a range of local in-plane offsets $\rr_0$ and interlayer distances $d$, as discussed in detail in Appendix \ref{app:DFT_configs}.

In these DFT calculations, P-stacked bilayers were placed in a periodic three-dimensional box with a separation of 30~\AA~between the mean planes of the repeated bilayer images, to ensure that no interaction occurred between them. For P-stacked bilayers, the lack of inversion symmetry in the monolayer means that, away from certain high-symmetry configurations, layer interchange is not a symmetry operation. Therefore, it is possible for layer-asymmetric interband hybridisation to give rise to some interlayer charge transfer
for XM$'$ and MX$'$ bilayers \cite{Li2017}, resulting in a potential jump across the WSe$_2$ bilayer, for XM$'$ stacking, shown in Fig. \ref{fig:pot_diff}. We therefore construct supercells containing two P-stacked bilayers separated by a large vacuum, with the second supercell mirror-reflected with respect to the first: this avoids the need to artificially resolve the potential mismatch at the supercell boundary\cite{WeakFerroSam}.

\begin{figure}
	\includegraphics[width = \columnwidth]{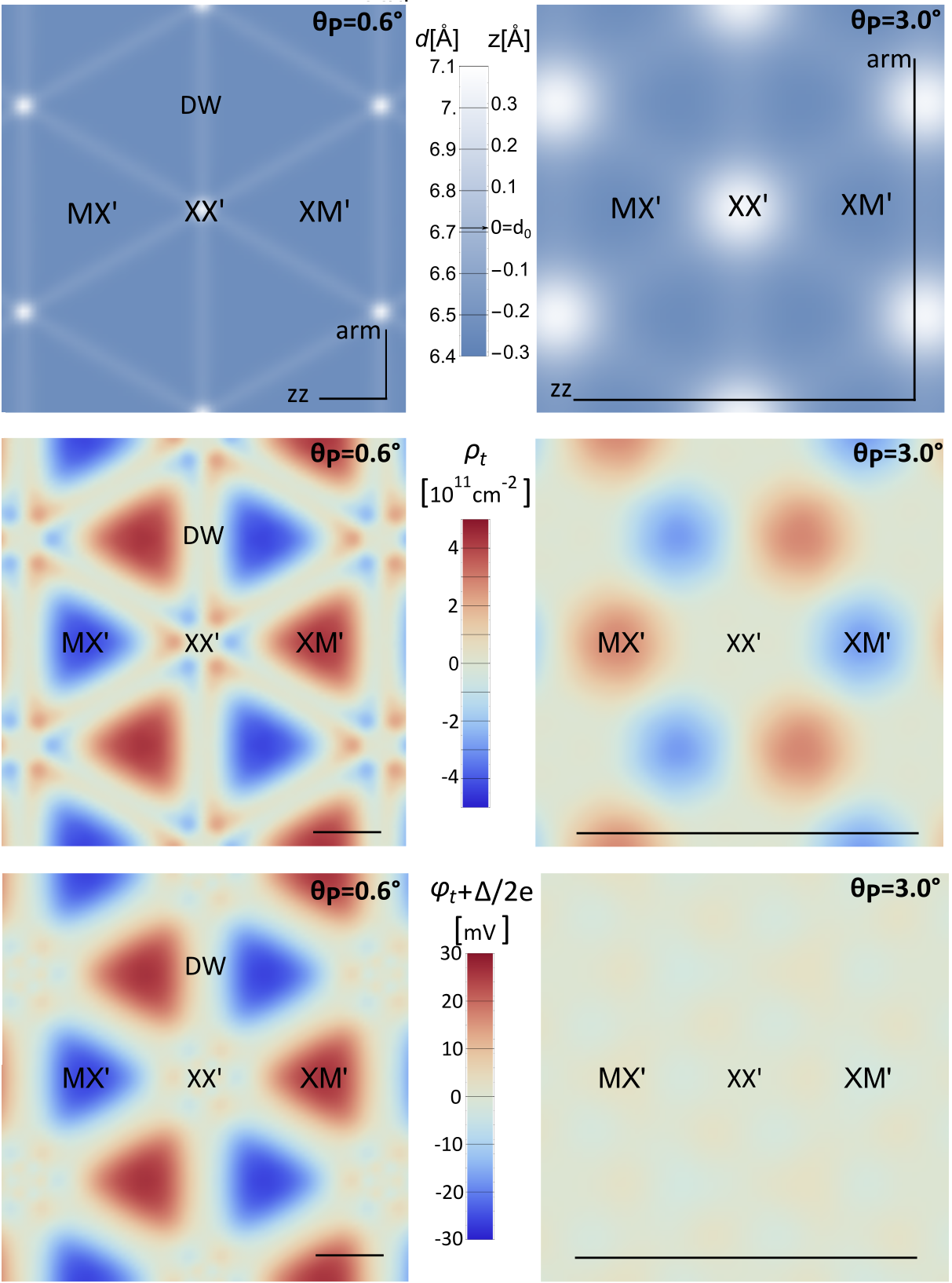}
	\caption{
		\label{Fig:P_relaxed} 
		Maps of the interlayer distance $d(\bm{r})=d_0+z_P(\bm{r}_0(\bm{r}))$ (top), sum of the piezo- and ferroelectric charge densities in top layer (middle), and their total potential (bottom) for twisted P-WSe$_2$ bilayers at different twist angles $\theta_P=0.6^\circ$ (left panels) and $\theta_{AP}=3^\circ$ (right panels). In the bottom layer the charges and potential have opposite signs. For the interlayer distance maps we used the parameters discussed in the caption of Fig.\ \ref{Fig:AP_relaxed_bands}. The scale bar is 10\,nm in all panels. The monolayer zigzag (zz) and armchair (arm) crystallographic axes are shown in the top panels; all maps are shown at the same crystallographic orientation.}
\end{figure}

The $\bm{r}_0$- and $z$-dependences of the electron potential energy jump across the WSe$_2$ bilayer can be described as 
\begin{equation}\label{Eq:potential_jump}
\Delta^{P}(\bm{r}_0,z) = \Delta_a^{P}(z)f_a\left(\bm{r_0}\right), 
\end{equation}
where the $z$-dependent function $\Delta_a(z)$ is fitted by a simple exponential (see Fig. \ref{fig:pot_diff} and Sec.\ \ref{sec:hybrydisation_Hams}A). The magnitude of the jump is maximal for XM${}'$ and XM${}'$ stackings, reaching 66\,meV. This is produced by the charge double layer located between the inner chalcogen sublayers, characterised by the areal polarisation density (in CGS units)
\begin{equation}\label{eq:PfromDelta}
    P \approx \frac{\Delta^{P}}{4\pi e}.
\end{equation}
The ferroelectric polarization is opposite in MX$'$ and XM$'$ domains, attaing a value (see Table \ref{tab_fit_splitting})
\begin{equation*}
P^{\rm XM'}=-P^{\rm MX'}=3.7\times 10^{-3} \frac{e}{\rm nm}.
\end{equation*} 

In Fig.\ \ref{Fig:P_relaxed} we show maps of interlayer distance, sum of piezo- and ferro-charge densities, and electric potential in the top layer for the two reconstruction regimes. For marginal twist angles $\theta_P<\theta^{*}_P$, ferrocharges determine the polarization of XM$'$ and MX$'$ domains, while their effect is compensated by piezocharges along domain walls, giving the domain corners a charge opposite to that of the main body. For larger twist angles $\theta_P\gtrsim\theta^{*}_P$, piezo- and ferrocharges almost completely suppress each other, leading to vanishingly small total interlayer charge polarization (see Fig.\ \ref{Fig:P_relaxed}). 

\begin{figure*}
    \centering
    \includegraphics[width = 0.9\linewidth]{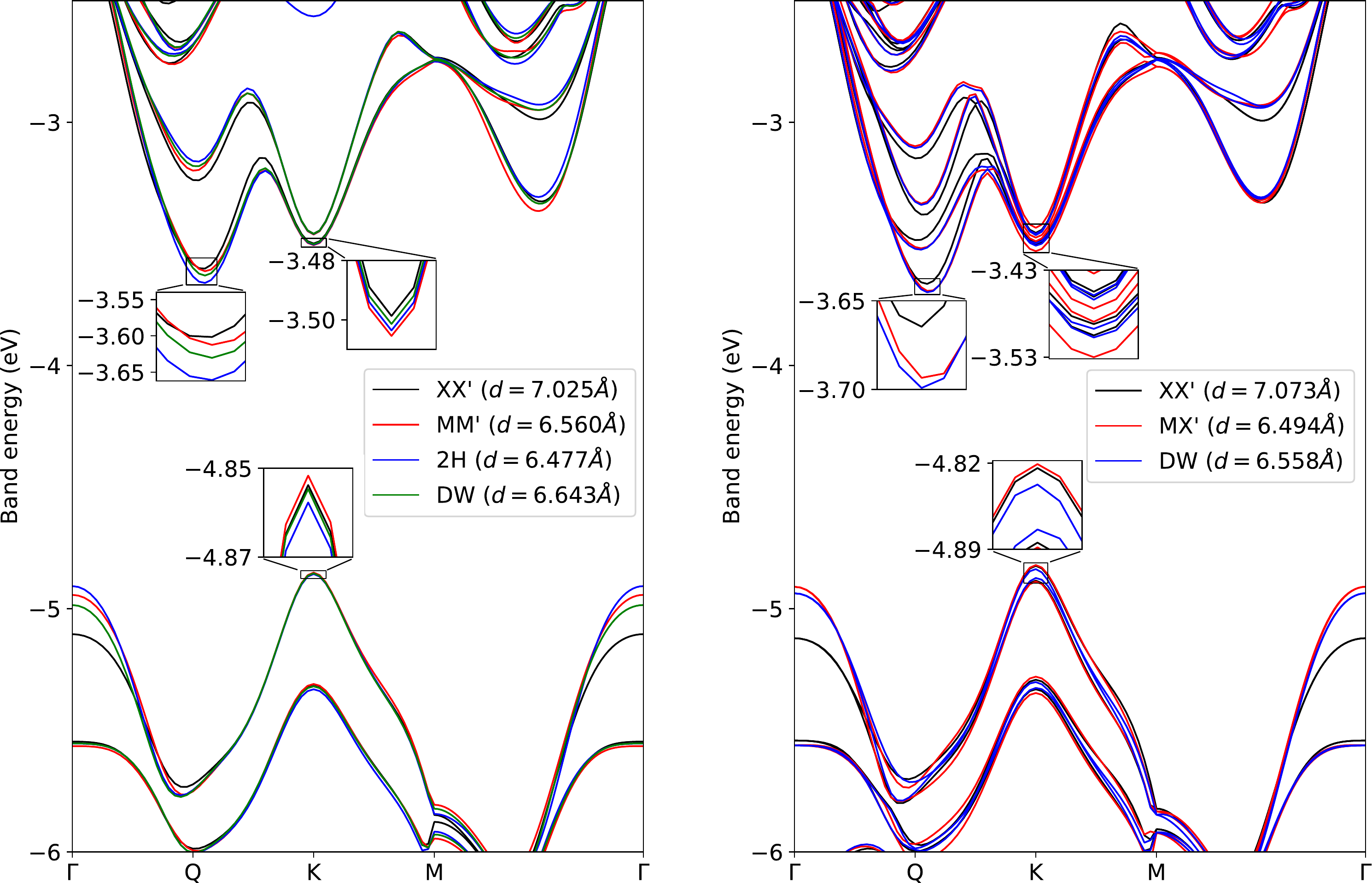}
    \caption{Band energies for high-symmetry configurations of AP- (left panel) and P- (right panel) stacked WSe$_2$ bilayers. The vacuum level is set to 0~eV (for MX$'$ stacking in the P case, the mean of the two vacuum levels at either side of the bilayer). The interlayer distances $d$ are given by Eq.\ (\ref{Eq:interlayer_distance}) with a reference distance $d_0$. The structure parameters for the monolayer are taken from experiment.}
    \label{fig:stacking_bands}
\end{figure*}

In Fig.\ \ref{fig:stacking_bands}, we show DFT-calculated bands for high-symmetry stacking configurations, choosing the optimal interlayer distance for each configuration, as set out in Sec.\ \ref{sec:adhesion}. The band energies are aligned with respect to their corresponding vacuum levels, using the mean of the two vacuum energies at either side of a P-stacked bilayer. The $\Gamma$-point valence-band edge shows strong variation with stacking because interlayer hybridisation at $\Gamma$ is strong, and therefore sensitive to changes in stacking configuration and interlayer distance. By contrast, the stacking-dependent variation of the $K$-point valence band edge is weaker, but still present.

\section{Minimal models for resonant hybridisation of conduction/valence states in bilayers}\label{sec:hybrydisation_Hams}

In a reconstructed twisted bilayer, the stacking orders and interlayer distances, and hence the interlayer hybridisation and band-edge state energies, vary continuously across the moir\'e supercell. To determine the band energies in the different regions of the twisted bilayer, it is therefore necessary to interpolate and understand the $(\mathbf{r}_0,z)$-dependence seen in the DFT results. We have developed and applied models for interlayer coupling between the relevant conduction- and valence band edge states described in the following sections, taking into account the competing effects of changes in stacking order and interlayer distance, and revealing the underlying symmetries and physical mechanisms responsible for the demonstrated behavior. Since in P-bilayers the interlayer charge transfer, induced by non-resonant hybridisation between filled valence- ang empty conduction band states, is intrinsically taken into account in DFT computations, the weak ferroelectric effect in the P-orientation is captured in the models below. Due to the non-uniform strain patterns that arise from atomic reconstruction, a complete picture of the varying band energies in the moir\'e superlattice must also include a contribution that takes piezoelectric effects into account.

In Sections \ref{sec:GammaModels} and \ref{sec:Khyb} we present the resulting resonant hybridisation Hamiltonians for $\Gamma$-point states in the valence band, and $K$-point states in the valence- and conduction bands for P- and AP-aligned bilayers. The corresponding Hamiltonian for $Q$-point states in the conduction band is discussed in Sec.\ \ref{sec:Qpointmodel}. In each of these cases we offer interpolation formulae applicable to both P- and AP orientations of the bilayers and illustrate the resulting variation of the corresponding band edges throughout the moir\'e supercells using numerically computed maps for both $\theta_{P/AP}<\theta^*_{P/AP}$  and $\theta_{P/AP}<\theta^*_{P/AP}$.

\subsection{$\Gamma$-point valence band for P- and AP-bilayers}\label{sec:GammaModels}

\begin{figure}
	\includegraphics[width = 1\linewidth]{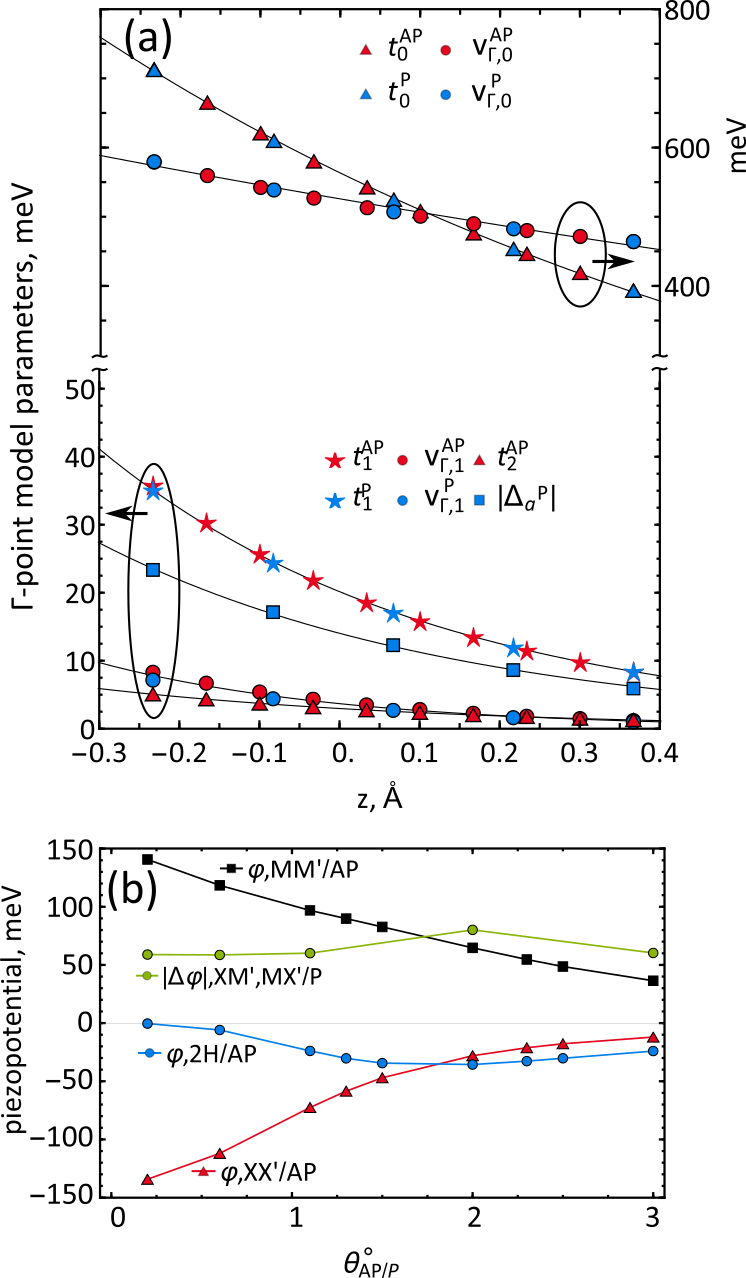}
	\caption{
		\label{Fig:Gamma_coefs_main}
		 (a) Interlayer distance dependence of the parameters in the effective Hamiltonian (\ref{Eq:effHam_Gamma}) describing hybridisation of $\Gamma$-states in P- and AP-bilayers, extracted from DFT data. The analysis shows that $t_{0,1}^{AP}(d)=t_{0,1}^P(d)$ and $v_{\Gamma;0,1}^{AP}(d)=v_{\Gamma;0,1}^P(d)$. (b) Twist-angle dependence of piezopotential (for AP) and difference of piezopotentials in two layers (for P) in various areas of the moir\'e supercell. Comparing the piezopotential magnitudes with the values of major terms in the model (\ref{Eq:effHam_Gamma}) ($t_0$ and $v_{\Gamma,0}$, shown in top panel), we conclude that accounting for the piezopotential is essential to correctly establish the form of the $\Gamma$-point valence band edge at marginal twist angles $\theta_{AP}\lesssim\theta_{AP}^*$, whereas in the weak reconstruction regime $\theta_{AP}\gtrsim\theta^*_{AP}$ the effect of piezopotential is small.   
	}
\end{figure}

Hybridisation between the local $\Gamma$-point VBM of two WSe$_2$ monolayers can be described by the following Hamiltonian: 
\begin{equation}\label{Eq:effHam_Gamma}
H^{P/AP}_{\Gamma,VB}=H_{\Gamma}^{P/AP} + \delta H_{\Gamma}^{P/AP}.
\end{equation}
Here, the dominant contribution to the coupling reads
\begin{equation}\label{Eq:H0_Gamma}
\begin{split}
H_{\Gamma}^{P/AP}=&\,\varepsilon_{\Gamma}^{P/AP}\Lambda_0 +  T_{\Gamma}^{P/AP} \Lambda_x - \frac{S^{P/AP}_{\Gamma}}{2}\Lambda_z, \\
\delta H_{\Gamma}^{P/AP} =&\,\delta\varepsilon_{\Gamma}^{P/AP}\Lambda_0  + \delta T_{\Gamma}^{P/AP}(\bm{r}_0, d)\Lambda_x
\end{split}
\end{equation}
where $\Lambda_0$ is a $2\times2$ unit matrix, and $\Lambda_{x,y,z}$ are the Pauli matrices acting on the layer subspace. The matrix elements are
\begin{equation}\label{Eq:H0_matrix_elements}
\begin{split}
\varepsilon_{\Gamma}^{P/AP}(\bm{r}_0,z)& = \varepsilon_{A'} +  v_{\Gamma,0}^{P/AP}(z),\\
T_{\Gamma}^{P/AP}(\bm{r}_0,z) & =  \frac{t_0^{P/AP}(z)}{2}+\frac{t_1^{P/AP}(z)}{2}f_s\left(\bm{r}_0\right), \\
S_{\Gamma}^{P}(\bm{r}_0,z) & = \Delta^P(\bm{r}_0,z),\,S_{\Gamma}^{AP}(\bm{r}_0,z)=0,\\
\delta\varepsilon_{\Gamma}^{P/AP}(\bm{r}_0,z) &= v_{\Gamma,1}^{P/AP}(z)f_s\left(\bm{r}_0\right), \\
\delta T_{\Gamma}^{AP}(\bm{r}_0,z) &= \frac{t_2^{AP}(z)}{2}f_a\left(\bm{r}_0\right),\, \delta T_{\Gamma}^{P}(\bm{r}_0,z) = 0.
\end{split}
\end{equation}
Here, $T_{\Gamma}^{P/AP}$ and $\delta T_{\Gamma}^{P/AP}$ describe resonant hybridisation of the monolayer states, whereas $\varepsilon_{\Gamma}^{P/AP}$ and $\delta\varepsilon_{\Gamma}^{P/AP}$ are due to coupling of one monolayer's top valence band at the $\Gamma$-point with remote bands in the opposite layer. Functions $t_{0,1}^{P/AP}(z)$ and $v_{\Gamma;0,1}^{P/AP}(z)$, characterising the interlayer distance dependence of the matrix elements, are found to be the same for P- and AP- configurations (i.e., $t_0^{P}=t_0^{AP}\equiv t_0$, $t_1^{P}=t_1^{AP}\equiv t_1$, $v_{\Gamma,0}^{P}=v_{\Gamma,0}^{AP}\equiv v_{\Gamma,0}$, $v_{\Gamma,1}^{P}=v_{\Gamma,1}^{AP}\equiv v_{\Gamma,1}$) from analysis of the DFT results (see Fig. \ref{Fig:Gamma_coefs_main}), allowing us to remove the P and AP superscripts in the following discussion. We find that these functions can be described by exponential functions, $A(z)=Ae^{-q_Az}$, and parametrise them in Table \ref{tab_fit_splitting}. In Eq.\ \eqref{Eq:H0_matrix_elements}, $\varepsilon_{A'}$ is the energy of the top valence band state at the BZ center of an isolated WSe$_2$ monolayer, and $S_{\Gamma}^P$ describes the electron energy jump due to the interlayer charge transfer introduced in the previous section. Note that this term vanishes in AP-bilayers as they are centrosymmetric.

Decomposition of $H^{P/AP}_{\Gamma,VB}$ into $H^{P/AP}_{\Gamma}$ and $\delta H^{P/AP}_{\Gamma}$ reflects the hierarchy of these two contributions to the model. Comparing $t_{0}$ and $t_1$ with $t_2^{AP}$, and $v_{\Gamma,0}$ with $v_{\Gamma,1}$, shown in Fig.\ \ref{Fig:Gamma_coefs_main}(a), we conclude that 
\begin{equation}
    \begin{split}
    t_{0}\gg t_{1}\gg t_{2}^{AP},\\
    v_{\Gamma,0}\gg v_{\Gamma,1}.
    \end{split}
\end{equation}
 Therefore,  $H^{P/AP}_{\Gamma}$ gives the dominant effect of interlayer hybridisation, while $\delta H^{P/AP}_{\Gamma}$ characterises only its fine features (see Appendix \ref{app:hamiltonianG}). For this reason, it is the variation of interlayer distance across the moir\'e supercell that is mainly responsible for the position dependence of the $\Gamma$-point band edge energy in the supercell of twisted WSe$_2$ bilayers (see Section \ref{sec:Twisted}) 
\footnote{The final expression in Eqs.\ \eqref{Eq:H0_matrix_elements} follows from even parity of the Hamiltonian (\ref{Eq:effHam_Gamma}) with respect to $\bm{r}_0\to-\bm{r}_0$, for P-aligned bilayers (see Appendix \ref{app:hamiltonianG}). We note here that for AP-bilayers the asymmetry of the Hamiltonian (\ref{Eq:effHam_Gamma}) with respect to $\bm{r}_0$ is only expressed in a small term $\propto t_2^{AP}$.}.

\begin{table}
	\caption{
		{Fitting parameters for interlayer-distance-dependent functions in Hamiltonian (\ref{Eq:effHam_Gamma}) for $\Gamma$-point state hybridisation (each function was fitted by $Ae^{-q_Az}$), and ferroelectric parameters for preferential stacking domains. \label{tab_fit_splitting}
		}
	}
	\begin{center}
	\begin{tabular}{P{2.0cm} Q{1.3cm}@{.}R{1.7cm} Q{1.3cm}@{.}R{1.7cm}}
		\hline\hline
		&  \multicolumn{2}{c}{$A$} & \multicolumn{2}{c}{$q_A$, \AA$^{-1}$} \\
		\hline
		\hline
		$t_0$ &  0&5\,eV  & 1&0\\ 

		$t_1$  & 20&3\,meV  & 2&4\\ 

		$t_2^{AP}$  & 2&1\,meV  & 2&3 \\ 

		$v_{\Gamma,0}$  & 0&5\,eV &  0&4 \\ 

		$v_{\Gamma,1}$ & 3&3\,meV & 3&3 \\ 
		$\Delta_a$ & 14&1\,meV & 2&2 \\ 
		$\varepsilon_{A'}$ & -5&8\,eV & N&A. \\
		\hline
		\hline
		\multicolumn{5}{c}{$\Delta^{\rm MX'}=-\Delta^{\rm XM'} = 66.0\,{\rm meV}$}\\
		\multicolumn{5}{c}{$P^{\rm MX'}=-P^{\rm XM'} = 3.7\times 10^{-3}\,e/{\rm nm}$}\\
		\hline
		\hline
	\end{tabular}
	\end{center}
\end{table}

\subsection{Interlayer hybridisation at the $K$-point}\label{sec:Khyb}

Unlike the spin-degenerate $\Gamma$-point states considered above, the valence and conduction states of monolayers at the $K$-points are split by the atomic spin-orbit (SO) interaction. This leads to spin-valley locking of the $K$-valley states \cite{PRL_Xiao} and, consequently, to different hybridisation between them in P- and AP-bilayers. For P-bilayers, the local valence- and conduction band edge states at the $\tau K$-point ($\tau=\pm 1$) are formed by resonantly coupled monolayer states with spin projection $s=-\tau$ and $s=\tau$, respectively, whereas for AP-bilayers hybridisation at the $\tau K$-point is off-resonance, because same-spin valence and conduction band-edge states in opposite layers are shifted in energy due to the SO splitting of monolayer states.

Nonetheless, an effective Hamiltonian describing hybridisation of the monolayer band-edge states in the $\tau K$-point of P/AP-bilayers can be represented in the form of Eq.\ (\ref{Eq:effHam_Gamma})  ($\alpha = {\rm CB} ,\,{\rm VB} $ for conduction- and valence band, respectively):
\begin{equation}\label{eq:HKgeneral}
\begin{split}
H_{\alpha,\tau K}^{P/AP}=&\,\varepsilon_\alpha^{P/AP}\Lambda_0-\frac{S_\alpha^{P/AP}}{2}\Lambda_z\\
 &+ \frac{T_{\alpha,\tau}^{P/AP}}{2}\Lambda_+ +\frac{T_{\alpha,\tau}^{P/AP*}}{2}\Lambda_- .
\end{split}
\end{equation}
Here, $\Lambda_{\pm}=\Lambda_x\pm i\Lambda_y$, and the matrix elements are
\begin{equation}\label{eq:allmelemsK}
\begin{split}
   T_{VB,\tau}^P(\rr_0,z)=& t^P_{VB}(z)T_\tau(x_0,y_0),\\
   T_{\rm VB,\tau}^{AP}(\rr_0,z)=& t_{VB}^{AP}(z)T_\tau(x_0,y_0+\tfrac{a}{\sqrt{3}}),\\
   T_{CB,\tau}^P(\rr_0,z)=& t^P_{CB}(z)T_\tau(x_0,y_0),\\
   T_{\rm CB,\tau}^{AP}(\rr_0,z)=& t^P_{CB}(z)T_\tau(x_0,y_0),\\
   S_{\rm CB/VB}^P(\rr_0,z) =& \Delta^P(\rr_0,z), \\
   S_\alpha^{AP}(\rr_0,z)=& \lambda_\alpha\Big[\Delta_{\alpha}^{SO} + \tilde{\Delta}_{\alpha,1}^{SO}(z)f_s\left(\mathbf{r}_0 \right)\\
   & +\tilde{\Delta}_{\alpha,2}^{SO}(z)f_a\left(\mathbf{r}_0 \right)\Big],\\
	\varepsilon_{\alpha}^{P/AP}(\rr_0,z)=& \varepsilon_{\alpha} - v_{0}(z) - v_{\alpha,1}^{P/AP}(z)\Big[\cos(\chi_{P/AP})f_s(\bm{r}_0)\\
	 & - \lambda_{\alpha}\sin(\chi_{P/AP})f_a(\bm{r}_0)\Big],
\end{split}    
\end{equation}
where $\chi_P=0$, $\chi_{AP}=\pi/4$, $\lambda_{\rm VB}=1$, $\lambda_{\rm CB}=-1$, and we have defined the function
\begin{equation*}
    T_\tau(x,y) = e^{i\tau\tfrac{4\pi x}{3a}}+2e^{-i\tau\tfrac{2\pi x}{3a}}\cos{\left(\tfrac{2\pi y}{a\sqrt{3}}\right)}.
\end{equation*}

\begin{figure}
    \centering
    \includegraphics[width=\columnwidth]{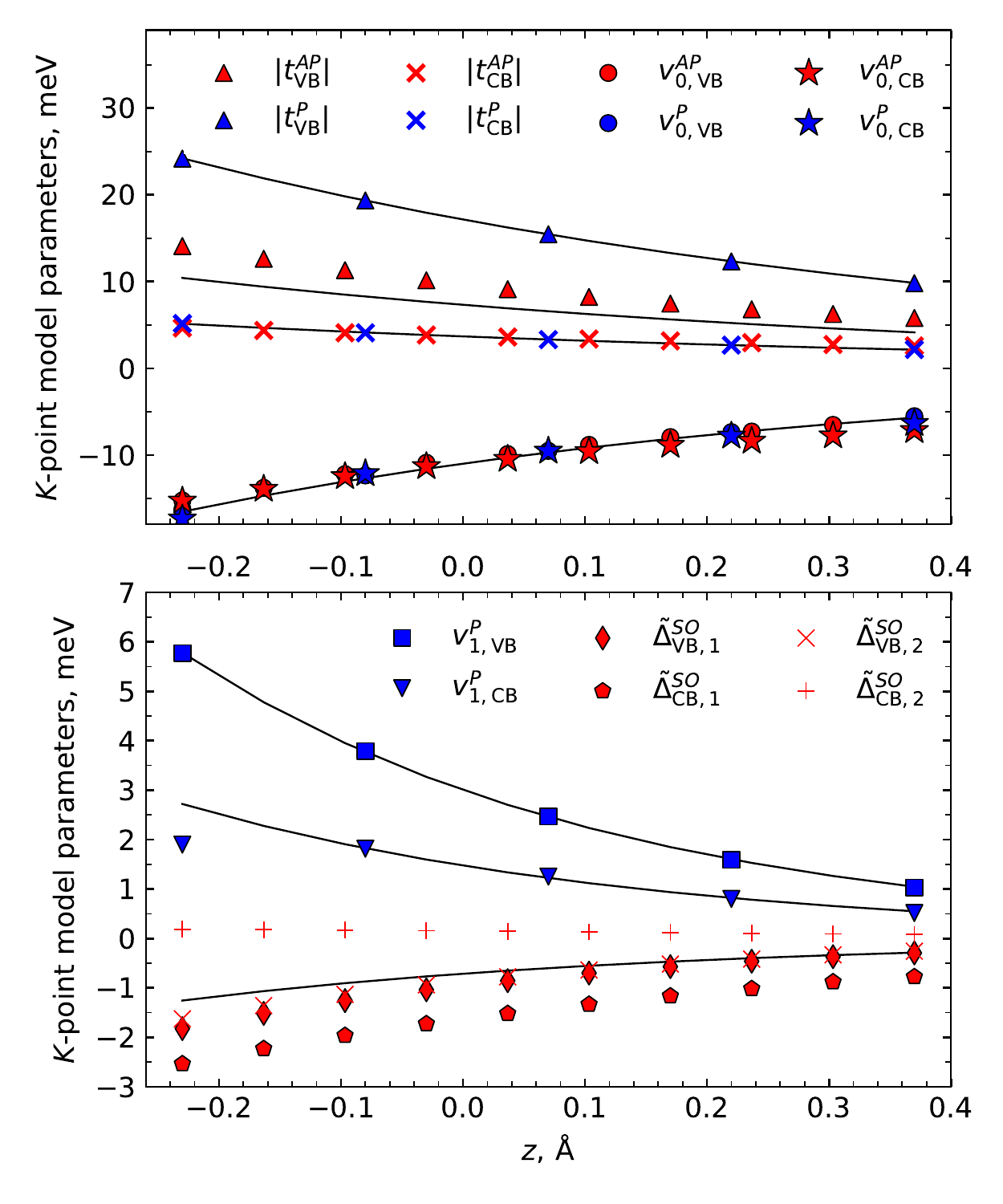}
    \caption{Interlayer distance dependence of the parameters in the effective Hamiltonian \eqref{eq:HKgeneral} describing hybridisation of the highest valence and lowest conduction bands at the $K$ point. DFT results are shown with symbols, whereas lines show the interpolation described in Table \ref{tab:KpointAll}, with the definition $\tilde{\Delta}_{\rm CB,1}^{SO}=\tilde{\Delta}_{\rm VB,1}^{SO}=\tilde{\Delta}_{\rm VB,2}^{SO}\equiv \tilde{\Delta}_{1}^{SO}$.}
    \label{fig:KfitAll}
\end{figure}

In Eq.\ \eqref{eq:allmelemsK}, $t_\alpha^{P/AP}(z)$ is the tunnelling parameter between bands $\alpha={\rm CB,\,VB}$ of the two layers. For P-bilayers, $S_\alpha^P$ accounts for the potential energy drop caused by the interlayer charge transfer, while for AP-bilayers $S_\alpha^{AP}$ represents the SO splitting in the corresponding band, containing the monolayer SO splitting $\Delta_\alpha^{SO}$, as well as small $z$-dependent corrections $\tilde{\Delta}_1^{SO}$ and $\tilde{\Delta}_{\alpha,2}^{SO}$. The DFT analysis displayed in Fig.\ \ref{fig:KfitAll} shows that $|t_{CB}^P|=|t_{CB}^{AP}|\equiv |t_{CB}|$, and also that $\tilde{\Delta}_{\rm CB,1}^{SO}=\tilde{\Delta}_{\rm VB,1}^{SO}=\tilde{\Delta}_{\rm VB,2}^{SO}\equiv \tilde{\Delta}_{1}^{SO}$ and  $|\Delta_{1}^{SO}|,|\Delta_{CB,2}^{SO}|\ll|\Delta^{SO}_{CB/VB}|$. In the last line of Eq.\ \eqref{eq:allmelemsK}, $\varepsilon_{\rm CB}$ and $\varepsilon_{\rm VB}$ are the monolayer $K$-point conduction- and valence band edge energies, whereas $v^{P/AP}_{\rm CB/VB,0}$ and $v^{P/AP}_{\rm CB/VB,1}$ take into account hybridisation with remote bands. Through comparison with the DFT results, we also find that $v^{P}_{0,\rm CB}=v^{AP}_{0,\rm CB}=v^{P}_{0,\rm VB}=v^{AP}_{0,\rm VB}\equiv v_0$, see Fig.\ \ref{fig:KfitAll}. Based on DFT computation, we find that all $z$-dependent functions that appear in Eq.\ (\ref{eq:allmelemsK}) can be described using exponential functions, $A(d)=Ae^{-qz}$, with the parameter values listed in Table \ref{tab:KpointAll}.

{\setlength{\tabcolsep}{0.1\columnwidth}
\setlength\extrarowheight{2pt}
\renewcommand{\arraystretch}{1.2}
\begin{table}[b]
	\caption{
		{Fitting parameters for the interlayer-distance-dependent functions in the Hamiltonian \eqref{eq:HKgeneral} for $K$-point state hybridisation. Each function was fitted as $A(d)=Ae^{-q_A z}$.\label{tab:KpointAll}
		}
	}
	\begin{tabular}{l r@{.}l r@{.}l}
		\hline\hline
		&  \multicolumn{2}{c}{$A$} & \multicolumn{2}{c}{$q$, \AA$^{-1}$} \\
		\hline
		\hline
		$|t_{\rm CB}|$         &  3&7 meV      & 1&5 \\
		$|t_{\rm VB}^P|$       &  17&2 meV     & 1&5 \\
		$|t_{\rm VB}^{AP}|$    &  9&8 meV      & 1&5 \\
		$ \tilde{\Delta}_{1}^{SO}$   &  -1&14 meV     & 2&5 \\
		$ \tilde{\Delta}_{{\rm CB},2}^{SO}$   &   0&1 meV     & 1&3 \\
		
		$v_{0}$                           &  11&0 meV    & 1&8 \\
		$v_{{\rm CB},1}^P$                &  1&5 meV     & 2&7 \\
		$v_{{\rm VB},1}^P$                &  3&0 meV     & 2&9 \\
		$v_{{\rm CB},1}^{AP}$             &  0&5 meV      & 2&7 \\
		$v_{{\rm VB},1}^{AP}$             &  1&2 meV      & 3&1 \\
		$\Delta_{\rm CB}^{SO}$ &  41&0 meV     & N&A. \\
		$\Delta_{\rm VB}^{SO}$ &  459&0 meV     & N&A. \\
		$\varepsilon_{\rm CB}$ &  -3&491 eV    & N&A. \\
		$\varepsilon_{\rm VB}$ &  -4&846 eV    & N&A. \\
		\hline
		\hline
	\end{tabular}
\end{table}
}

\subsection{Interlayer hybridisation for conduction band states at the $Q$-points}\label{sec:Qpointmodel}

The conduction band of TMD crystals possesses additional minima at the six inequivalent $Q$-points of the Brillouin zone, located approximately half way between the $K$- and $\Gamma$-points. For TMD monolayers, the global conduction band minimum is at the $K$-point. However, for TMD homobilayers the larger interlayer hybridisation between $Q$-point states may shift the global conduction band minima to the $Q$-valleys. By constrast to the $K$- and $\Gamma$-valleys, hybridisation of $Q$-valley states is anisotropic. In this section we introduce a model Hamiltonian describing interlayer hybridisation of Q$_1$-valley states ($\mathbf{Q}_1\approx\mathbf{K}/2$), and provide the rules for applying the model to the five remaining $Q$-valleys: $-\mathbf{Q}_1$, $\pm\mathbf{Q}_2 =\pm  C_3\mathbf{Q}_1$ and $\pm\mathbf{Q}_3 = \pm C_3^2\mathbf{Q}_1$. The effective model reads     
\begin{equation}\label{Eq:Q_point_model}
	\begin{split}
	    H^{P/AP}_{Q}=	\varepsilon^{P/AP}_{Q}\Lambda_0 - \frac{S^{P/AP}_Q}{2}\Lambda_z \\
	    + T^{P/AP}_{Q}\Lambda_{+} + T^{P/AP*}_{Q}\Lambda_{-}.
	\end{split}
\end{equation}
Here, the matrix elements are expressed as follows:
\begin{align}\label{Eq:Q_tunneling_matrix}
	T_{Q}(\bm{r}_0) & = |t_0| + |t_1|e^{-i\bm{G}_1\cdot\bm{r}_0+i\phi_1} + |t_2|e^{i\bm{G}_2\cdot\bm{r}_0+\phi_2} \nonumber\\ 
	&+ |t_{3}|e^{i\bm{G}_3\cdot\bm{r}_0+i\phi_{3}}+|t_{3}|e^{-i\bm{G}_3\cdot\bm{r}_0-i\phi_{3}},\\
	\varepsilon_{Q}(\bm{r}_0) & = \varepsilon_Q + v_{0} \nonumber \\
	&+ \sum_{j=1,2,3}\left[v_{j}^s\cos(\bm{G}_j\cdot\bm{r}_0)+v_{j}^a\sin(\bm{G}_j\cdot\bm{r}_0)\right], 
\end{align}
where we suppressed the $P/AP$ superscript in every term to shorten notations. The rest of the matrix elements are
\begin{align}\label{Eq:Q_tunneling_matrix_2}
	S^P_Q & = \Delta^Q_a\sum_{j=1,2,3}\sin(\bm{G}_j\cdot\bm{r}_0).\\
	S^{AP}_Q & = \Delta^Q_{SO}.
\end{align}
To fit the parameters of the $Q$-point model \eqref{Eq:Q_point_model}, which are gathered in Table \ref{Tab:Q_tab}, we used an additional set of configurations (lateral offsets) as in Ref.\ \cite{APLarxiv}. The matrix element $|t_0|$ gives the dominant contribution to resonant hybridisation, whereas the $|t_{1,2,3}|$-terms are necessary to describe the stacking-dependent variation of the $T_{Q}(\bm{r}_0)$ matrix element. $\varepsilon_{Q}(\bm{r}_0)$ characterises hybridisation with remote bands having odd terms ($\propto\sin(\bm{G}_{1,2,3}\cdot\bm{r}_0)$) only for AP-bilayers. We also mention that $Q$-point states are formed by a mixture of the orbital species forming the the band edges at the $K$- and $\Gamma$-valleys \cite{kormanyos2015k}. Therefore, the amplitude of the $S^P$ term, describing the potential jump for $Q$-point states due to interlayer charge transfer, slightly differs from that of the hybridisation Hamiltonians for the $K$- and $\Gamma$-valleys. For the point $-\mathbf{Q}_1$, related to $\mathbf{Q}_1$ by time reversal symmetry, the hybridisation Hamiltonian is the complex conjugate of Eq.\ \eqref{Eq:Q_point_model}, whereas the Hamiltonians for the $\pm 120^{\circ}$- rotated $Q$-points ($\mathbf{Q}_2$ and $\mathbf{Q}_3$, respectively) can be obtained by applying the corresponding $\pm 120^{\circ}$ rotation to the reciprocal vectors $\bm{G}_{1,2,3}$ in all the matrix elements.

\begin{table}
	\caption{Fitting parameters for the interlayer-distance-dependent functions in the Hamiltonian \eqref{Eq:Q_point_model} for $Q$-valley. Each function was fitted as $Ae^{-qz}$, while $\varphi$ is the phase of corresponding parameter. $\Delta^Q_{SO}\approx 0.214$\,eV. \label{Tab:Q_tab}}
	\begin{tabular}{cccc}
        \hline\hline
		 & $A$, meV & $q$, \AA$^{-1}$ &  $\varphi$ \\
		\hline
		\hline
		$|t_0^P|/|t_0^{AP}|$ & 168/179.5 & 0.69/0.66 & N/A   \\
		$|t_1^P|/|t_1^{AP}|$ & 11.3/6.6 & 2.2/1.6 & 0/-0.46$\pi$ \\
		$|t_2^{P}|/|t_2^{AP}|$ & 11.3/6.6 & 2.2/1.6 & 0/0.46$\pi$ \\
		$|t_{3}^{P}|/|t_{3}^{AP}|$ & 2.4/2.4 & 1.98/1.98 & 0/0.3$\pi$ \\
		$v_1^{(s)P}/v_1^{(s)AP}$ & -3.8/-5 & 2.9/2.45 & N.A.\\
		$v_2^{(s)P}/v_2^{(s)AP}$ & -3.8/-5 & 2.9/2.45 & N.A. \\
		$v_3^{(s)P}/v_3^{(s)AP}$ & -1.8/1 & 2.3/0.8 & N.A. \\
		$v_1^{(a)P}/v_1^{(a)AP}$ & N.A./2.5 & N.A./3.45  & N.A. \\
		$v_2^{(a)P}/v_2^{(a)AP}$ & N.A./2.5 & N.A./3.45 & N.A. \\
		$v_3^{(a)P}v_3^{(s)AP}$ & N.A./2.4 & N.A./2.77 & N.A. \\
		$v_0^P/v_0^{AP}$ & -1.4/-3.3 & 2.0/2.5 & N.A. \\
		$\varepsilon_Q^{P}/\varepsilon_Q^{AP}$ & $-3.52$/$-3.41$ $\,\times 10^3$ & N.A. & N.A.\\
		$\Delta^{Q}_a$ & 21.5 & 2.3 & N.A.\\
		\hline\hline
	\end{tabular}
\end{table}

\section{Band edge maps for twisted AP-bilayers}\label{sec:Twisted}

In this section we combine our results on atomic reconstruction in twisted AP-bilayers
with the interlayer hybridisation models introduced in Section \ref{sec:hybrydisation_Hams}. This analysis is performed separately for $\Gamma$-point states in the valence band, $K$-point states in the conduction- and valence bands, and $Q$-point conduction states. In particular, we identify where in the moir\'e supercell the minima for conduction band electrons and valence band holes would appear,
and produce confinement profiles of quantum dot potentials for each of these specific areas.

\subsection{Modulation of the valence band edge at the $\Gamma$-point}\label{sec:ModGamma}

To apply the model (\ref{Eq:effHam_Gamma}) to a twisted bilayer, we relate the local stacking vector $\rr_0$ at position $\rr$ to the twist angle $\theta_{P/AP}$ as $\bm{r}_{0}(\bm{r})=\theta_{P/AP} \hat{z}\times\bm{r}+\bm{u}^t-\bm{u}^b$, using also the local interlayer distance shift $z_{AP}(\bm{r})$ in the model parameters, and supplementing the diagonal matrix elements with the electron piezoelectric potential energy, $-e\varphi$, equal in the top and bottom layers due to inversion symmetry. As a result, the spatial modulation of the top valence band energy at the $\Gamma$-point is expressed as
\begin{equation}\label{Eq:effHam_Gamma_with_piezo}
E_{\Gamma}^{AP}(\bm{r}) = -e\varphi(\bm{r}) + \varepsilon_{\Gamma}(\bm{r})+\delta \varepsilon_{\Gamma}(\bm{r}) + \left|T_{\Gamma}(\bm{r}) + \delta T_{\Gamma}(\bm{r})\right|,
\end{equation}
where the notation,
\begin{align}
\begin{split}
\label{Eq:local_stacking_dependence}
\varepsilon_{\Gamma}(\bm{r})&=\varepsilon_{\Gamma}\left[\bm{r}_0(\bm{r}),z_{AP}(\bm{r})\right],\\ \delta\varepsilon_{\Gamma}(\bm{r})&=\delta \varepsilon_{\Gamma}\left[\bm{r}_0(\bm{r}),z_{AP}(\bm{r})\right], \\ T_{\Gamma}(\bm{r})&=T_{\Gamma}\left[\bm{r}_0(\bm{r}),z_{AP}(\bm{r})\right],\\
\delta T_{\Gamma}(\bm{r})&=\delta T_{\Gamma}\left[\bm{r}_0(\bm{r}),z_{AP}(\bm{r})\right],
\end{split}
\end{align}
is used to describe the dependence of the matrix elements on local stacking and interlayer distance within the moir\'e supercell. Therefore, to determine the position of the $\Gamma$-point valence band edge in twisted bilayers at $0^{\circ}<\theta_{AP}\leq 4^{\circ}$ we take into account both in-plane reconstruction (inducing the piezopotential) and relaxation of interlayer distances (affecting the hybridisation magnitude).

For marginal twist angles $\theta_{AP}\lesssim 1^{\circ}$, the $\Gamma$-point valence band edge is located at three equivalent corners (labeled 2H$_{\rm c}$ in Fig. \ref{Fig:AP_Top_valence_band}) of 2H-domains around XX$'$ areas, which is due to the superposition of the largest splitting in 2H domains with the highest piezopotential energy  (see inset in Fig. \ref{Fig:AP_relaxed_bands}). For twist angles $\theta_{AP}> 1^{\circ}$ the piezopotential amplitude decays, and the valence band edge shifts toward the middle of the 2H domains (Fig. \ref{Fig:AP_Top_valence_band}).  Thus, at marginal twist angles the valence band edge at $\Gamma$ forms triple quantum dots for holes around the XX$'$ corner of the 2H domains.

\subsection{Modulation of the valence band edge at the $K$ point}\label{sec:KpointAPQD}

\begin{figure}
	\includegraphics[width = \linewidth]{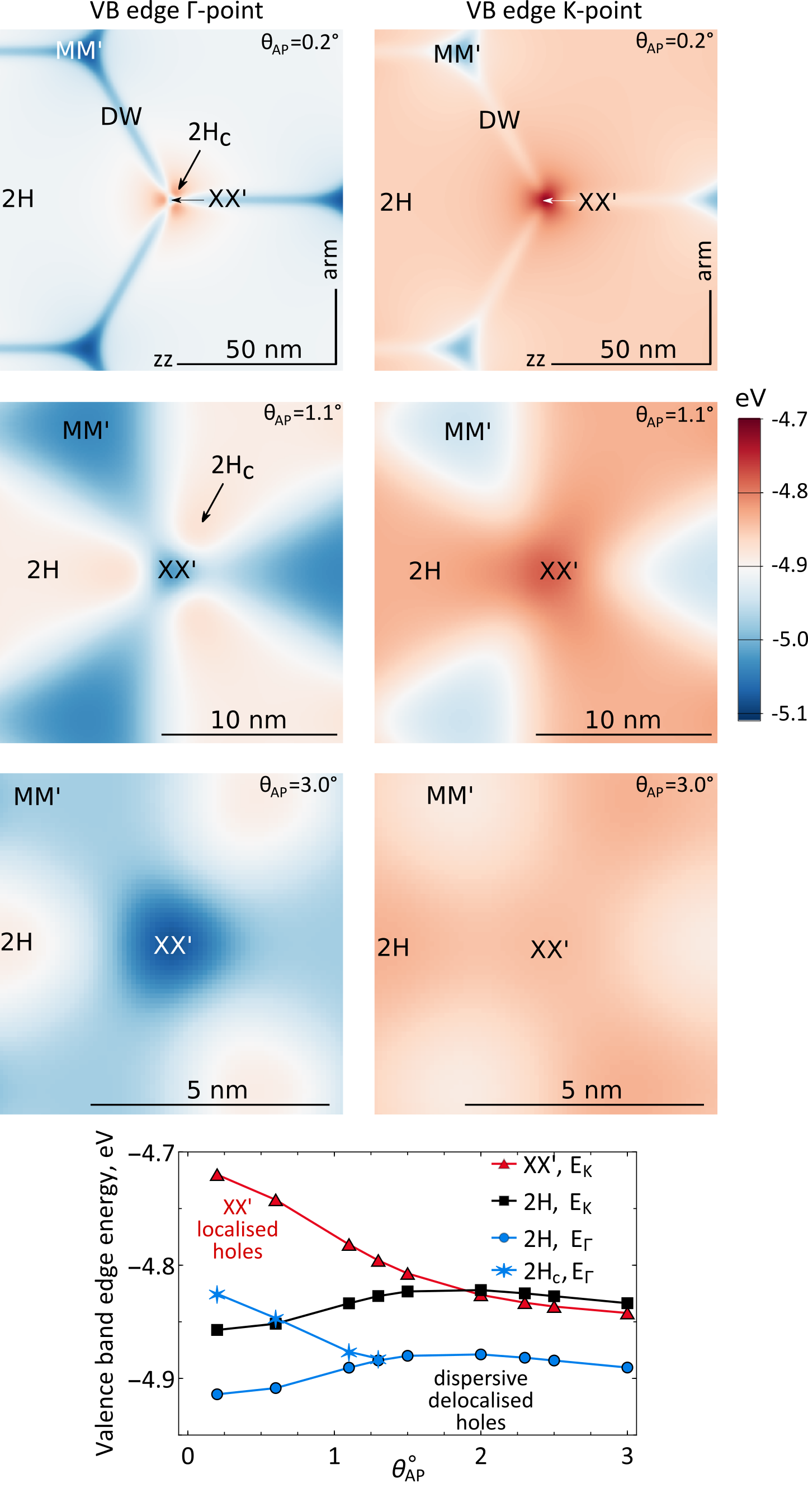}
	\caption{
		\label{Fig:AP_Top_valence_band}
		  Maps of the local $\Gamma$- (left column) and $K$-point (right column) valence band (VB) edges across twisted AP-WSe$_2$ bilayers for the labeled twist angles. The vacuum level is set to 0~eV. At marginal twist angles $\theta_{AP}<1^{\circ}$ the $\Gamma$-point band edge is around the corners of 2H domains labeled by 2H$_c$. This is prescribed by the combined effects of the piezopotential, which is strongest at the XX$'$ domain corners (see Fig.\ \ref{Fig:AP_relaxed_bands}), and the interlayer coupling, which is strongest in 2H domains. For $\theta_{AP}\gtrsim 1^{\circ}$ the $\Gamma$-point band edge is more homogeneously distributed inside 2H domains because of the weaker contribution from the piezopotential. The $K$-point band edge position is mainly determined by the piezopotential, as the interlayer coupling of the $K$-states is an order of magnitude smaller than that of the $\Gamma$-states. Zigzag and armchair crystallographic directions in constituent layers marked as zz and arm, respectively, on the top panels, are the same for all maps. The bottom panel shows the twist-angle dependence of the $\Gamma$- and $K$ valence band edges in high symmetry regions of the moir\'e supercell.     
	}
\end{figure}

\begin{figure}
    \centering
    \includegraphics[width=1.0\linewidth]{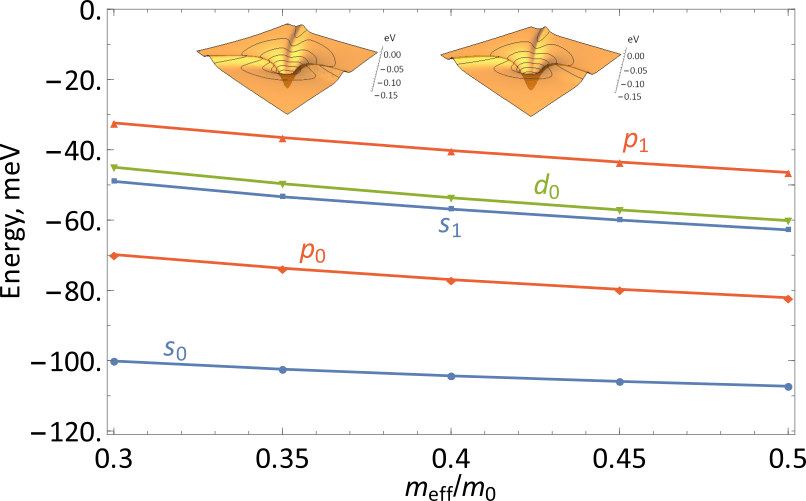}
    \caption{\label{Fig:QD_potential} Lowest energy levels of localised hole states in the quantum dot formed in XX$'$ areas of marginally twisted AP-bilayers. The energies, counted from continuum in 2H domains, results from solution of Eq. (\ref{Eq:QD_equation}) and are labeled by orbital momentum $s, p, d \dots$ for $|l|=0,1,2,\dots$, with subscript indicating radial quantum number $n_r=0,1,2,\dots$. $p$ and $d$ states are double degenerate in the sign of orbital momentum. On insets we show that discrepancy between calculated (left) and fitted (right) quantum dot potentials does not exceed 5\,meV in the whole vicinity of XX$'$ area.}
\end{figure}

Applying the rules described in the previous section to Eq.\ (\ref{eq:HKgeneral}), we obtain the following expression for the $K$-point valence band edge variation in the moir\'e superlattice:
\begin{equation}\label{Eq:Kv_edge_AP_with_piezo}
E_{\tau\mathbf{K}}^{AP}(\bm{r})=\varepsilon_{VB}^{AP}(\mathbf{r}) - e\varphi(\bm{r}) + \sqrt{\left|T_{VB,\tau}^{AP}(\mathbf{r})\right|^2 +\frac{\left(\Delta^{SO}_{VB}\right)^2}{4} },
\end{equation}
where we have used the same shorthand notation for local matrix elements as in Eq.\ \eqref{Eq:local_stacking_dependence}. Since interlayer coupling of $K$-point states is an order of magnitude weaker than that for $\Gamma$-point states, the piezopotential energy plays a key role in establishing the $K$-point valence band edge (Fig. \ref{Fig:AP_Top_valence_band}). For small twist angles $\theta_{AP}\lesssim 1^{\circ}$, the $K$-point valence band maximum represents attractive quantum dots for holes located at XX$'$ stacking regions, with a depth exceeding $100$\, meV at marginal twist angles (see inset in Fig. \ref{Fig:AP_Top_valence_band}). By contrast, for larger twist angles $\theta_{AP}\gtrsim 1.5^{\circ}$ the valence band edge shifts toward the 2H domains, following the minima of the piezopotential (Fig. \ref{Fig:AP_relaxed_bands}). 

For the whole range of twist angles, the energy of the valence band edge at $K$ is higher than that at $\Gamma$. Therefore, for marginal twist angles ($\theta_{AP}\lesssim  1^{\circ}$) the band edge will be dominated by hole states localised in QDs at XX$'$ areas. Due to the large intralayer SO splitting \cite{zeng2013optical} of $K$-point states in WSe$_2$ monolayers (Table \ref{tab:KpointAll}), the quantised states will belong to the higher spin-split band  at the $\tau\mathbf{K}$ valley [see Eq.\ (\ref{Eq:Kv_edge_AP_with_piezo})]. To compute the quantum dot states, we solve the Schr\"{o}dinger equation
\begin{equation}\label{Eq:QD_equation}
\left[\frac{\hat{\bm{p}}^2}{2m_{\rm VB}} - E^{AP}_{\tau\mathbf{K}}(\bm{r})\right]\Psi=E\Psi,
\end{equation}
where $m_{\rm VB}>0$ is the monolayer valence band effective mass at the $K$-point, and $\hat{\bm{p}}=-i\hbar(\partial_x,\partial_y)$. We approximate the hole potential energy around the XX$'$ region as
\begin{equation}\label{Eq:XX_band_edge_AP}
\begin{split}
E^{AP}_{\tau\mathbf{K}}(\bm{r})\approx& -V(r) - V_3(r,\phi),\\  
    V(r)=&\frac{V_0}{1+\frac{r^2}{\rho_0^2}}, \\
    V_3(r,\phi)=&\frac{V_1-V_2\sqrt{\beta^2+\frac{r^2}{l^3}\left[\rho_1-\frac{r\cos3\phi}{\sqrt{1+\frac{r^2}{\rho_1^2}}}\right] }}{\left(1+\frac{r^2}{l^3}\left[\rho_1-\frac{r\cos3\phi}{\sqrt{1+\frac{r^2}{\rho_1^2}}}\right]\right)^{3/2}},
\end{split}
\end{equation}
where $V$ is the axial-symmetric part of the potential, and $V_3$ describes trigonal warping\footnote{The form of $V_3$ is chosen to describe at large distances the potential of three 120$^{\circ}$-rotated domain walls intersecting at the same XX$'$ region. For $r\gg\rho_1$ and $|\delta\phi|=|\phi-2\pi n/3|\ll 1$ ($n=1,2,3$), we can characterize the potential cross-section of a single domain wall as
$V_3\approx \left(V_1-V_2\sqrt{\beta^2+9\xi^2/2l^2}\right)/\left(1+9\xi^2/2l^2\right)^{1.5} $, where $\xi=r\delta\phi$. Then, the relation between parameters $V_{1,2}$ and $\beta$ is governed by the condition that the total potential across a domain wall vanishes, \emph{i.e.}, $\int_{-\infty}^{+\infty} V_3d\xi=0$.}, $r=\sqrt{x^2+y^2}$ is the in-plane distance measured from the middle of the XX$'$ area, and $\phi$ is the polar angle. From fitting, we find that $V_0=-155$\,meV, $\rho_0=4.3$\,nm, $V_1\approx21$\,meV, $V_2=-21$\,meV, $\beta=0.015$, $\rho_1=0.79$\,nm, $l=4$\,nm. Although the particular parameter values for $V$ and $V_3$ were fitted for $\theta_{AP}=0.2^{\circ}$, we believe that the final results can be applied to twist angles $0^{\circ}< \theta_{AP}\leq 0.3^{\circ}$, for which the potential amplitudes are essentially unchanged. In Eq.\ (\ref{Eq:XX_band_edge_AP}) we measure the potential energy relative to its value inside the 2H domains, and demonstrate the quality of our fitting in the insets of Fig.\ \ref{Fig:QD_potential}.

We solve Eq.\ (\ref{Eq:XX_band_edge_AP}) using perturbation theory over $V_3$, which is much smaller than the axially symmetric potential $V$. In the zeroth-order approximation, the wave functions are eigen-functions of angular momentum $l=0,\pm1,\pm2,\dots$, \emph{i.e.} $\psi=\chi_{l}(r)e^{il\phi}$. Then, we perturbatively take into account the weak coupling between the lowest energy states with $|l|=1,2$, given by the trigonal warping term $V_3$. In Fig.\ \ref{Fig:QD_potential} we plot the energies of the lowest levels in a single quantum dot, labeling them according to the largest component of the orbital angular momenta. We note that, together with the double degeneracy of states with opposite orbital angular momentum sign (for $l\neq 0$), each QD level has an additional twofold Kramers degeneracy with the state at the opposite valley.

Since the decay lengths ($\approx 2$\,nm) of the lowest-energy states $s_0$ and $p_0$ are much smaller than the moir\'e superlattice period for the marginal twist angles, there is only a weak overlap between states in neighboring quantum dots (XX$'$ regions), giving a realisation of the Hubbard model when interactions are considered\cite{PhysRevLett.121.026402}. Therefore, we anticipate a Mott insulating state in marginally twisted $p$-doped AP-WSe$_2$ bilayers, similar to those observed at larger twist angles in P-WSe$_2$ bilayers\cite{Zhang2020}, and in twisted bilayer graphene\cite{cao2018unconventional,cao2018correlated,Codecidoeaaw9770,Pan2020}.

\subsection{Modulation of the conduction band edge at the $K$-point}

\begin{figure}
    \centering
    \includegraphics[width=0.97\columnwidth]{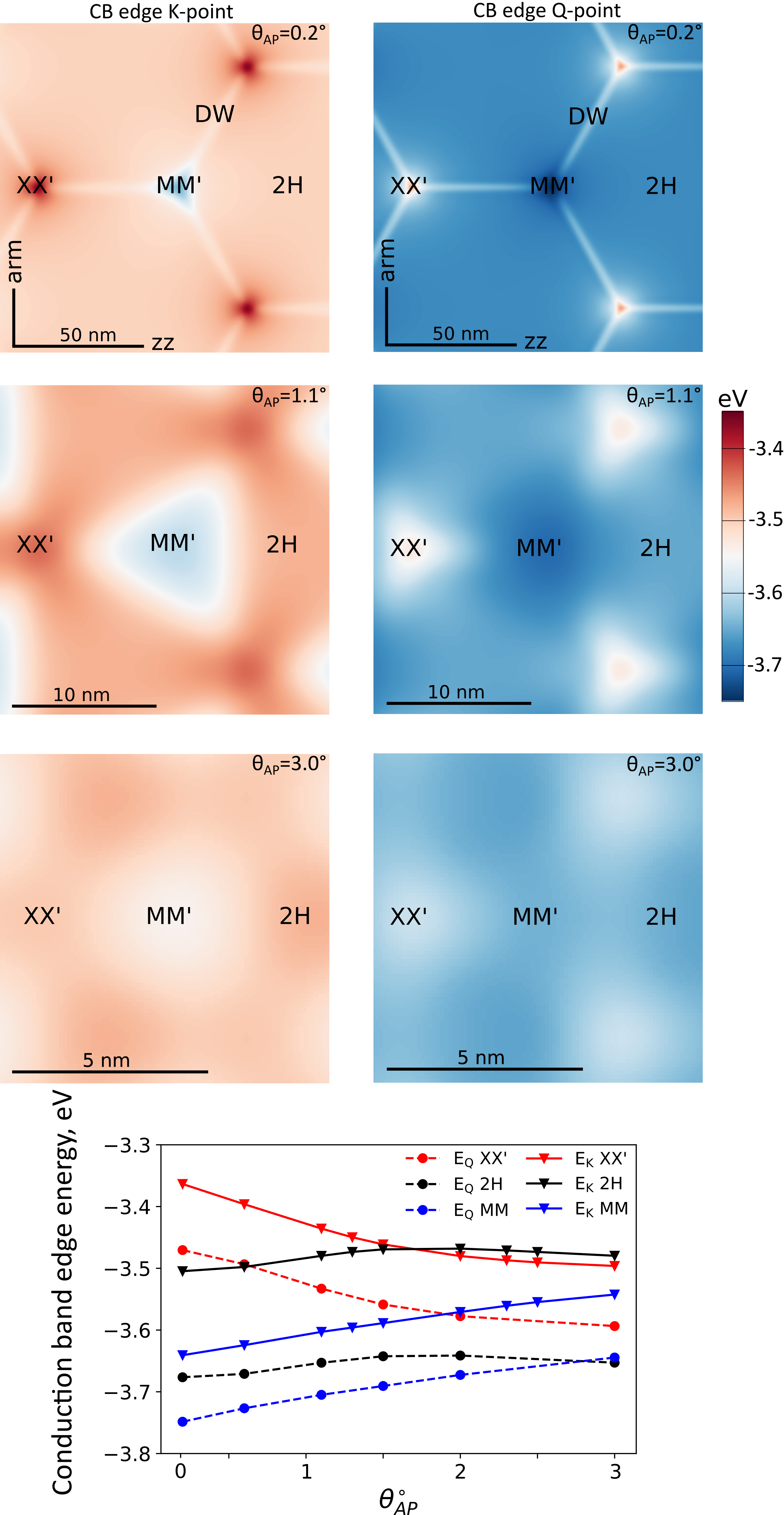}
    \caption{Piezopotential-induced modulation of the $K$- (left) and $Q$-point (right) conduction band edges, which creates quantum-dot-like potentials for electrons in MM$'$ areas of the supercell of twisted AP-WSe$_2$ bilayers. The zig-zag (zz) and armchair (arm) crystallographic directions shown in the top panels are the same for all maps. The vacuum level is set to 0~eV. The bottom panel shows the twist-angle dependences of the $K$-point conduction band edge in the labeled areas of the supercell.}
    \label{Fig:K_point_CB_AP}
\end{figure}

Modulation of the $K$-point conduction-band edge is determined by the same equation \eqref{Eq:Kv_edge_AP_with_piezo}, replacing VB$\to$CB in the matrix elements of Eq.\ \eqref{eq:HKgeneral}. Similarly to the valence band edge case, the conduction band edge variation across the moir\'e supercell is dominated by the piezopotential energy of electrons (Fig.\ \ref{Fig:K_point_CB_AP}). 

Unlike holes, for $K$-point electrons the piezopotential energy minimum (as well as the band edge) appears inside MM$'$ areas for the range of twist angles $0^{\circ}\leq \theta_{AP}\lesssim 3^{\circ}$. Therefore, we expect the formation of localised electron states in MM$'$ regions, which are split from the continuum of conduction band states, with the energy distance between levels tuned by the twist angle magnitude (see bottom panel in Fig.\ \ref{Fig:Gamma_coefs_main}). 

\subsection{Modulation of the conduction band edge at the Q$_1$-point}

The conduction-band edge at the Q$_1$ point is described by the lower eigenvalue of the Hamiltonian \eqref{Eq:Q_point_model}, supplemented by the piezopotential, which is the same in both layers:
\begin{equation}\label{Eq:Q_point_band_edge}
    E^{AP}_Q(\bm{r}) = -e\varphi(\bm{r})+\varepsilon_Q^{AP}(\bm{r})-\sqrt{\left|T_{Q}^{AP}(\bm{r})\right|^2 + \frac{\left(\Delta_{SO}^Q\right)^2}{4}}.
\end{equation}
At marginal twist angles, $\theta_{AP}<1^{\circ}$, the $Q$-point band edge is also dominated by the piezopotential modulation, which produces quantum dot potentials for electrons in MM$'$ corners of the hexagonal domain wall network (Fig.\ \ref{Fig:K_point_CB_AP}). Therefore, for $Q$-valleys we also can expect the formation of localised electron states with a discrete spectrum in MM$'$ regions, similar to the $K$-valley. However, unlike the $K$-valley levels, $Q$-valley levels with angular momenta of opposite sign are not degenerate due to the anisotropy of the effective-mass tensor\cite{kormanyos2015k}. For larger twist angles $\theta_{AP}\gtrsim 2.5^{\circ}$ the band edge gradually shifts toward 2H regions. Here, we also mention that despite the explicit lack of $C_3$ symmetry in the model \eqref{Eq:Q_point_model}, this symmetry approximately persists for the band edge, mainly due to the zig-zag orientation of the domain walls resulting from lattice reconstruction. Therefore, the conduction-band edge modulation for the $\pm 120^{\circ}$-rotated $Q_2$ and $Q_3$ valleys are the same as in Fig.\ \ref{Fig:K_point_CB_AP}.

\section{Band edge variation in twisted P-bilayers}\label{sec:twisted_P}

In this section we combine the details of atomic reconstruction in twisted P-bilayers obtained in Sec.\ \ref{sec:adhesion}, with the hybridisation analysis of Sec.\ \ref{sec:hybrydisation_Hams}. Here, we separate the analysis of valence band modulation at $\Gamma$ and $K$ from that of the $K$- and $Q$-point conduction band edges.

\subsection{Valence band edge modulation at the $\Gamma$- and $K$-points}\label{sec:GammaVBEdgeMapsP}

\begin{figure*}
    \centering
    \includegraphics[width = \textwidth]{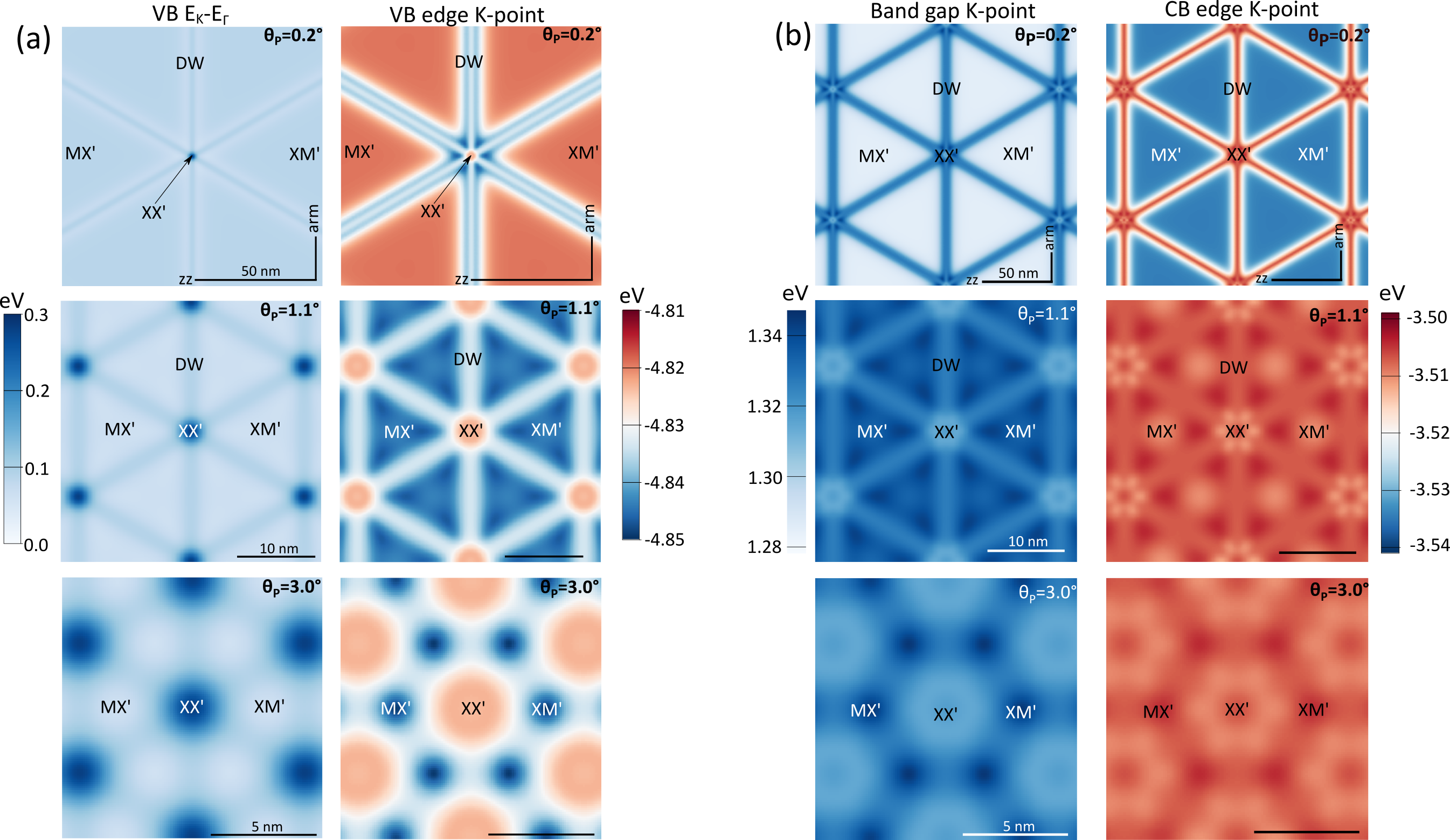}
    \caption{
    \label{Fig:P_valence_band_edge}
     (a) Maps of the valence band edge difference $E_{{\rm\bm{ K}}}^{P}-E_{\Gamma}^{P}$ (left column) and $K$-point valence band edge (right column), for twisted P-WSe$_2$ bilayers at the labeled twist angles. (b) Modulation of the direct band gap (left) and of the conduction band edge (right) at the $K$-point, for twisted P-WSe$_2$ bilayers with the labeled twist angles. At $\theta_{P} \lesssim 1.0^{\circ}$, the conduction band edge and the smallest band gap lie inside MX$'$ and XM$'$ domains, while at larger angles $\theta_P\gtrsim 1.0^{\circ}$ the former possesses weak minima around XX$'$ areas in the shape of benzene molecules. All maps were calculated with the vacuum level set to 0~eV, and are displayed for the same orientations of zigzag (zz) and armchair (arm) crystallographic directions, shown in the upper panels.}
\end{figure*}

Unlike AP-bilayers, in P-bilayers the piezocharges induced by lattice reconstruction have opposite signs in the opposite layers, leading to a vanishing total piezopotential over the bilayer, but a nonzero layer-asymmetric contribution $\varphi_t(\bm{r})=-\varphi_b(\bm{r})\equiv \Delta\varphi/2$. As a result, in  the hybridisation models (\ref{Eq:effHam_Gamma}) and (\ref{eq:HKgeneral}) the piezopotential contributes to the splitting of the coupled states rather than to an overall energy shift. Diagonalizing the Hamiltonians (\ref{Eq:effHam_Gamma}) and (\ref{eq:HKgeneral}) with the substitution of local matrix elements discussed in Sec.\ \ref{sec:hybrydisation_Hams}, we obtain expressions for the local valence band edge energies at the $\Gamma$- and $K$-points across the moir\'e superlattice:
\begin{equation}\label{Eq:G_edge_P_bilayer}
E^{P}_{\Gamma}(\bm{r})=\varepsilon_{\Gamma}(\bm{r}) + \delta\varepsilon_{\Gamma}(\bm{r}) + \sqrt{T_{\Gamma}^2(\bm{r}) + \frac{\left[\Delta^P(\bm{r})+e\Delta\varphi(\bm{r})\right]^2}{4}},
\end{equation}
\begin{equation}\label{Eq:K_edge_P_bilayer}
E^{P}_{{\rm VB},\tau{\rm \bm{K}}}(\bm{r}) = \varepsilon_{VB}^P(\bm{r}) +\sqrt{\left|T_{VB,\tau}^P(\bm{r})\right|^2 + \frac{\left[\Delta^P(\bm{r})+e\Delta\varphi(\bm{r})\right]^2}{4}},
\end{equation}
where local matrix elements are defined as in Eqs.\ (\ref{Eq:effHam_Gamma_with_piezo}) and (\ref{Eq:Kv_edge_AP_with_piezo}). For the full range of twist angles, the valence band edge at $K$ lies more than $50$\,meV higher than that at the $\Gamma$-point across a moir\'e supercell, see Fig. \ref{Fig:P_valence_band_edge}(a). This behavior of the valence band maximum for WSe$_2$ bilayers is peculiar, since for other semiconducting TMDs homobilayers (WS$_2$, MoS$_2$, and MoSe$_2$) interlayer hybridisation of the monolayer $\Gamma$-point states, formed by $p_z$ and $d_{z^2}$ orbitals of chalcogens and metals, respectively, pushes the $\Gamma$-point valence band edge more than $100$~meV higher in energy than the highest valence band state at the K-point  \cite{JinPRL2013,Yang2018,zhang2014direct}. The real-space location of the $K$-point band edge in the moir\'e supercell depends on the twist angle because of different spatial variations of $T_{VB,\tau}^P$, $\Delta^P$ and $e\Delta\varphi$ in Eq.\ (\ref{Eq:K_CB_edge_P_bilayer}) across the moir\'e supercell. The latter two possess opposite signs, leading to a decrease of the band edge energy (\ref{Eq:K_edge_P_bilayer}) in areas where both contributions exist. For marginal twist angles $\theta_P\lesssim 1^{\circ}$, the competition between the piezo- an ferropotential leads to a reduction of the band edge energy along domain walls, but pushing it up in MX$'$/XM$'$ domains where $\Delta^P$ is maximal and $e\Delta\varphi$ and $T_{VB,\tau}^P$ vanish. At the same time, the valence band energy inside MX$'$/XM$'$ domains appears only $\approx 5$\, meV higher than that in XX$'$ areas given merely by $T_{VB,\tau}^P$. This makes the valence band edge landscape rather shallow in marginally twisted P-WSe$_2$ bilayers.

With the reduced domain sizes at larger twist angles $\theta_P \geq 1^{\circ}$, the piezopotential extends inside XM$'$ and MX$'$ domains, leading to a cancellation of the potential jump $\Delta^P$ that lowers the VB edge energy inside them, shifting it toward XX$'$ areas. 

\subsection{Modulation of the conduction band edge and vertical band gap at $K$}  

Similarly to the valence band, for the conduction band at $K$ we can substitute the local lateral offset and interlayer distance into the Hamiltonian \eqref{eq:HKgeneral}. Then, taking the piezopotential into account leads to the following expression for the $K$-point conduction band edge:
\begin{equation}\label{Eq:K_CB_edge_P_bilayer}
E^{P}_{{\rm CB},\tau{\rm \bm{K}}}(\bm{r}) = \varepsilon_{\rm CB}^P(\bm{r}) -\sqrt{\left|T_{{\rm CB},\tau}^P(\bm{r})\right|^2 + \frac{\left[\Delta^P(\bm{r})+e\Delta\varphi(\bm{r})\right]^2}{4}}.
\end{equation} 

For marginal twist angles $\theta_{P}\lesssim 1^{\circ}$, the $K$-point conduction band edge lies in MX$'$ and XM$'$ domains (Fig. \ref{Fig:P_valence_band_edge}b), in a similar manner to the valence band edge (Fig.\ \ref{Fig:P_valence_band_edge}a), placing the direct band gap at the $K$ point inside XM$'$/MX$'$ regions.

For larger twist angles, the conduction band edge shifts toward the corners of XM$'$ and MX$'$ domains, forming a weak localising ($\lesssim 10$\,meV) potential around XX$'$ areas with the shape of benzene molecules. 

Thus, at marginal twist angles the optoelectronic properties of P-WSe$_2$ bilayers will be similar to the aligned WSe$_2$ bilayers corresponding to 3R-polytypes of bulk TMD crystals, while at larger angles the band edge modulation across the supercell is vanishingly small. 

\subsection{Optical selection rules for ground state bright $\tau \bm{K}$-valley interlayer excitons in XM$'$ and MX$'$ domains}

Here, we establish optical selection rules for ground states of the bright ($K$-$K$) interlayer excitons formed by the electrons and holes inside XM$'$ domains. The corresponding selection rules for MX$'$ domains can then be obtained by mirror reflection with respect to the $xy$ plane. Using the hybridisation model \eqref{eq:allmelemsK} for XM$'$-stacking ($\bm{r}_0=(0,a/\sqrt{3})$), we obtain a vanishing resonant interlayer hybridisation ($T_{\rm CB,VB}=0$), giving electron/hole states that are layer-polarised in the conduction/valence bands of the top/bottom layers, respectively, due to the ferroelectric potential. The momentum matrix element, which characterises coupling of the K-K interlayer excitons with light, transforms under $C_3$ rotations as\footnote{\mbox{$C_3\psi_{{\rm VB}, \tau \bm{K}}(\bm{r})=e^{\tau\bm{K}\cdot C_3^{-1}\bm{r}}u_{{\rm VB},\tau\bm{K}}(C_3^{-1}\bm{r})$} \mbox{$=e^{i\tau \frac{4\pi}{3}}e^{i\tau(\bm{K}+\bm{G}_2)\cdot\bm{r}}u_{{\rm VB},\tau\bm{K}}(\bm{r})$}; \mbox{$C_3\psi_{{\rm CB}, \tau \bm{K}}(\bm{r})=e^{i\tau\bm{K}\cdot(C_3^{-1}\bm{r}-\bm{r}_0)}u_{{\rm CB},\tau\bm{K}}(C_3^{-1}\bm{r}-\bm{r}_0)=$} \mbox{$=e^{i\tau(\bm{K}+\bm{G}_2)\cdot(\bm{r}-\bm{r}_0+\bm{a}_1)}u_{{\rm CB},\tau\bm{K}}(\bm{r}-\bm{r}_0+\bm{a}_1)=$} \mbox{$= e^{i\tau \frac{2\pi}{3}+i\tau\bm{G}_2\cdot(\bm{r}-\bm{r}_0)}e^{i\tau\bm{K}\cdot(\bm{r}-\bm{r}_0)}u_{{\rm CB},\tau\bm{K}}(\bm{r}-\bm{r}_0)=$} \mbox{$= e^{i\bm{G}_2\cdot\bm{r}}e^{i\tau\bm{K}\cdot(\bm{r}-\bm{r}_0)}u_{{\rm CB},\tau\bm{K}}(\bm{r}-\bm{r}_0)$}, where in the last equality we substituted $\mathbf{r}_0=(0,a/\sqrt{3})$, corresponding to XM$'$ stacking. }
\begin{multline}\label{Eq:optics}
    \langle C_3 \psi_{{\rm VB},\tau \bm{K}}|C_3\left(p_x\pm ip_y\right)C_3^{-1}|C_3 \psi_{{\rm CB},\tau \bm{K}}\rangle =\\
    = e^{i\frac{2\pi}{3}(\tau\mp 1)} \langle \psi_{{\rm VB},\tau \bm{K}}|p_x\pm ip_y| \psi_{{\rm CB},\tau \bm{K}}\rangle.
\end{multline}
Here, \mbox{$\psi_{{\rm VB}, \tau \bm{K}}=e^{i\tau\bm{K}\bm{r}}u_{{\rm VB},\tau \bm{K}}(\bm{r})$} is the Bloch function of the bottom-layer valence state ($u_{{\rm VB},\tau \bm{K}}(\bm{r})$ transforms as \mbox{$(x-i\tau y)^2$} \cite{kormanyos2015k}), and \mbox{$\psi_{{\rm CB}, \tau \bm{K}}=e^{i\tau\bm{K}(\bm{r}-\bm{r}_0)}u_{{\rm CB},\tau \bm{K}}(\bm{r}-\bm{r}_0)$} is the top-layer conduction-band Bloch function, shifted by the offset \mbox{$\bm{r}_0=(0,a/\sqrt{3})$} ($u_{{\rm CB},\tau \bm{K}}(\bm{r}-\bm{r}_0)$ transforms as $z^2$). The matrix element of dipole transitions couples $p_x\pm ip_y$ with $A_x \mp iA_y$, where $A_x$ and $A_y$ are the components of the electromagnetic vector potential. Based on \eqref{Eq:optics}, this leads to circularly polarised luminescence with counter-clockwise polarisation ($\sigma_-$) for $+\bm{K}$-valley excitons, and clockwise polarisation ($\sigma_+$) for $-\bm{K}$-valley excitons.           

\subsection{Modulation of the conduction band edge at Q$_1$}  

\begin{figure}
    \centering
    \includegraphics[width = 0.7\linewidth]{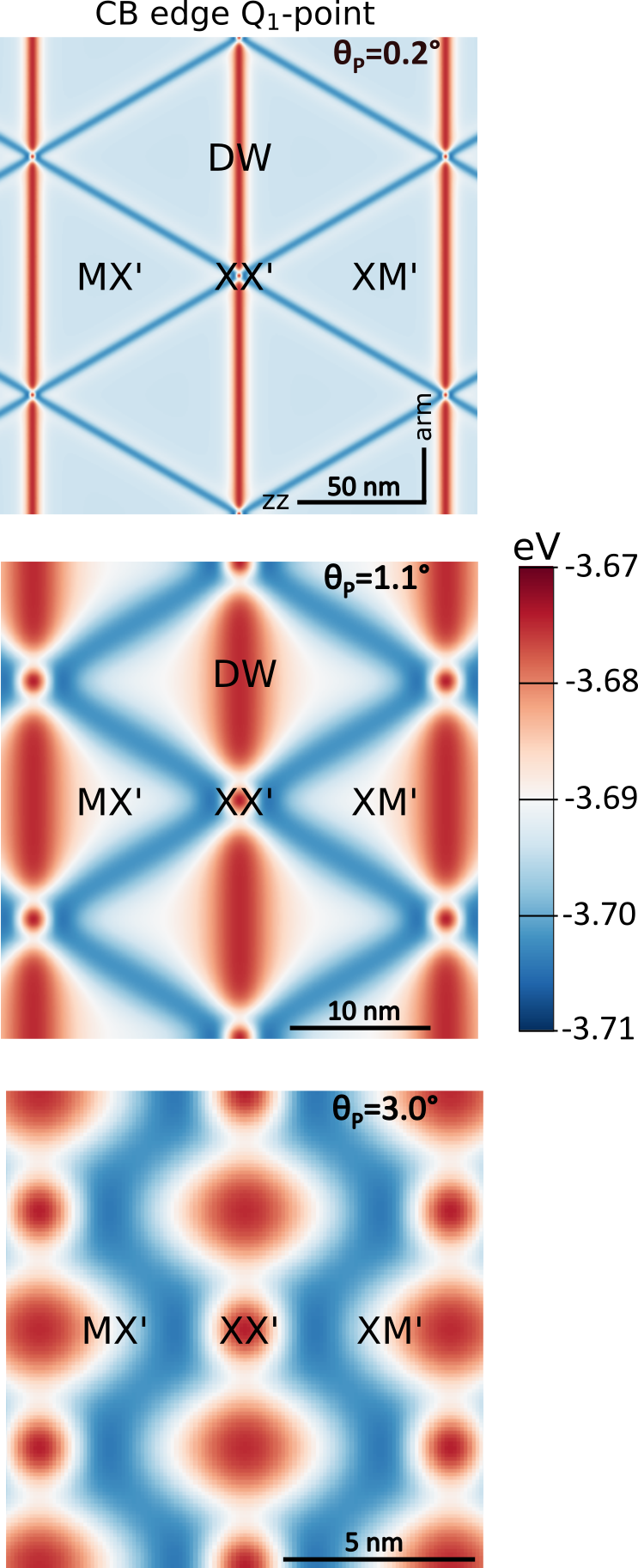}
    \caption{
    \label{Fig:P_Q_CB_edge}
      Modulation of the conduction band edge at the $\pm$Q$_1$-points, for twisted P-WSe$_2$ bilayers with the labeled twist angles. At $\theta_{P} \lesssim 2.0^{\circ}$, the conduction band edge consists of one-dimensional channels of $\approx 10$\,meV depth along two of three DWs orientations, transforming to zigzagging landscape for higher twist angles. Conduction band edge variations for $\pm$Q$_{2,3}$ can be obtained applying $\pm120^{\circ}$-rotation, respectively.}
\end{figure}

The conduction band edge at the Q$_1$-point is described by the lower eigenvalue of the Hamiltonian \eqref{Eq:Q_point_model}, substituting the local lateral offset $\mathbf{r}_0(\mathbf{r})$ and supplemented with the piezopotential, which has opposite sign in the top and bottom layers:
\begin{equation}\label{Eq:Q_point_band_edge}
    E^P_Q(\bm{r}) = \varepsilon_Q^P(\bm{r})-\sqrt{\left|T_{Q}^P(\bm{r})\right|^2 + \frac{\left[S^P_Q(\bm{r})+e\Delta\varphi(\bm{r})\right]^2}{4}}.
\end{equation}
The conduction-band edge maps for Q$_1$ states, shown in Fig.\ \ref{Fig:P_Q_CB_edge}, are not symmetric under 120$^{\circ}$ rotations, because of the anisotropy of the resonant interlayer coupling term $T^P_Q$ resulting from low-symmetry of the Q-states. At small twist angles $\theta_P\lesssim 2.0^{\circ}$, the anisotropic coupling, along with the piezopotential that is substantial in the vicinity of DWs, place the band edge in one-dimensional channels with $\approx10$\,meV depth, forming along two of three domain wall orientations. Although for larger twist angles the domain wall structure disappears, the conduction band edge still occurs along channels with a zig-zag shape. Band edge landscapes for the $\mathbf{Q}_{2,3}$-points are obtained by $\pm120^{\circ}$-rotation of the maps in Fig.\ \ref{Fig:P_Q_CB_edge}, respectively, while those for $-{\bf Q}_{1,2,3}$-states are the same as for ${\bf Q}_{1,2,3}$.  

\begin{figure*}[p!]
    \centering
    \includegraphics[width=2\columnwidth]{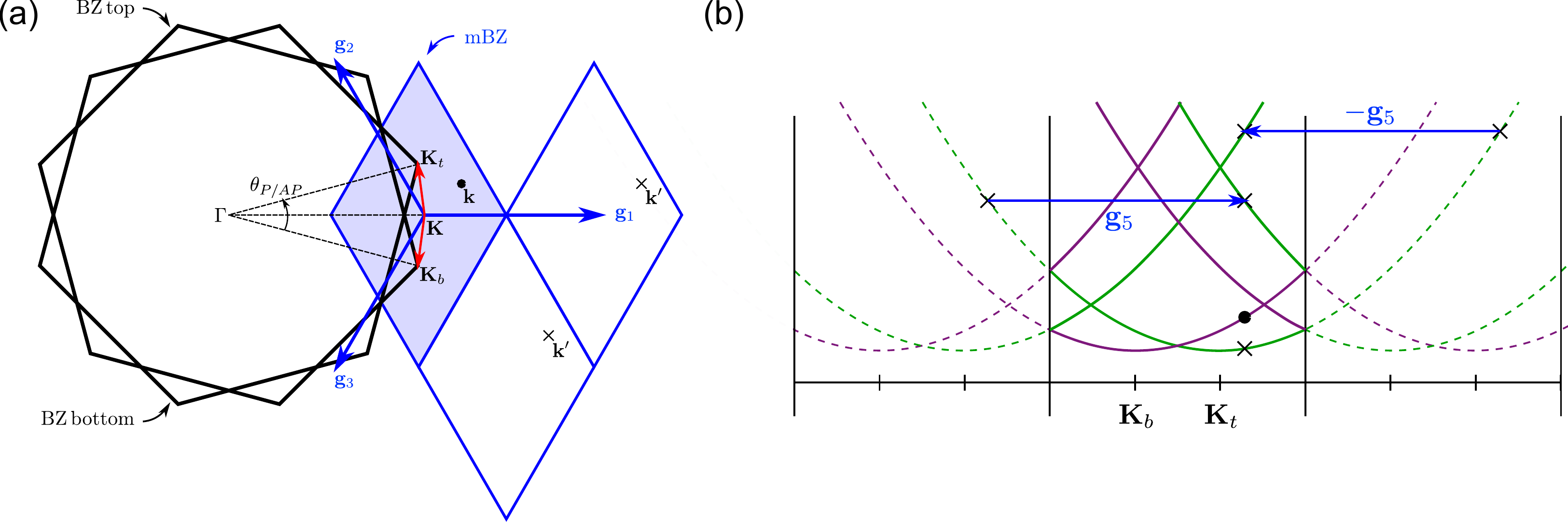}
    \caption{Zone folding scheme for the moir\'e miniband calculations. (a) The top- and bottom layer Brillouin zones appear rotated by angles $\pm \theta_{P/AP}/2$, such that their corners $\KK_t$ and $\KK_b$ are shifted from $\KK$ by $\pm \Delta\KK/2$ (red arrows), respectively. The wave vector $\kk$ shown ($\bullet$) is equivalent to the wave vectors $\kk'=\kk-\gG_2$ and $\kk'=\kk+\gG_1$ ($\times$), in the sense that they are tunnel-coupled by the first term of Eq.\ \eqref{eq:H12}. This defines the first mBZ, shaded blue in the figure. (b) Top- and bottom layer electron dispersions along the $\overline{\KK_b\,\KK_t}$ axis in an extended mBZ scheme. Superlattice momentum states outside the first mBZ appear as ``folded'' minibands. The bottom-layer state marked with a $\bullet$ symbol tunnel-couples to the top-layer states marked as $\times$, including those separated by second-star moir\'e vectors $\pm \gG_5 = \pm (\gG_2 - \gG_3)$, by the second term in Eq.\ \eqref{eq:H12}.}
    \label{fig:moireschematic}
\end{figure*}

\section{Moir\'e superlattice minibands for wave vectors near the $\Gamma$, $K$ and $Q$ valleys}\label{sec:Kminibandstogether}

In-plane lattice reconstruction is weak for misalignment angles larger than $\theta_{P/AP}^*$, such that $\mathbf{u}^t$ and $\mathbf{u}^b$ can be neglected and the local stacking vector approximated by $\rr_0(\rr) \approx \theta_{P/AP} \hat{\mathbf{z}}\times \rr$. This spatial modulation of the interlayer registry results in a moir\'e superlattice with periodicity $a_M$ related to the lattice constant $a$ as $a_M = a\, \theta_{P/AP}^{-1}$. In reciprocal space, the superlattice is described by a set of moir\'e Bragg vectors $\mathbf{g}_j \equiv \GG^b_{j}-\GG_{j}^{t}$, where $\GG^b_{j}$ and $\GG^t_{j}$ are reciprocal lattice vectors of the individual rotated layers, given by
\begin{equation}\label{eq:moirebragg}
    \GG^b_{j}=\mathcal{R}_{-\theta_{P/AP}/2}\GG_j,\,\quad \GG^t_{j}=\mathcal{R}_{\theta_{P/AP}/2}\GG_j,
\end{equation}
where $\mathcal{R}_\phi$ rotates by an angle $\phi$ about axis $\hat{\mathbf{z}}$, and $\GG_j$ are the reciprocal lattice vectors before rotation.

\textbf{Model for the mSL near the $\boldsymbol{\Gamma}$ point band edge.} The local Hamiltonian for $\Gamma$-point valence electrons at position $\rr$ is obtained from the registry-dependent model \eqref{Eq:H0_Gamma} with local $\rr_0\to\rr_0(\rr)$, supplemented by the piezoelectric potentials $\varphi_{b} = \varphi_t\equiv  \varphi$ for AP- and $\varphi_{b} = -\varphi_t\equiv - \varphi$ for P-bilayers. This gives the Hamiltonian [see Eq.\ \eqref{Eq:H0_Gamma}]
\begin{equation}
\begin{split}
    \mathcal{H}_{{\rm VB}, \Gamma}^{P/AP}(\rr) =&\, \frac{\hbar^2(-i\nabla)^2}{2m_\Gamma}\Lambda_0 + H_{\Gamma}^{P/AP}[\rr_0(\rr),z(\rr)] \\
    &-\frac{e\varphi(\rr)}{2}\left[(1\mp 1)\Lambda_0 + (1 \pm 1)\Lambda_z \right],
\end{split}
\end{equation}
acting on plane-wave states of the form $\psi(\rr)=(e^{i\kk\cdot\rr},\,e^{i\kk\cdot\rr})^T$. Note that we have considered also the spatial dependence of the coefficients \eqref{Eq:H0_matrix_elements} coming from the interlayer distance modulation described by Eq.\ (\ref{Eq:interlayer_distance}), in which we have used the first 10-20 stars of moir\'e harmonics characterised by Fourier coefficients presented in Appendix \ref{sec:minibands}. Similarly, we have written the piezopotential in terms of the Fourier expansion (see Appendix \ref{sec:minibands})
\begin{equation}\label{eq:piezoharmonics}
    \varphi(\rr)=\sum_{j=1} \varphi_{j}\cos{\left( \gG_j\cdot\rr\right)}.
\end{equation}
The superlattice Hamiltonian for $\Gamma$-point electrons takes on the form
\begin{equation}\label{eq:HminiG}
\begin{split}
    \mathcal{H}_{\Gamma,VB}^{P/AP}(\rr) =& \left[\tfrac{\hbar^2k^2}{2m_\Gamma}+\varepsilon_\Gamma^{P/AP}(\rr) \right]\Lambda_0 - \tfrac{S_\Gamma^{P/AP}(\rr)}{2}\Lambda_z\\
    &- \frac{e\varphi(\rr)}{2}\left[(1\mp1)\Lambda_0 + (1\pm1) \Lambda_z \right]\\
    &+ \mathrm{Re}\,T_\Gamma^{P/AP}(\rr)\Lambda_x - \mathrm{Im}\,T_\Gamma^{P/AP}(\rr)\Lambda_y,
\end{split}
\end{equation}
where the spatial dependence comes through $\rr_0(\rr)$ and $z(\rr)$, as in Eq.\ \eqref{Eq:local_stacking_dependence}.

\textbf{Model for the mSL near the $\mathbf{K}$ point band edge.} To describe $K$-point conduction- and valence electrons one must consider that the top- and bottom layer $K$ valleys are also rotated as $\KK_{t/b} = \mathcal{R}_{\pm\theta_{P/AP}/2}\KK$, with $\KK$ the valley vector before rotation, introducing a valley mismatch $\Delta\KK=\KK_t-\KK_b$, see Fig.\ \ref{fig:moireschematic}a. This is included in the local Hamiltonian for $\tau \KK$ valley electrons by applying a unitary transformation $\mathcal{U}_{\theta_{P/AP}}^\tau(\Delta \KK)$ that adjusts the wave vectors upon rotation of the reciprocal lattices. The superlattice Hamiltonian at valley $\tau\KK$ for band $\alpha={\rm CB},\,{\rm VB}$, for conduction- and valence band, respectively, is given by
\begin{equation}\label{eq:Hmini0}
\begin{split}
    &\mathcal{H}_{\alpha, \tau\KK}^{P/AP}(\rr) =  \mathcal{U}_{\theta_{P/AP}}^\tau(\Delta\KK)\Big[H_{\alpha,\tau \KK}^{P/AP}(\rr) \\
    &+\frac{\hbar^2}{2m_\alpha}  \frac{(-i\boldsymbol{\nabla}-\tau \KK_t )^2+(-i\boldsymbol{\nabla}-\tau \KK_b )^2}{2}\Lambda_0\\
    &+ \frac{\hbar^2}{2m_\alpha}\frac{\KK_t+\KK_b + 2i\tau \boldsymbol{\nabla}}{2} \cdot\Delta\KK\Lambda_z \\ &-\frac{e\varphi(\rr)}{2}\left[(1\mp1)\Lambda_0 + (1\pm1)\Lambda_z \right]\Big]\mathcal{U}_{\theta_{P/AP}}^{\tau\,-1}(\Delta\KK),\\
    &\mathcal{U}_{\theta_{P/AP}}^\tau(\Delta\KK) = \begin{pmatrix} e^{i\tau \Delta \KK\cdot\rr/2} & 0 \\ \\ 0 & e^{-i\tau \Delta \KK\cdot\rr/2} \end{pmatrix},
\end{split}
\end{equation}
and takes the final form
\begin{equation}\label{eq:Hmini}
\begin{split}
    \mathcal{H}_{\alpha,\tau\KK}^{P/AP}(\rr) =& \frac{\hbar^2}{2m_\alpha}\left(\kk\Lambda_0 - \frac{\Delta\KK}{2}\Lambda_z \right)^2\\
    &+ \varepsilon_\alpha^{P/AP}(\rr)\Lambda_0-\tfrac{S_\alpha^{P/AP}(\rr)}{2}\Lambda_z\\
    &-\frac{e\varphi(\rr)}{2}\left[(1\mp1)\Lambda_0 + (1\pm1)\Lambda_z \right]\\
    &+\mathrm{Re}\left[e^{i\tau\Delta\KK\cdot\rr}\,T_{\alpha,\tau}^{P/AP}(\rr)  \right]\Lambda_x\\
    &- \mathrm{Im}\left[e^{i\tau\Delta\KK\cdot\rr}\,T_{\alpha,\tau}^{P/AP}(\rr)  \right]\Lambda_y.
\end{split}
\end{equation}

\textbf{Model for the mSL near the $\mathbf{Q}_1$ point band edge.} The $Q_1$ valley case is analogous to that of the $K$ valley. The top- and bottom-layer $Q_1$ valleys are rotated as $\mathbf{Q}_{1,t/b}=\mathcal{R}_{\pm \theta_{P/AP}/2}\mathbf{Q}_1$, with $\mathbf{Q}_1$ the valley vector before rotation. This results in a $Q_1$-valley mismatch $\Delta \mathbf{Q}=\mathbf{Q}_{1,t}-\mathbf{Q}_{1,b}$, as well as a relative rotation of the wave vectors, implemented by a unitary transformation $\mathcal{U}_{\theta_{p/AP}}(\Delta\QQ)$ that acts on the plane-wave states as
\begin{equation}\label{eq:rotQ}
    \mathcal{U}_{\theta_{p/AP}}(\Delta\QQ) \begin{pmatrix}
    e^{i(\QQ_1+\kk)\cdot\rr} \\ e^{i(\QQ_1+\kk)\cdot\rr}
    \end{pmatrix} = \begin{pmatrix}
    e^{i(\QQ_1+\tfrac{\Delta\QQ}{2}+\mathcal{R}_{\theta_{P/AP}/2}\kk)\cdot\rr} \\ e^{i(\QQ_1-\tfrac{\Delta\QQ}{2}+\mathcal{R}_{-\theta_{P/AP}/2}\kk)\cdot\rr}
    \end{pmatrix}.
\end{equation}
Note that, unlike at the $K$ valleys, the monolayer conduction band dispersions at $Q_1$ are anisotropic, with distinct masses $m_x$ and $m_y$ along and perpendicular to the $\overline{\Gamma\,K}$ line. Therefore, the wave vectors $\mathcal{R}_{\pm\theta_{P/AP}/2}\kk$ in Eq.\ \eqref{eq:rotQ} will rotate the monolayer dispersions. The resulting $Q_1$-point Hamiltonian is
\begin{widetext}
\begin{equation}\label{eq:QHmini}
\begin{split}
    \mathcal{H}_{Q}^{P/AP}(\rr) =& \frac{\hbar^2}{2m_x}\left(k_x\Lambda_0 - \frac{\theta_{P/AP}}{2}k_y\Lambda_z \right)^2
    + \frac{\hbar^2}{2m_y}\left(k_y\Lambda_0 + \left[\frac{\theta_{P/AP}}{2}k_x-\tfrac{|\Delta\QQ|}{2}\right]\Lambda_z \right)^2
    + \varepsilon_Q^{P/AP}(\rr)\Lambda_0-\tfrac{S_Q^{P/AP}(\rr)}{2}\Lambda_z\\
    &-\frac{e\varphi(\rr)}{2}\left[(1\mp1)\Lambda_0 + (1\pm1)\Lambda_z \right]
    +\mathrm{Re}\left[e^{i\Delta\QQ\cdot\rr}\,T_{Q}^{P/AP}(\rr)  \right]\Lambda_x
    - \mathrm{Im}\left[e^{i\Delta\QQ\cdot\rr}\,T_{Q}^{P/AP}(\rr)  \right]\Lambda_y.
\end{split}
\end{equation}
To show the spatial dependence of the matrix elements, we take the $K$-point Hamiltonian interlayer hybridisation term as an example. From Eq.\ \eqref{eq:allmelemsK} and Eqs.\ \eqref{eq:TappVB} and \eqref{eq:TappCB} in Appendix \ref{app:hamiltonianK} we have
\begin{equation}\label{eq:howtominibands}
\begin{split}
    &e^{i\tau\Delta\KK\cdot\rr}T_{\alpha,\tau}^{P/AP}=e^{i\tau\Delta\KK\cdot\rr}t_\alpha^{P/AP}e^{-q_{t_\alpha^{P/AP}}z(\rr)}\sum_{\mu=0}^2e^{-i\tau C_3^\mu \Delta\KK\cdot \rr}e^{i\tfrac{4}{3}\tau\mu\chi_{P/AP}(\lambda_\alpha+1)}\\
    &\approx e^{i\tau\Delta\KK\cdot\rr}\tilde{t}_\alpha^{P/AP}\left[1-q_{t_\alpha^{P/AP}} \sum_{j=1}^{\mathcal{N}}z_j^s\cos{(\mathbf{g}_j\cdot\rr )}-q_{t_\alpha^{P/AP}} \sum_{j=1}^{\mathcal{N}}z_j^a\sin{(\mathbf{g}_j\cdot\rr )} \right]\sum_{\mu=0}^2e^{-i\tau C_3^\mu \Delta\KK\cdot \rr}e^{i\tfrac{4}{3}\tau\mu\chi_{P/AP}(\lambda_\alpha+1)}\\
    &\approx \tilde{t}_\alpha^{P/AP}\left[1+e^{i\tau\mathbf{g}_2\cdot \rr}e^{i\tfrac{4}{3}\tau\chi_{P/AP}(\lambda_\alpha+1)} + e^{-i\tau\mathbf{g}_1\cdot \rr}e^{i\tfrac{8}{3}\tau\chi_{P/AP}(\lambda_\alpha+1)}\right]\\
    &-\sum_{\nu=-1,+1}\sum_{j=1}^{\mathcal{N}}q_{t_\alpha^{P/AP}}\frac{\tilde{t}_\alpha^{P/AP}}{2}\left(z_j^s -i\nu z_j^a \right)\left[e^{i\nu\mathbf{g}_j \cdot \rr}+e^{i(\tau\mathbf{g}_2+\nu \mathbf{g}_j )\cdot \rr}e^{i\tfrac{4}{3}\tau\chi_{P/AP}(\lambda_\alpha+1)} + e^{i(-\tau\mathbf{g}_1+\nu\mathbf{g}_j )\cdot \rr}e^{i\tfrac{8}{3}\tau\chi_{P/AP}(\lambda_\alpha+1)}\right],
\end{split}
\end{equation}
where we have introduced $\tilde{t}_{\alpha}^{P/AP}\equiv \exp{(-q_{t_\alpha}^{P/AP}z_0)}t_{\alpha}^{P/AP}$, approximated the exponential dependence on interlayer distance modulation by its first-order expansion in the coefficients $z_j$ for $j\ge1$, and used the fact that $(1-C_3)\Delta\KK=\mathbf{g}_2$ and $(1-C_3^2)\Delta\KK=-\mathbf{g}_1$, with $C_3=\mathcal{R}_{2\pi/3}$. The interlayer tunnelling matrix element is then
\begin{equation}\label{eq:H12}
\begin{split}
    &\int d^2r\,\psi_{\kk'}^\dagger(\rr)e^{i\tau\Delta\KK\cdot\rr}T_{\alpha,\tau}^{P/AP}(\rr)\psi_{\kk}(\rr)=\tilde{t}_\alpha^{P/AP}\left[\delta_{\kk,\kk'} + \delta_{\kk-\kk',-\tau\mathbf{g}_2}e^{i\tfrac{4}{3}\tau\chi_{P/AP}(\lambda_\alpha+1)} + \delta_{\kk'-\kk,\tau\mathbf{g}_1}e^{i\tfrac{8}{3}\tau\chi_{P/AP}(\lambda_\alpha+1)} \right]\\
    &-\sum_{\nu=-1,+1}\sum_{j=1}^{\mathcal{N}}q_{\tilde{t}_\alpha^{P/AP}}\frac{t_\alpha^{P/AP}}{2}\left(z_j^s-i\nu z_j^a \right)\left[\delta_{\kk-\kk',-\nu\gG_j }+\delta_{\kk-\kk',-\tau\mathbf{g}_2-\nu \mathbf{g}_j }\,e^{i\tfrac{4}{3}\tau\chi_{P/AP}(\lambda_\alpha+1)} + \delta_{\kk-\kk',\tau\mathbf{g}_1-\nu\mathbf{g}_j }\,e^{i\tfrac{8}{3}\tau\chi_{P/AP}(\lambda_\alpha+1)}\right].
\end{split}
\end{equation}
We point out that \eqref{eq:H12} is a direct generalisation of the hybridisation model presented in Ref.\ \cite{moire_bands_prb_2019}, which considered a rigid rotation of the TMD layers, and neglected out-of-plane relaxation.
\end{widetext}

Like \eqref{eq:H12}, all matrix elements of Eqs.\ \eqref{eq:HminiG} and \eqref{eq:Hmini}  contain a new momentum conservation rule $\kk = \kk' + \gG_j$, consequence of the moir\'e superlattice periodicity, whereby wave vectors are only conserved up to a moir\'e vector $\gG_j$. This determines the moir\'e Brillouin zone (mBZ) shown in Fig.\ \ref{fig:moireschematic}(a) as a blue-shaded rhombus, defined by the first-star moir\'e vectors $\gG_j$.  Wave vectors $\kk'$ outside this region of reciprocal space are ``folded'' into the mBZ as $\kk'=\kk\pm\gG_j$, and treated as part of distinct minibands that couple vertically at wave vector $\kk$, according to the matrix elements in Eq.\ \eqref{eq:Hmini}. For instance, Fig.\ \ref{fig:moireschematic}(b) shows a bottom-layer state of wave vector $\kk$ near the $\KK$ valley, and three out of the multiple top-layer states to which it couples by interlayer tunnelling through the second term of \eqref{eq:H12}.

In the following sections we show the low-energy $\Gamma$- and $K$-point electronic spectra of twisted P and AP bilayers for twist angles $\theta_{P/AP}>\theta_{P/AP}^*$, computed by direct diagonalisation of Eqs.\ \eqref{eq:HminiG} and \eqref{eq:Hmini}. The plane-wave basis used for the numerical calculations is large enough to provide convergence for several of the lowest conduction (highest valence) minibands.

\subsection{Miniband structures near the $\Gamma$ point for twisted AP and P bilayers}\label{sec:Gminibands}
\begin{figure*}
    \centering
    \includegraphics[width=\columnwidth]{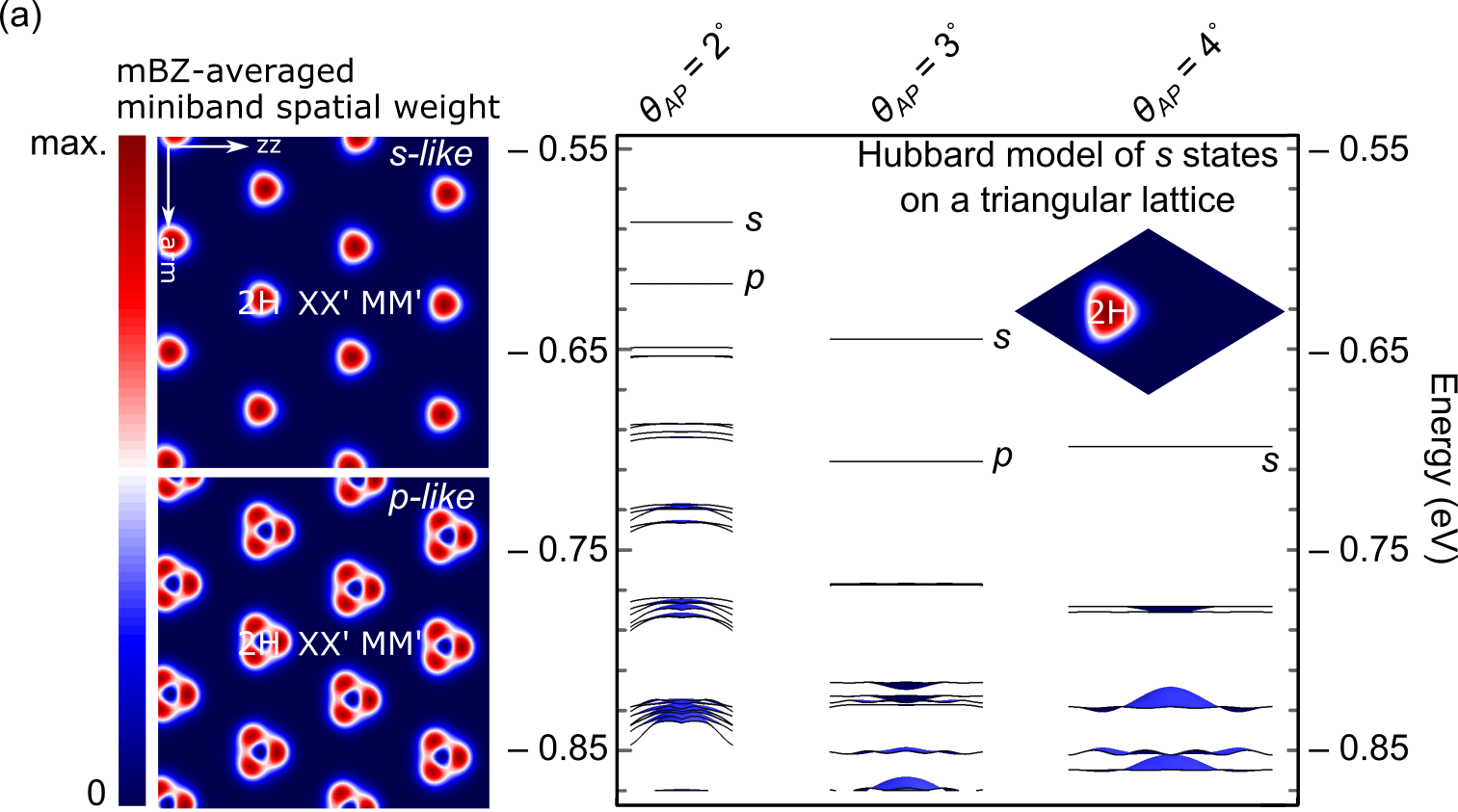}    \includegraphics[width=\columnwidth]{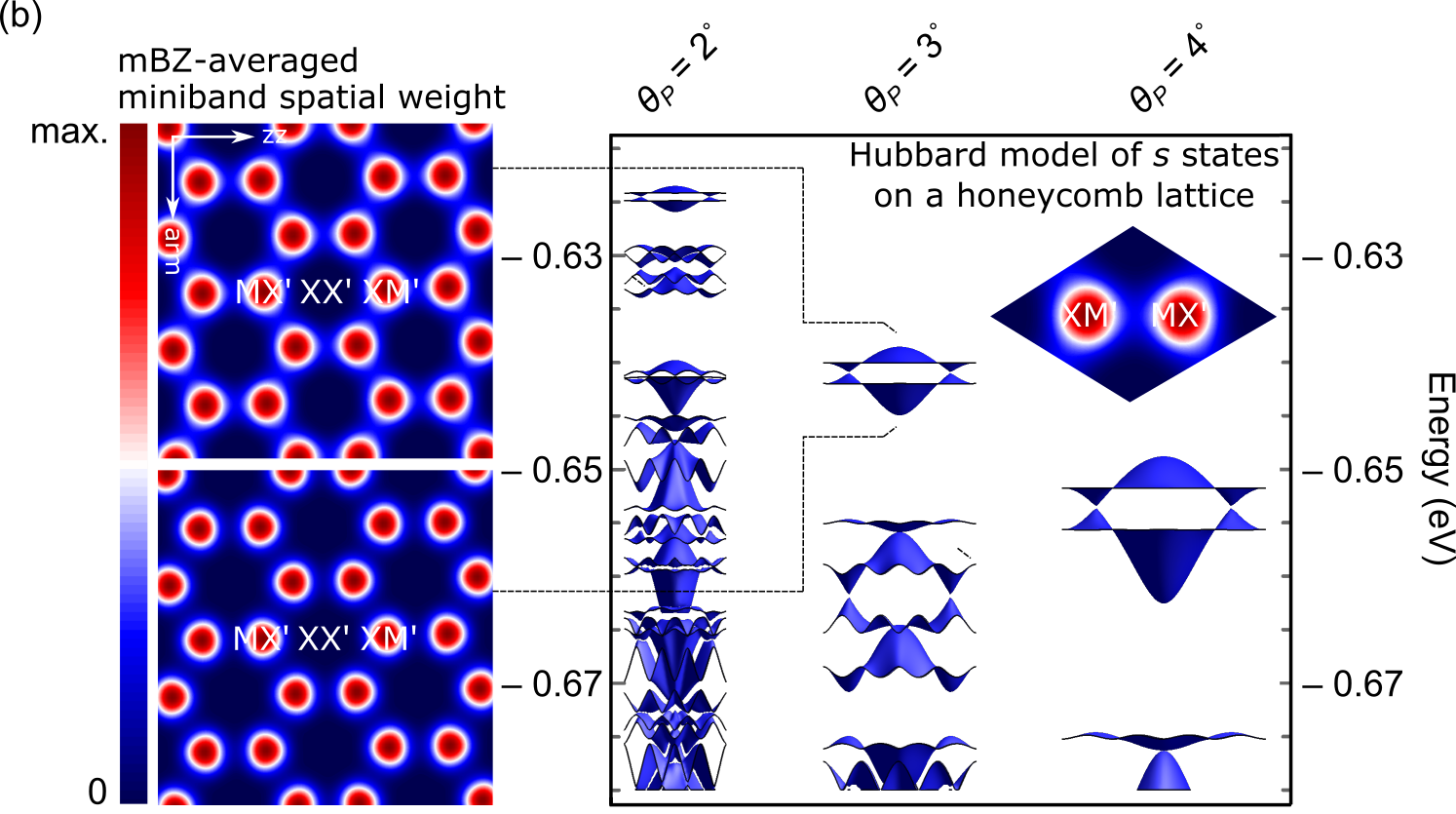}
    \caption{$\Gamma$-point moir\'e valence miniband spectra of twisted AP (a) and P (b)  WSe${}_2$ bilayers, at large twist angles $\theta_{P/AP}=2^\circ,\,3^\circ$ and $4^\circ$. The left panels show the mBZ-averaged modulus squared of the wavefunction, representing the probability of finding a valence electron at a given position in the superlattice. AP bilayers exhibit flat minibands corresponding to a triangular lattice of quantum dots at 2H regions. The unit cell is shown in the inset. P bilayers exhibit graphene-like dispersions with two gapless Dirac cones inside the mBZ, and bandwidths ranging from $\sim 1\,{\rm meV}$ at $2^\circ$ to $\sim 10\,{\rm meV}$ at $4^\circ$. The graphene-like bands originate from a honeycomb lattice of states localised at MX$'$ and XM$'$ sites, coupled by interlayer tunnelling. All energies are measured with respect to the monolayer WSe${}_2$ VB edge.}
    \label{fig:GPointMinibands}
\end{figure*}

Figures \ref{fig:GPointMinibands}(a) and \ref{fig:GPointMinibands}(b) show valence minibands for twisted AP and P bilayers, respectively, computed numerically by the zone folding method described in Sec.\ \ref{sec:Kminibandstogether}. Each miniband is spin-degenerate due to time reversal symmetry of the $\Gamma$ point.

\textbf{AP structures} exhibit extremely flat minibands ($\lesssim 1\,\mathrm{meV}$ bandwidth) for twist angles up to $4^\circ$. To gain insight into the origin of those minibands, we plot the modulus squared of their wavefunctions averaged over the entire mBZ in the left panels of Fig.\ \ref{fig:GPointMinibands}(a). These show an array of QD states localised at 2H-stacking sites in the moir\'e supercell, where the valence band maximum for $\Gamma$ point states was predicted in Sec.\ \ref{sec:ModGamma}. The highest miniband represents an array of trigonally warped $s$-like states\footnote{More precisely, these states correspond to the $A_1$ representation of the $C_{3v}$ point group of the band-edge energy about 2H sites\cite{PhysRevB.102.195403}.} (see Fig.\ \ref{Fig:AP_Top_valence_band}), giving a realisation of the SU${}_2$ Hubbard model for $\Gamma$-point holes. Similarly, the next highest states are a $p$-like doublet formed by the trigonally warped $p$ orbitals shown in the bottom-left panel of Fig.\ \ref{fig:GPointMinibands}(a)\footnote{These states belong to the $E$ representation of group $C_{3v}$, as discussed in Ref.\ \cite{PhysRevB.102.195403}.}.

\textbf{P-bilayer minibands} are shown in Fig.\ \ref{fig:GPointMinibands}(b) for $2^\circ$, $3^\circ$ and $4^\circ$ twist angles. At $\theta_{P}=4^\circ$ the top two minibands exhibit a graphene-like dispersion of $\approx 15\,{\rm meV}$ bandwidth with two Dirac cones at the inequivalent mBZ points $\boldsymbol{\kappa}=\Delta \KK/2$ and $\boldsymbol{\kappa}'=-\Delta \KK/2$. As the twist angle decreases, the bandwidth of this miniband pair drops considerably, reaching values of $\sim 1\,{\rm meV}$ at $2^\circ$, but the gapless Dirac dispersion persists.

The real-space distribution of the top two miniband wave functions is shown in the left panels of Fig.\ \ref{fig:GPointMinibands}(b) to consist of arrays of trigonally warped $s$-like orbitals centred at MX$'$ and XM$'$ stacking regions of the moir\'e supercell. These correspond, respectively, to bottom- and top-layer $\Gamma$-point valence states localised by the effective moir\'e potential (see Sec.\ \ref{sec:GammaVBEdgeMapsP}). Together, the bottom- and top-layer states form a bipartite triangular (honeycomb) lattice [Fig.\ \ref{fig:GPointMinibands}(b) inset] whose sites are coupled by interlayer tunnelling, constituting a graphene analog with mesoscopic-scale inter-site distances\cite{AngeliPNAS}.

Overall, for both P and AP structures, the top valence states at the $\Gamma$ point can be described by mesoscale lattice models involving arrays of $s$- or $p$-like orbitals. Then, Coulomb interactions will be significant for these sites, giving rise to mesoscopic realisations of the Hubbard model on a triangular (AP structures) or a honeycomb lattice (P structures), potentially leading to strongly correlated ground states.\cite{HubbardSim2020,Wang2020,CorrelatedInsulating2020}

\subsection{Miniband structures near the $K$ point for twisted AP bilayers}\label{sec:KminibandsAP}
\begin{figure*}
    \centering
    \includegraphics[width=\columnwidth]{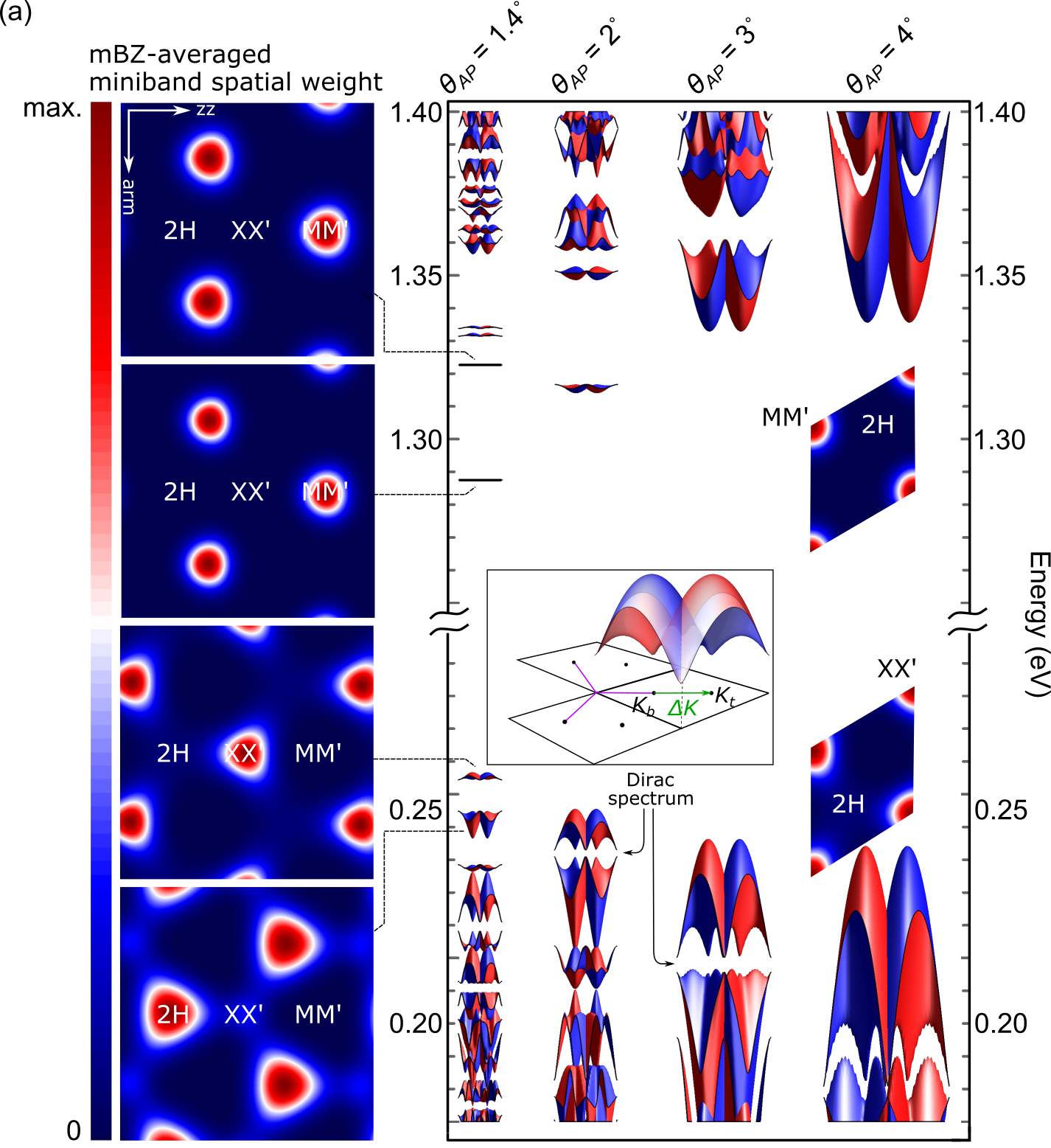}    \includegraphics[width=\columnwidth]{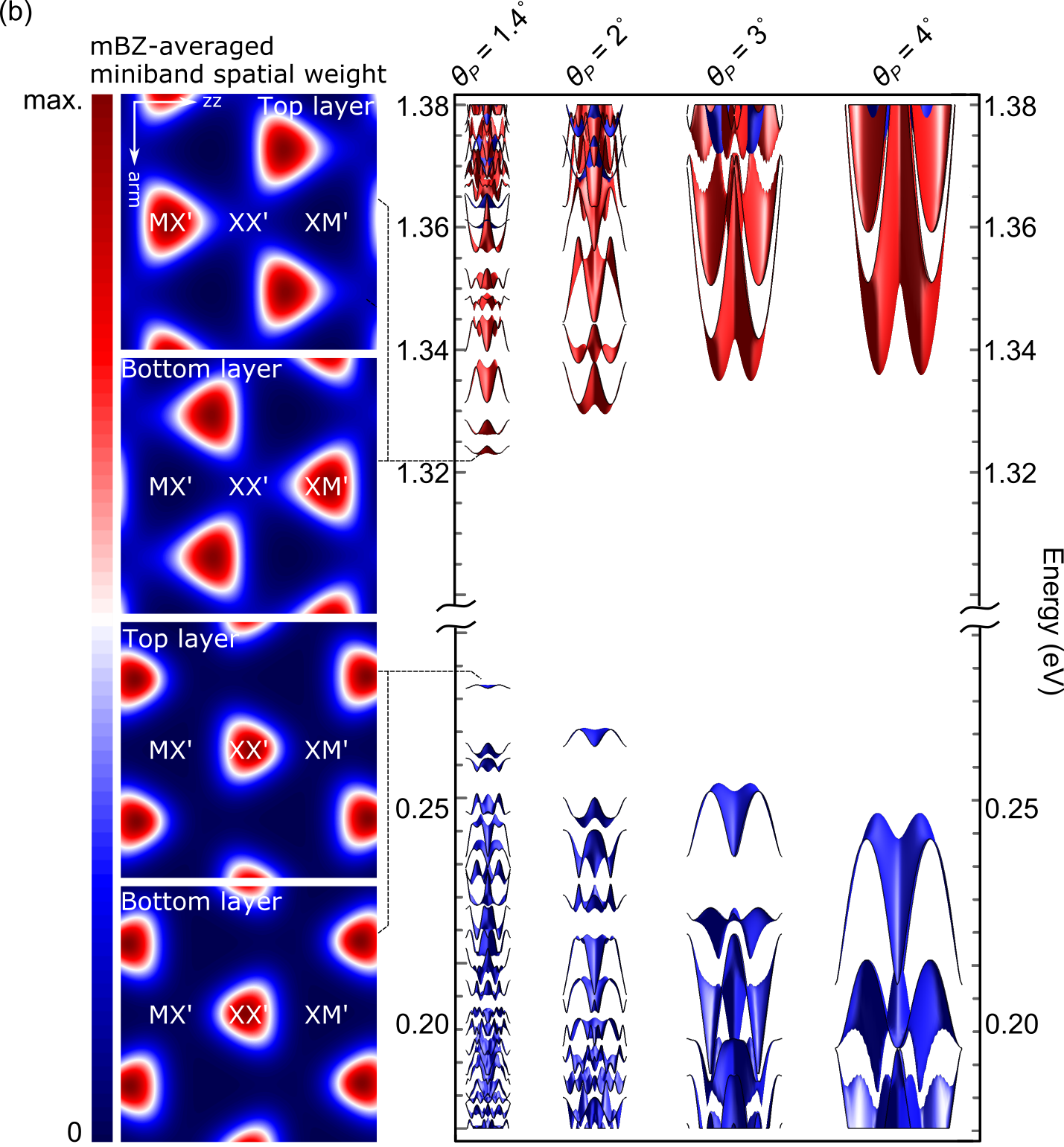}
    \caption{$K$-point moir\'e miniband spectra of twisted AP (a) and P (b)  WSe${}_2$ bilayers, at large twist angles $\theta_{P/AP}=1.4^\circ,\,2^\circ,\,3^\circ$ and $4^\circ$. Spin-up and spin-down minibands are shown in red and blue, respectively. (a) For twisted AP bilayers, the bottom (top) two flat conduction minibands appearing for $\theta_{AP}=1.4^\circ$ correspond to conduction (valence) states localised at MM$'$ (XX$'$) sites in the supercell, as shown in the top (bottom) left panels. The conduction- and valence-band edges can be effectively described by mesoscale triangular lattices with unit cells depicted in the top-right and bottom-right insets, respectively. The left inset shows part of the extended mBZ to illustrate the three minibands responsible for the Dirac-like dispersion observed at the mBZ corner for $\theta_{AP} > 2^\circ$. (b) For twisted P bilayers with twist angles $\theta_P=1.4^\circ$ and $2^\circ$, top- (bottom-) layer conduction states localise at MX' (XM') sites, as shown by the layer-resolved probability density maps in the left panels, forming two nearly decoupled triangular lattices. By contrast, valence states localise at XX$'$ regions, forming a single mesoscale triangular lattice. All energies are measured with respect to the monolayer WSe$_2$ VB edge, for direct comparison with the $K$-point miniband spectra of Fig.\ \ref{fig:GPointMinibands}.}
    \label{fig:KPointMinibands}
\end{figure*}

Figure \ref{fig:KPointMinibands}(a) shows both valence and conduction $K$-point minibands for twisted AP bilayers of WSe${}_2$, where the moir\'e superlattice is dominated by the piezoelectric potential. By contrast to the $\Gamma$ point, electronic states at the $K$ point are spin-split by the SO interaction. Spin-up (-down) states are shown in red (blue) in Fig.\ \ref{fig:KPointMinibands}. Band hybridisation takes place exclusively between same spin bands in opposite layers, which in this configuration are separated by the SO splittings. For both the valence and conduction bands, the SO splittings are a whole order of magnitude larger than the corresponding tunnelling energies (see Table \ref{tab:KpointAll}), resulting in weak interlayer hybridisation. As a consequence, the interlayer distance modulation \eqref{Eq:z_local_P} is also negligible in this case, since it only enters the hybridisation Hamiltonian.

\textbf{Valence band.} For $\theta_{AP}>2^\circ$, the top valence states are delocalised, forming two degenerate spin-polarised parabolic minibands with maxima at the mBZ wave vectors $\boldsymbol{\kappa}$ and $\boldsymbol{\kappa}'$, respectively. Note that this and the next highest miniband are separated by a gapped Dirac dispersion at the mBZ corner. This is caused by three degenerate bottom-layer plane-wave states, $|\boldsymbol{\kappa} - \boldsymbol{\Delta\KK} \rangle$, $|\boldsymbol{\kappa} - \boldsymbol{\Delta\KK} + \mathbf{g}_3\rangle$ and $|\boldsymbol{\kappa} - \boldsymbol{\Delta\KK} - \mathbf{g}_1\rangle$, with $\boldsymbol{\kappa} - \Delta\KK$ the mBZ corner, that fold upon each other and then get split by the first harmonic of the moir\'e perturbation in Eq.\ \eqref{eq:Hmini}.

In the lowest harmonics approximation, the resonant mixing of those three states and the folded plane-wave states with close wave numbers $\kk=\boldsymbol{\kappa}-\Delta\KK+\qq$ ($|\boldsymbol{\kappa} - \boldsymbol{\Delta\KK}+\qq \rangle$, $|\boldsymbol{\kappa} - \boldsymbol{\Delta\KK} + \mathbf{g}_3+\qq\rangle$ and $|\boldsymbol{\kappa} - \boldsymbol{\Delta\KK} - \mathbf{g}_1+\qq\rangle$) is described (up to an overall energy shift) by the Hamiltonian
\begin{equation}\label{eq:HcornerAP}
\begin{split}
    &H_{\rm corner}^{AP}(\mathbf{q}) = \begin{pmatrix}
    \delta_1(\mathbf{q}) & -e\varphi_1 & -e\varphi_1^* \\
    -e\varphi_1^* & \delta_{2,+}(\mathbf{q}) & -e\varphi_1 \\
    -e\varphi_1 & -e\varphi_1^* & \delta_{2,-}(\mathbf{q})
    \end{pmatrix},\\
    &\delta_1 = 2\tfrac{\hbar^2 |\Delta\KK|}{m_{{\rm VB},K}} q_x,\,\delta_{2,\pm} = -\tfrac{\hbar^2|\Delta\KK|}{m_{{\rm VB},K}}(q_x \pm \sqrt{3}q_y).
\end{split}
\end{equation}
Here, $m_{{\rm VB},K}$ is the valence-band effective mass, $\varphi_1$ is the amplitude of the first piezopotential moir\'e harmonic (see Appendix \ref{sec:minibands}), and we point out that $\mathrm{Re}\varphi_1>0$, $\mathrm{Im}\varphi_1>0$.

Diagonalising the matrix in Eq.\ \eqref{eq:HcornerAP} at $\mathbf{q}=\boldsymbol{0}$ gives the energy levels $\varepsilon_0=-2e\mathrm{Re} \varphi_1$ and $\varepsilon_\pm =  e\mathrm{Re} \varphi_1 \mp e\sqrt{3}\mathrm{Im}\varphi_1$, with $\varepsilon_{\pm}>\varepsilon_0$, and the eigenvector matrix $U$. Shifting the energy reference to $e\mathrm{Re} \varphi_1$, applying the similarity transformation $\tilde{H}_{\rm corner}^{AP}=U^{-1}H_{\rm corner}^{AP}(\mathbf{q})U$ and projecting out the level $\varepsilon_0$, we obtain the effective Hamiltonian
\begin{equation}
\begin{split}
    &\tilde{H}_{\rm corner}^{AP}(\qq)\approx\begin{pmatrix}
    \sqrt{3}\mathrm{Im}\varphi_1  & \hbar v_D q_- \\
    \hbar v_D q_+ & -\sqrt{3}\mathrm{Im}\varphi_1
    \end{pmatrix},\\
    &q_\pm = e^{\pm i \tfrac{2\pi}{3}}(q_x \pm iq_y),\,
    v_{D} = \tfrac{\hbar|\Delta\KK|}{m_{{\rm VB},K}}.
\end{split}
\end{equation}
Here, we used an expansion up to linear order in $\mathbf{q}$ to highlight the Dirac-like features identified in Fig.\ \ref{fig:KPointMinibands}(a), with $v_D$ the effective Fermi velocity, and $2\sqrt{3}\mathrm{Im}\varphi_1$ the ``Dirac mass''.

At angles $\theta_{AP}^*<\theta_{AP}<2^\circ$, the valence band edge consists of a flat ($2\,{\rm meV}$ bandwidth at $\theta_{AP}=1.4^\circ$), spin-degenerate miniband. The mBZ-averaged probability densities shown in the bottom-left panels of Fig.\ \ref{fig:KPointMinibands}(a) indicate that this miniband is formed by an array of weakly coupled states localized at superlattice sites where the piezopotential is maximum (XX$'$ areas). These are the $s_0$ QD states discussed in
Section \ref{sec:KpointAPQD}, whereas QD orbitals delocalise already at $\theta = 1.4^\circ$. Accounting for spin- and valley degeneracies, the QD states forming the top valence miniband give a realisation of the SU${}_4$ Hubbard model on a triangular lattice\cite{MITSU4}.

\textbf{Conduction band.} Remarkably, the conduction band edge consists of a spin-degenerate doublet of extremely flat ($<0.1\,{\rm meV}$ bandwidth at $\theta_{AP}=1.4^\circ$) minibands, formed by $s$-type QD states localised at MM$'$ areas of the superlattice, where the piezopotential is minimum [top-left panels of Fig.\ \ref{fig:KPointMinibands}(a)]. As a result, for twist angles $\theta_{AP} < 2^\circ$, the conduction band edge states are well described by a periodic array QD states with the supercell shown in the top-right inset of Fig.\ \ref{Fig:K_point_CB_AP}(b), also giving a realisation of the SU${}_4$ Hubbard model on a triangular lattice.

\subsection{Miniband structures near the $K$ point for twisted P bilayers}\label{sec:KminibandsP}

In contrast to the $K$-point states in AP structures, in P bilayers hybridisation is resonant, hence strong. Another difference from AP structures is that, as discussed in Sec.\ \ref{sec:twisted_P}, the valence band edge is dominated by interlayer hybridisation at XX$'$ regions, whereas for the conduction band a weaker interlayer hopping (see Table \ref{tab:KpointAll}) shifts the band edge toward  MX$'$ and XM$'$ regions for small twist angles. Figure \ref{fig:KPointMinibands}(b) shows the miniband spectra of twisted P bilayers at twist angles $\theta_P=1.4^\circ,\,2^\circ,\,3^\circ$ and $4^\circ$.

\textbf{Valence band.} For $\theta_P>2^\circ$ the top valence minibands have relatively large bandwidths ($\sim 50\,{\rm meV}$ for $\theta=4^\circ$), corresponding to delocalised carriers. However, narrow minibands appear for $\theta_P=1.4^\circ$ and $2^\circ$, corresponding to valence electrons localised at XX$'$ regions of the supercell. This is highlighted in the bottom-left panels of Fig.\ \ref{fig:KPointMinibands}(b), where we plot the layer-resolved mBZ-averaged modulus squared of the top miniband wavefunction. Note that, due to resonant hybridisation, the states are evenly spread between the two layers.

\textbf{Conduction band.} Similarly to the valence band case, $K$-point conduction electrons are delocalised for $\theta>2^\circ$, whereas for $\theta_P=1.4^\circ,\,2^\circ$ the states become localised. In the conduction band case, however, electrons are localised by the combined piezo- and ferroelectric potential at XM$'$ and MX$'$ sites, forming two separate mesoscale triangular lattices, shown in the top-left panels of Fig.\ \ref{fig:KPointMinibands}(b). Given the weak interlayer hybridisation between conduction bands (see Table \ref{tab:KpointAll}), these two lattices couple only weakly, and a graphene-like spectrum does not develop.

\subsection{Miniband structures near the $Q$ point for twisted AP and P bilayers}\label{sec:Qminibands}
\begin{figure*}[t!]
    \centering
    \includegraphics[width=\columnwidth]{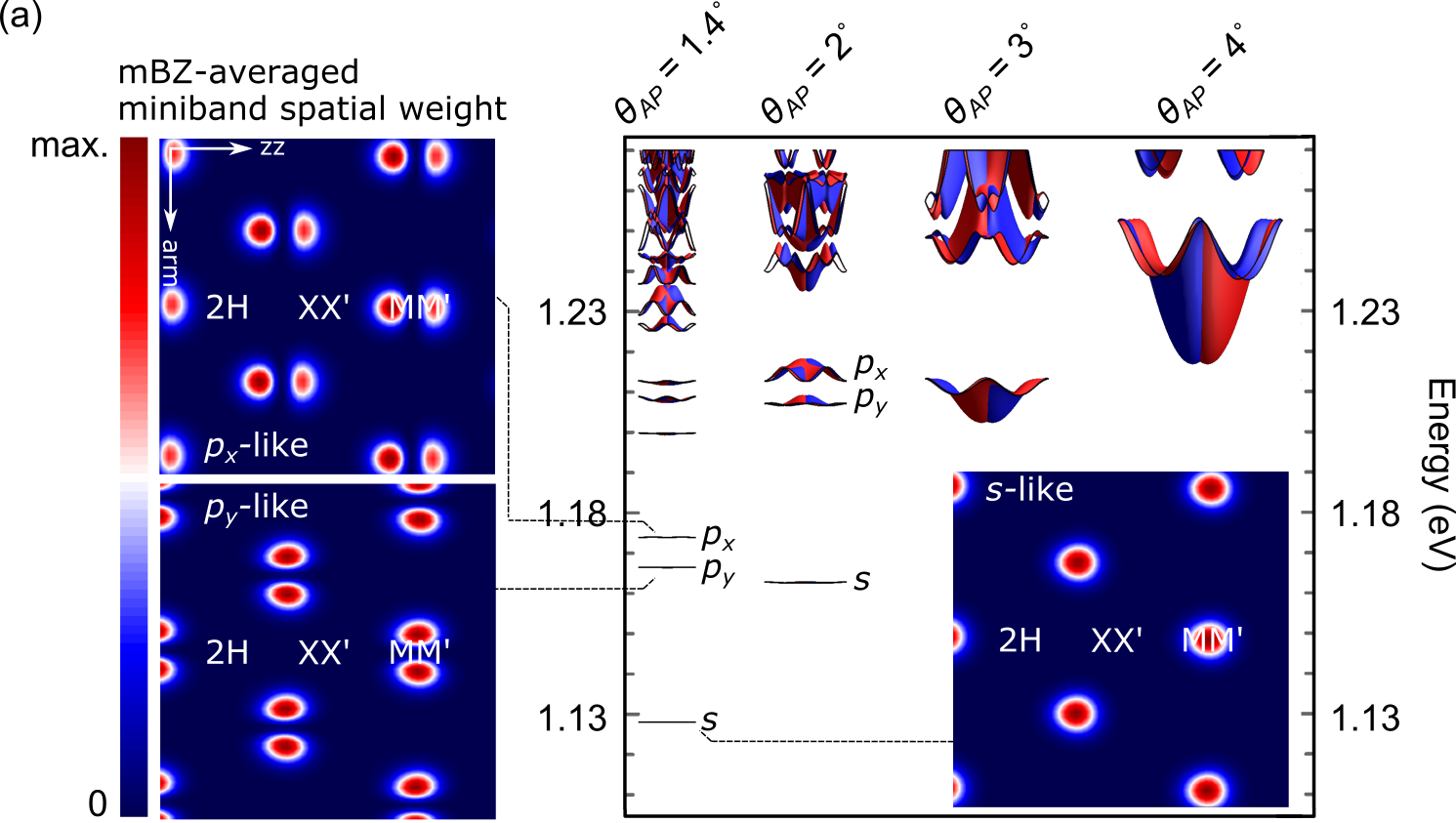}    \includegraphics[width=\columnwidth]{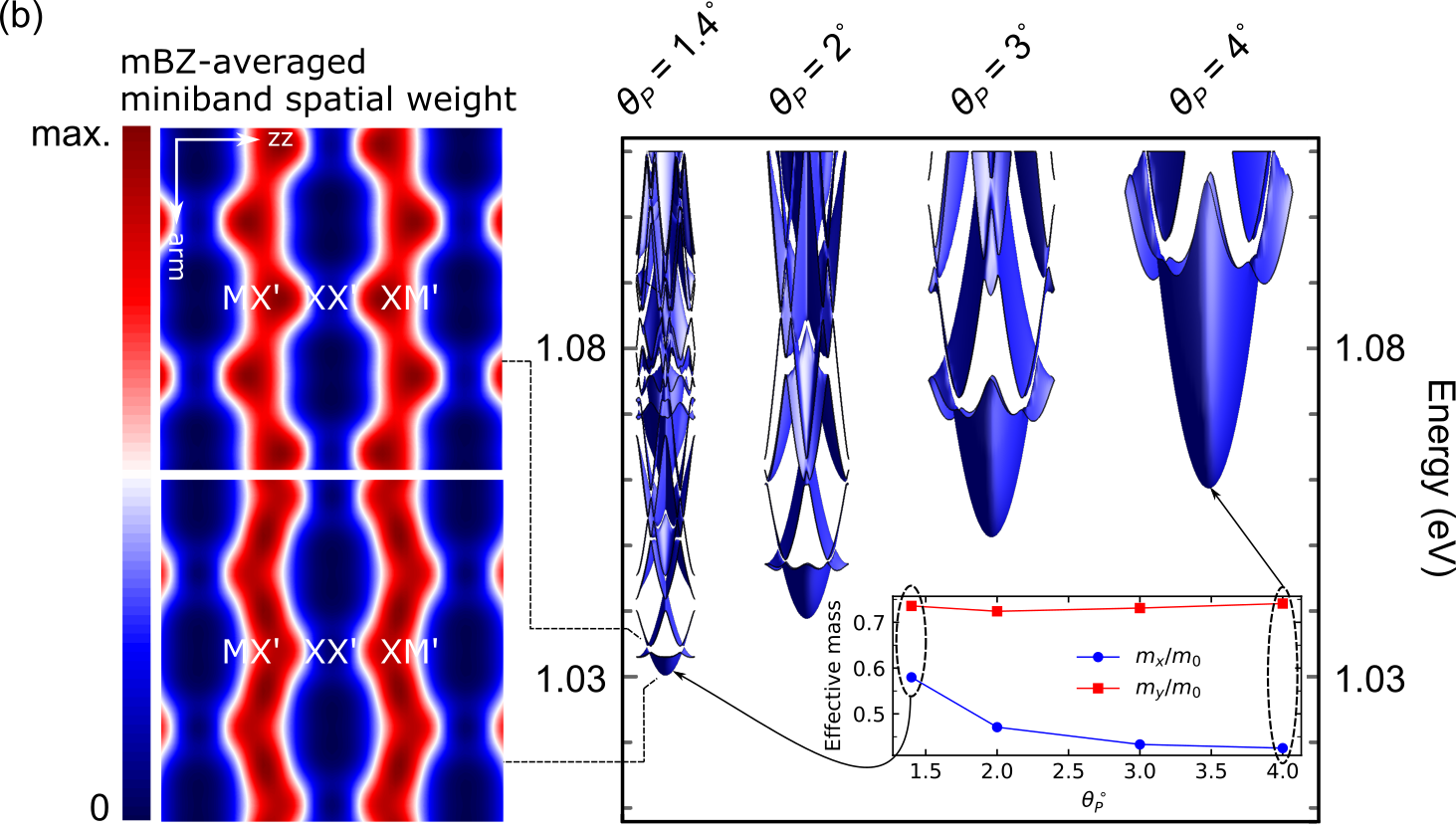}
    \caption{$Q$-point moir\'e conduction miniband spectra of twisted AP (a) and P (b)  WSe${}_2$ bilayers, at large twist angles $\theta_{P/AP}=2^\circ,\,3^\circ$ and $4^\circ$. For P bilayers, the in-plane relaxation vectors $\mathbf{u}^t(\rr)$ and $\mathbf{u}^b(\rr)$ are considered to first order in the minibands calculation, as discussed in the main text. The inset in (b) shows the effective masses (top panel) and the ratio of the root-mean-squared group velocities (bottom panel) in the $x$ (zig-zag) and $y$ (armchair) directions of the lowest miniband, for each of the twist angles considered. The left panels in (a) and (b) and the inset of (a) show the mBZ-averaged modulus squared of the wavefunction, representing the probability of finding a valence electron at a given position in the superlattice. All energies are measured with respect to the monolayer WSe${}_2$ VB edge.}
    \label{fig:QPointMinibands}
\end{figure*}

Figures \ref{fig:QPointMinibands}(a) and \ref{fig:QPointMinibands}(b) show the $Q_1$-point conduction minibands of twisted AP and P bilayers, respectively, for large twist angles $\theta_{P/AP}=1.4^\circ,\,2^\circ,\,3^\circ$ and $4^\circ$.

\textbf{AP structures} display spin-degenerate, flat minibands for twist angles $\theta_{AP}^*<\theta_{AP}\lesssim 2^\circ$ [Fig.\ \ref{fig:QPointMinibands}(a)], corresponding to arrays of localised states at MM$'$ regions of the superlattice. $Q_1$ point electrons are localised by the piezopotential and interlayer hybridisation, which combine to produce deep potential wells at these sites. The inset and two left panels of Fig.\ \ref{fig:QPointMinibands}(a) show the mBZ-averaged moduli squared of the three lowest miniband wave functions for $\theta_{AP}=1.4^\circ$. The lowest of these is an $s$-type state with slight trigonal warping, followed by a $p_y$-like state, and then by a $p_x$-like state at higher energy. The splitting of the two $p$ states is caused by the mass anisotropy of the monolayer states, and their order is a consequence of the fact that $m_y > m_x$.

In addition to spin degeneracy, each of these localised states is also degenerate with the inequivalent $-Q_1$ and $\pm Q_{2}$ and $\pm Q_{3}$ valleys, giving a total degeneracy factor of 12. Therefore, in the presence of interactions, we predict that $n$-doped twisted WSe${}_2$ constitutes a realisation of the SU${}_N$ Hubbard model with large-$N$ for the smallest twist angles \cite{Affleck_largeN}.

In the case of \textbf{P structures} with twist angles $\theta_P < 2^\circ$, we find that the $Q_1$-point minibands are more sensitive to neglecting in-plane relaxation of the lattice. Therefore, we expanded the relaxation field in moir\'e harmonics as
\begin{equation}\label{eq:ufield}
    \mathbf{u}^t(\rr)-\mathbf{u}^b(\rr) = \sum_{j}\mathbf{u}_j\,\sin{\left(  \mathbf{g}_{j} \cdot \rr\right)}.
\end{equation}
The expansion coefficients $\mathbf{u}_j$ are reported in Appendix \ref{sec:minibands}. Then, with a procedure analogous to that leading to Eq.\ \eqref{eq:howtominibands}, we approximate the $Q_1$-point interlayer hybridisation term as ($\mathbf{V}_0=\boldsymbol{0}$,\,$\mathbf{V}_1=-\GG_1$,\,$\mathbf{V}_2=\GG_2$,\,$\mathbf{V}_3=\GG_3$,\,$\mathbf{V}_4=-\GG_3$)
\begin{widetext}
\begin{equation}\label{eq:QPointRelax}
\begin{split}
    e^{i\Delta\QQ \cdot \rr}T_Q^P(\rr)=&e^{i\Delta\QQ \cdot \rr}\sum_{\ell=0}^4t_{Q,\ell}^P e^{-q_{t_{Q,\ell}}z(\rr)}e^{i\mathbf{V}_\ell\cdot\left[\theta_P \hat{\mathbf{z}}\times\rr + \mathbf{u}^t(\rr)-\mathbf{u}^b(\rr) \right]}\\
    =&e^{i\Delta\QQ \cdot \rr}\sum_{\ell=0}^4\tilde{t}_{Q,\ell}^P e^{-q_{t_{Q,\ell}}\sum_{j=1}z_j^s\cos{\left(\gG_j\cdot\rr \right)}}e^{i\mathbf{v}_\ell\cdot\rr}e^{i\sum_{j}\mathbf{V}_\ell\cdot \mathbf{u}_j\,\sin{\left(\gG_{j} \cdot \rr \right)}}\\
    \approx&e^{i\Delta\QQ \cdot \rr}\sum_{\ell=0}^4\Big[ \tilde{t}_{Q,\ell}e^{i\mathbf{v}_\ell\cdot\rr} + \sum_{j=1}\frac{\tilde{t}_{Q,\ell} \mathbf{V}_\ell\cdot \mathbf{u}_{j}}{2}\left(e^{i(C_3^\mu\gG_{j} + \mathbf{v}_\ell)\cdot\rr} - e^{-i(C_3^\mu\gG_{j} - \mathbf{v}_\ell)\cdot\rr} \right)\\
    &\qquad\qquad\qquad\qquad\qquad- \frac{q_{t_{Q,\ell}}\tilde{t}_{Q,\ell}}{2}\sum_{j}z_j^s \left(e^{i(\gG_j+\mathbf{v}_\ell)\cdot\rr} + e^{-i(\gG_j-\mathbf{v}_\ell)\cdot\rr} \right) \Big],
\end{split}
\end{equation}
\end{widetext}
where $\tilde{t}_{Q,\ell}\equiv e^{-q_{t_{Q,\ell}}z_0}t_{Q,\ell}$; the moir\'e vectors $\mathbf{v}_0=\boldsymbol{0}$, $\mathbf{v}_1=-\gG_1$, $\mathbf{v}_2=\gG_2$, $\mathbf{v}_3=\gG_3$ and $\mathbf{v}_4=-\gG_3$ appear from the approximation $\GG_\ell \cdot (\theta \hat{\mathbf{z}}\times \rr) \approx \gG_\ell \cdot \rr$; and we have used the fact that, for P bilayers, all coefficients $z_j^a$ in Eq.\ \eqref{Eq:z_local_P} vanish (see Appendix \ref{sec:minibands}). The matrix elements are then obtained by integrating Eq.\ \eqref{eq:QPointRelax} between two plane-wave states, as in Eq.\ \eqref{eq:H12} above. The minibands shown in Fig.\ \ref{fig:QPointMinibands}(b) were computed based on these matrix elements, and their analogs for the intralayer term $\varepsilon_Q^P(\rr)$.

The P bilayer minibands exhibit dispersive bands for all twist angles considered, as shown in Fig.\ \ref{fig:QPointMinibands}(b). Miniband formation is dominated by the modulation of interlayer hybridisation across the lattice, caused by the variation of the interlayer distance and the in-plan relaxation field, and only weakly affected by the piezo- and ferroelectric potentials. The anisotropic features of the electron states are more clearly appreciated in the mBZ-averaged moduli squared of the wave functions, in the left panels of Fig.\ \ref{fig:QPointMinibands}(b). Mirroring the band edge landscapes of Fig.\ \ref{Fig:P_Q_CB_edge}, electrons in the bottom two minibands for $\theta_P=1.4^\circ$ are confined into channels that run roughly along the armchair direction, passing through MX${}'$ and XM${}'$ regions of the moir\'e superlattice, and avoiding XX${}'$ areas where interlayer hybridisation is weak.

For all twist angles considered, we quantify the lowest miniband anisotropy by computing the effective masses along the $x$ (zig-zag) and $y$ (armchair) directions, shown in the inset of Fig.\ \ref{fig:QPointMinibands}(b). The large monolayer mass anisotropy is recovered for large twist angles $\theta_P=4^\circ$, whereas in the small twist angles regime anisotropy is somewhat reduced by the interlayer hybridisation effect. In the presence of interactions, the anisotropic states of valley $Q_1$ discussed in Fig.\ \ref{fig:QPointMinibands} must be considered simultaneously with those of valleys $Q_2$ and $Q_3$, with which it is connected by $C_3$ rotations, and their corresponding time reversal partners $-Q_1$, $-Q_2$ and $-Q_3$.

\section{Tuning aligned bilayers by strain, pressure, electric field, and encapsulation}\label{sec:tuning_aligned_bilayers}
The TMDs feature in their valence/conduction bands competing local maxima/minima, known as `valleys'. These are the $\Gamma$ and K/K$'$ valleys in the valence band, and the Q and K/K$'$ valleys in the conduction band. Depending on the member of the TMD family involved, on the number of layers, and on external factors, which of these valleys form the band edges in an ultrathin TMD film can be tuned. 

From our discussions above, and from the band energies set out in Table \ref{tab_wfns}, we note that the energy differences between the local valence band maxima at $\Gamma$ and $K$ are small, with the band edge of 2H (MX$'$/XM$'$) bilayer WSe$_2$ at $K$ only around 40~meV(80~meV) above that at $\Gamma$. The conduction band also features two competing minima in the Brillouin zone, with the valleys at $Q$ and $K$ $\sim$ 130~meV apart in energy.

This opens up the possibility of controlling which valleys form the conduction and valence band edges using external parameters, such as displacement fields from gating and modification of the lattice parameters through strain\cite{Wang2014, Steinhoff2014, Roldn2015, PhysRevB.87.125415, PhysRevB.87.235434, Frisenda2017, PhysRevB.100.195126, Yan2020} and pressure (the latter of which may be induced by electrostatic attraction between top and back gates).

\begin{table*}
\caption{
{\color{black}
Wave function projection onto spherical harmonics in band edge states (Q approximated as being at K/2) of
aligned 2L-WSe$_2$ in 2H and MX$'$ stacking (M$_i$ and X$_i$ are the total metal and chalcogen
contributions on layer $i$), the expectation value of the electric dipole moment
in each wave function caclulated from the orbital projections ($d_z$), and the band energy (0~eV set to mean of vacuum levels). For 2H stacking, the band energies appear in spin-degenerate pairs due to inversion symmetry. At $\Gamma$, their wavefunctions are spread symmetrically over the layers. Elsewhere in the Brillouin zone within each degenerate pair, bands of opposing spin are polarized on opposing layers, with the polarization quantified by the dipole moment $d_z$. The spin-degeneracy of the bands in the 2H case is absent for MX$'$ stacking due to the inversion asymmetry of P bilayers.\label{tab_wfns}
}
}
\begin{tabular}{ccccccc}
 \hline
 \hline
 2H-WSe$_2$ &&&&&&\\
 band & M$_1$ &X$_1$ & M$_2$ &X$_2$ & $d_z (e \cdot \mathrm{\AA})$ & Energy (eV) \\
 \hline
VB,   $\Gamma$&  0.355&  0.127&  0.355&  0.127&  0.000& $-4.908$\\
VB-1, $\Gamma$&  0.355&  0.127&  0.355&  0.127&  0.000& $-4.908$\\
VB-2, $\Gamma$&  0.395&  0.104&  0.395&  0.104&  0.000&$-5.555$\\
VB-3, $\Gamma$&  0.395&  0.104&  0.395&  0.104&  0.000&$-5.555$\\
VB,   $K$     &  0.800&  0.175&  0.019&  0.007& $-3.070$& $-4.858$\\
VB-1, $K$     &  0.019&  0.019&  0.800&  0.175&  3.070& $-4.858$\\
VB-2, $K$     &  0.820&  0.165&  0.016&  0.002& $-3.130$& $-5.332$\\
VB-3, $K$     &  0.016&  0.002&  0.820&  0.165&  3.130& $-5.332$\\
CB,   $K$     &  0.931&  0.068&  0.002&  0.000&  $-3.229$& $-3.504$\\
CB+1, $K$     &  0.002&  0.000&  0.931&  0.068& 3.229&$-3.504$\\
CB+2, $K$     &  0.941&  0.060&  0.000&  0.000&  $-3.242$&$-3.460$\\
CB+3, $K$     &  0.000&  0.000&  0.941&  0.060& 3.242&$-3.460$\\
CB  , $Q$     &  0.230&  0.088&  0.461&  0.221&  1.194&$-3.634$\\
CB+1, $Q$     &  0.461&  0.221&  0.230&  0.088 & $-1.194$&$-3.634$\\
CB+2, $Q$     &  0.244&  0.102&  0.467&  0.176&  1.044&$-3.160$\\
CB+3, $Q$     &  0.467&  0.176&  0.244&  0.102& $-1.044$&$-3.160$\\
 \hline
 \hline
\end{tabular}
\quad
\begin{tabular}{ccccccc}
 \hline
 \hline
MX$'$-WSe$_2$ &&&&&&\\
 band & M$_1$ &X$_1$ & M$_2$ &X$_2$ & $d_z (e \cdot \mathrm{\AA})$ & Energy (eV)  \\
 \hline
VB,   $\Gamma$&  0.384&  0.151&  0.329&  0.137& $-0.215$&$-4.910$\\
VB-1, $\Gamma$&  0.384&  0.151&  0.329&  0.137& $-0.215$&$-4.910$\\
VB-2, $\Gamma$&  0.367&  0.100&  0.423&  0.111&  0.202&$-5.501$\\
VB-3, $\Gamma$&  0.367&  0.100&  0.423&  0.111&  0.202&$-5.501$\\
VB,   $K$     &  0.810&  0.180&  0.006&  0.004& $-3.186$&$-4.820$\\
VB-1, $K$     &  0.001&  0.001&  0.818&  0.178&  3.227&$-4.888$\\
VB-2, $K$    &  0.822&  0.170&  0.004&  0.002& $-3.209$&$-5.279$\\
VB-3, $K$     &  0.002&  0.001&  0.828&  0.169&  3.223&$-5.346$\\
CB,   $K$     &  0.002&  0.000&  0.931&  0.067&  3.232&$-3.530$\\
CB+1, $K$     &  0.001&  0.000&  0.940&  0.059&  3.240&$-3.489$\\
CB+2, $K$     &  0.932&  0.066&  0.002&  0.000& $-3.234$&$-3.474$\\
CB+3, $K$     &  0.939&  0.059&  0.002&  0.000& $-3.232$&$-3.433$\\
CB,   $Q$     &  0.249&  0.125&  0.418&  0.208&  0.816&$-3.647$\\
CB+1, $Q$     &  0.301&  0.121&  0.412&  0.167&  0.500&$-3.512$\\
CB+2, $Q$     &  0.429&  0.158&  0.263&  0.116& $-0.824$&$-3.323$\\
CB+3, $Q$     &  0.409&  0.157&  0.318&  0.118& $-0.431$&$-3.098$\\
 \hline
 \hline
\end{tabular}
\end{table*}

\subsection{Modulation of band edges by pressure and strain}
To investigate the effects of tuning via external parameters, we begin by exploiting the models presented above. We use their description of the interlayer distance dependence of valence band hybridisation to describe the change in the location of the valence band maximum (VBM) from $\Gamma$ to $K$ as the interlayer distance reduces under pressure. To do this, we make two approximations. 
First, we note that the variation in the $K$-point energy with interlayer distance is much smaller than that at $\Gamma$, so we approximate the change in $\Gamma$-$K$ splitting by the change in the $\Gamma$-point energy alone. 
Second, we keep only the dominant contributions to the interlayer hybridisation model, $H_{\Gamma}^{AP}$ from Eq.\ \eqref{Eq:H0_Gamma}. With these two approximations, the variation in the $\Gamma$-$K$ splitting can be written as
\begin{equation}
\label{eq:press_dependence}
    \frac{dE_{\Gamma-K}}{dz} \simeq \frac{dv^{AP}_{\Gamma,0}(z)}{dz}+\frac{1}{2}\left[\frac{dt^{AP}_{0}(z)}{dz}-\frac{3}{2}\frac{dt^{AP}_{1}(z)}{dz}\right],
\end{equation}
with the functions $v^{AP}_{\Gamma,0}(z)$, $t^{AP}_{0}(z)$, and $t^{AP}_{1}(z)$ defined and parametrized in Table \ref{tab_fit_splitting}. For 2H stacking we use $z=-0.23$~\AA, which gives $\frac{dE_{\Gamma-K}}{dz}\simeq -470$~meV/\AA. To convert this into a pressure sensitivity, we estimate from Eq. (\ref{Eq:Adhesion_energy}) that a 1\% change in $d$ for 2H stacking can be acheived with a pressure of 4.29\,kbar corresponding to a sensitivity to pressure of $\sim$7~meV/kbar.

\begin{figure}
\includegraphics[width=1\linewidth]{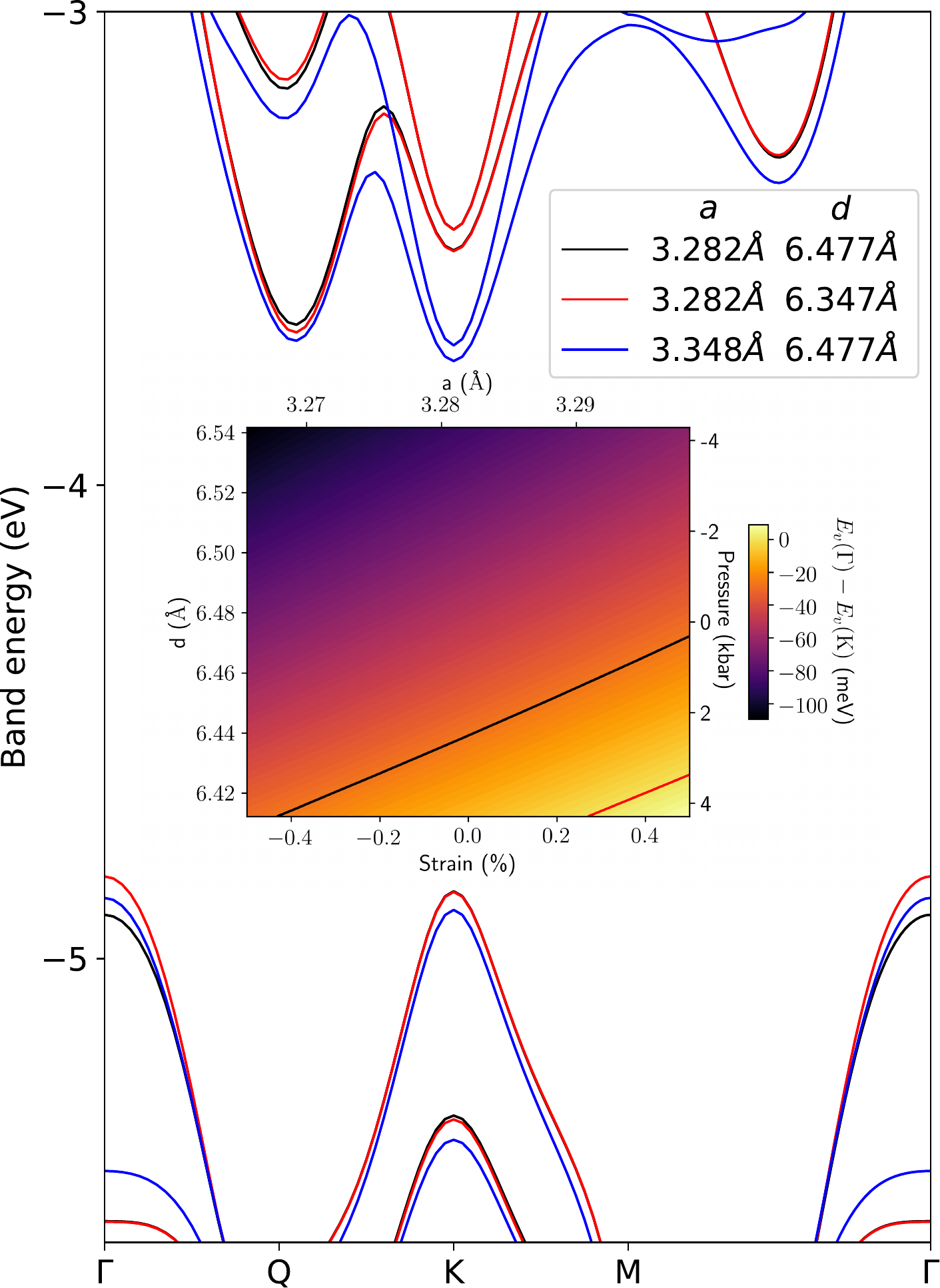}
\caption{
\label{Strain}
Band structure of 2H stacked 2L-WSe$_2$ for 0 strain and pressure (black lines), 8.6~kbar uniaxial pressure (red lines), and 2\% biaxial tensile strain (blue lines). Bands are aligned relative to vacuum. Inset: map of difference between $\Gamma$- and $K$-point valence band edge energies under strain and pressure. The red line shows the boundary across which the band edge moves from K to $\Gamma$, while the black line separates the regions in which the lowest energy exciton involves a hole at $K$ or at $\Gamma$.}
\end{figure}

We compare the results from those using Eq.\ \eqref{eq:press_dependence} with results directly calculated from DFT. Bands for bilayers of 2L-WSe$_2$ in
2H stacking calculated using DFT are exemplified in Fig.\ \ref{Strain},
for structural parameters of the monolayer taken
from experiments\cite{schutte1987crystal}.
We also show the band dispersions computed for a slightly smaller
interlayer separation (corresponding to a pressurized material). As found through the modelling above, a reduction in interlayer distance through pressure increases the interlayer hybridisation and band splitting at the $\Gamma$ point, reducing the difference between the valence band edge at $K$ and the local maximum at $\Gamma$.
The change in $\Gamma$-$K$ splitting with pressure found directly from DFT is approximately 9~meV/kbar, close to that found from the model keeping only the most dominant terms.

In Fig.\ \ref{Strain}, we also show DFT results using a slightly inflated lateral lattice constant (mimicking biaxial strain).
These show that strain may also be used to tune the location of the VBM in the BZ between the $K$-point and the $\Gamma$-point.
Quantitatively, the $\Gamma$-$K$ valence band splitting  in 2H bilayer WSe$_2$ varies, to linear order at a rate of $\sim$40~meV/strain(\%), from which we can predict that the crossover from the VBM being located at $K$ to being at $\Gamma$ would take place at $\sim$1\% strain. 

The conduction band edge of WSe$_2$ bilayers exhibits some tunability as well.
In the equilibrium structure, the conduction band minimum (CBM) is at the $Q$-point,
and there is a local minimum at $K$.
The spliting between the K- and Q-point CB minima ($\sim$130~meV) is much greater than that between K and $\Gamma$ in the VB, and the sensitivity of the energy splitting between the $Q$ and $K$ valleys to interlayer distance is weak, so changing the inter-layer spacing $d$
over the ranges considered here
does not affect the position of the CBM.
However, there is a strong sensitivity of the CB K-point energy to strain, with the K-Q splitting tunable by 100~meV/\% strain, such that $>1.5$\% strain can push the $K$-valley
below the $Q$-valley. This effect has been demonstrated experimentally in the onset of much stronger photoluminescence in strained bilayer WSe$_2$\cite{Desai2014} as the CBM moves to $K$. 
The tunability of the conduction band may be able to influence
superconductivity which can be engineered in
n-doped 2D crystals\cite{shi2015superconductivity}. Changing the position of the
CBM will impact the density of states, in part due to the change in effective masses,
but also because one valley has a 6-fold degeneracy in the Brillouin zone ($Q$) while
the other is only 2-fold degenerate ($K$).

We also use the DFT results to estimate
the effective mass of the valence band holes in bilayer WSe$_2$ and find very different
$m_{VB}(\Gamma) = -1.14\ m_e$
and $m_{VB}(K) = -0.37\ m_e$ masses 
(as well as conduction band masses of
$m_{CB}(K) = 0.41\ m_e$, $m_{CB}^{xx}(Q) = 0.45\ m_e$,
and $m_{CB}^{yy}(Q) = 0.62\ m_e$).
Then, we feed these values into an analytical interpolation formula based on
diffusion Monte Carlo calculations\cite{PhysRevB.96.075431} for 2D materials
with the Keldysh interaction and evaluate the exciton binding energies, where
we use the value of the screening length
$r_{*}=45.11$~\AA~ from $GW$ calculations \cite{PhysRevB.88.045318} for
monolayer WSe$_2$ which doubles for the bilayer, resulting in $r_{*}=90.22$~\AA.

With the help of the code provided in the Supplemental Material of
Ref. \citenum{PhysRevB.96.075431}, we find that the
binding energy for excitons comprising an electron at $Q$ with
a hole at $K$ or $\Gamma$
for 2L-WSe$_2$ encapsulated in hexagonal boron nitride
($\varepsilon_{hBN}=3.73$\cite{barth1998situ})
$E_b^{VB(\Gamma) \rightarrow CB(Q)}=162$~meV and
$E_b^{VB(K) \rightarrow CB(Q)}=135$~meV, respectively.
The 27~meV difference between the two exciton binding energies
promotes $\Gamma$-point hole excitons, which moves
the boundary between regions in which the lowest energy
exciton involves a hole at $K$ or $\Gamma$. 
For completeness, we also estimate exciton binding energies for
suspended bilayers, for which we obtain $E_b^{VB(\Gamma) \rightarrow CB(Q)}=334$ meV and
$E_b^{VB(K) \rightarrow CB(Q)}=299$ meV. In the conduction band, the difference between the exciton binding experienced by an electron at K compared with one at Q is smaller, by $\sim 6$~meV. Given the $\sim$100meV/\% dependence of the K-Q splitting on strain, the effect of the difference in exciton binding on the K/Q band edge crossover is a small one.

\begin{figure}
\includegraphics[width=1\linewidth]{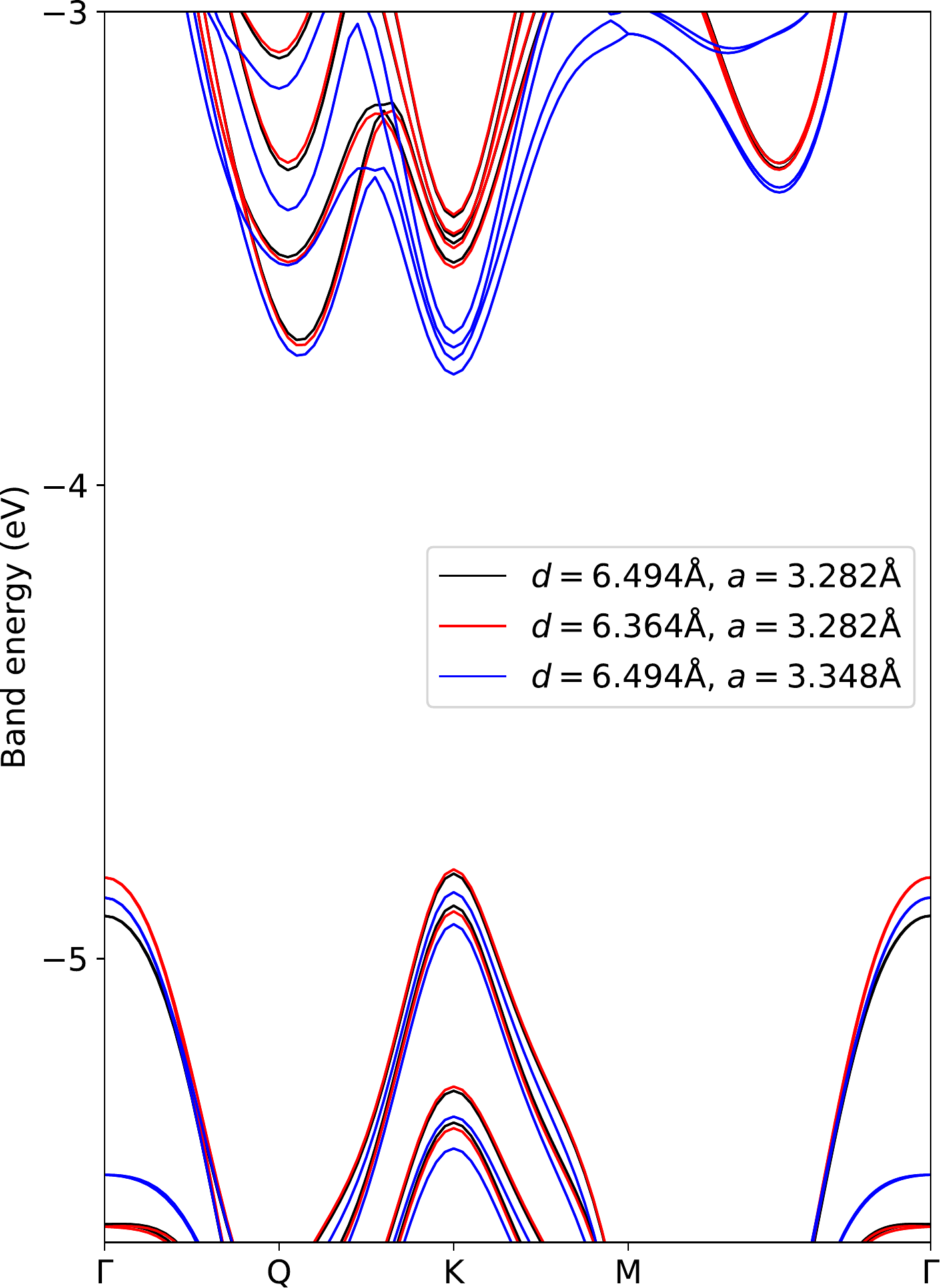}
\caption{
\label{Strain_3R}
Band structure of MX$'$/XM$'$ stacked 2L-WSe$_2$ for 0 strain and pressure (black lines), 8.6~kbar uniaxial pressure (red lines), and 2\% biaxial tensile strain(blue lines). Bands are aligned relative to mean of vacuum levels on either side of bilayer.}
\end{figure}

In MX$'$/XM$'$ P-stacked 2L-WSe$_2$, the small splitting in the band edges at the $K$-point described above increases the energy difference between the VBM itself and the local VBM at $\Gamma$, and reduces that between the conduction band edges at $Q$ and $K$. Since the behaviour of the band edge energies under pressure and strain, as shown in Fig. \ref{Strain_3R},
 remains very similar to that of the 2H-stacked bilayer, this results in a larger pressure and/or strain
being needed to realize a transition of the VBM to the $\Gamma$-point, while the crossover for the CBM from $Q$ to $K$ can be expected to happen at a smaller strain.
The exciton binding energies can be obtained for 3R stacking in the same way as set out for 2H stacking above. With the exception of the valence band effective mass at $\Gamma$ (which increases from 1.15\ $m_e$ in 2H stacking to
1.26\ $m_e$ in MX$'$/XM$'$ stacking) all band edge effective masses change by no more than 3\% between the two types of stacking, which leads to no more than
4 meV change in the values of exciton binding energies in MX$'$/XM$'$ stacking as compared to 2H.

\subsection{Modulation of band edges by electric displacement field}

A vertical electric field can also be used to tune the band edge alignment in the
bilayer, because it splits the non-hybridized $K$-point band edges
but does not change much the energy of the strongly hybridized layer-symmetric
and -antisymmetric states at $\Gamma$, and has an opposite effect to pressure and strain,
further promoting the VBM at the $K$-point. In Table \ref{tab_wfns} we give the electric dipole moment, $d_z$, of each wave function as obtained
from the orbital projections. Note that in the case of 2H stacking, the inversion
symmetry results in spin degenerate states where each spin component on its own
has a dipole moment (except at the $\Gamma$-point), where each component can 
be localized on a separate layer. Within each degenerate pair, the bands have the same magnitude of $d_z$  but with opposing signs, thus giving a band splitting away from $\Gamma$ of magnitude $2|E_zd_z|$, where $E_z$ is the perpendicular electric field across the WSe$_2$ bilayer. Since the $\Gamma$-point VB energy is not affected at linear order by an electric field, this has the effect of increasing the difference in energy between the $K$-point VBM and the VB at $\Gamma$ by an amount $|E_zd_z|$. For 2H WSe$_2$, an electric field $|E_z|=0.1$~V/nm would be expected to increase the energy of the top valence band at $K$ relative to that at $\Gamma$ by $\sim 30$~meV.

For MX$'$/XM$'$ stacking, as noted in the case of bilayer MoSe$_2$\cite{Sung2020} the energy splitting between the layer-polarized states discussed above will either increase or decrease, depending on the direction of the applied electric field relative to the orientation of the domain (that is, MX' or XM'). Where the applied field points in a direction opposite to the intrinsic field due to charge transfer, it will decrease the splitting of the top two valence bands at $K$ (so decreasing the difference between the VBM at $K$ and the VB at $\Gamma$) through the addition of $E_zd_z$ to the energies of VB and VB-1 (see Table \ref{tab_wfns}). This will lead to the two layer-polarized bands approaching each other when $E_z \sim +0.1$~V/nm, beyond which their splitting will increase once more.

In terms of the conduction band, the splitting at zero field between the $Q$- and $K$-point minima (130~meV and 117~meV for 2H and MX$'$/XM$'$ stacking, respectively) means that a much larger electric field would be required to move the band edge from the $Q$-point to the $K$-point. Furthermore, the $Q$-point conduction band also has a finite dipole moment, so it also shifts at linear order under an applied field, albeit at a slower rate. Taken together, this means that an applied field $E_z\sim 0.7$~V/nm would be required to bring the $Q$- and $K$-point minima to a similar energy.

\subsection{Encapsulation effect on the band edge alignment}
So far, except for strong perturbations from external influences such as strain and pressure, and independently of twist angle in twisted bilayers, the band edges in our modelling have been found to be at the $K$- and $Q$-points for the valence and conduction bands, respectively, of WSe$_2$ bilayers. Even so, we have considered maps and minibands associated with the valence $\Gamma$-point and conduction $K$-point states in this study. This is because, while the DFT results from which parameters for the models were obtained considered WSe$_2$ suspended in vacuum, real experimental devices will often be constructed featuring WSe$_2$ in contact with other 2D (or bulk) materials, which could affect the relative energies of the band edge states and other local maxima/minima.  

The construction of 2D material heterostructures can be associated with hybridisation between the bands of the material of interest with those of the encapsulating material, in particular where orbitals (such as $s$ and $p_z$) are concerned, where the wavefunctions extend from the 2D material surface and overlap at the interface between the encapsulated and encapsulating materials. Encapsulation of 2D materials by hexagonal boron nitride (hBN) is a common step undertaken in the fabrication of high-quality 2D material-based devices\cite{Dean2010, Lee2013, Kretinin2014, Bandurin2016}. hBN is chosen as an encapsulating material for reasons including its stability in air, as an atomically flat interface, and since its large band gap allows a type II band alignment with many materials of interest. Care must still be taken, however, to understand whether any of the hBN and 2D material bands may be aligned closely enough for the effects of interlayer hybridisation between them to become significant. 

\begin{figure}
\includegraphics[width=1\linewidth]{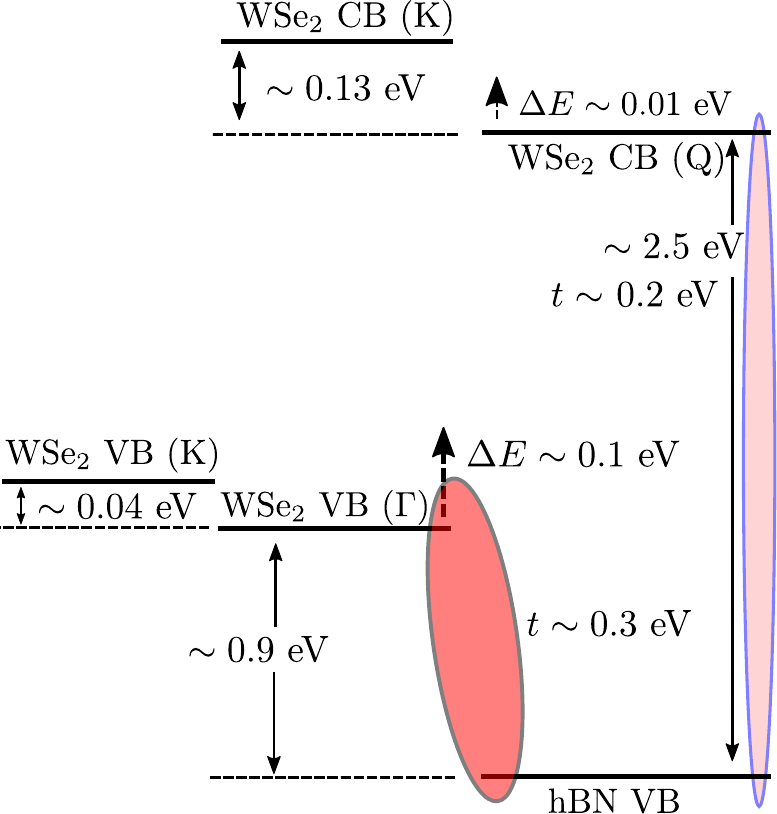}
\caption{
\label{fig:encaps_sketch}
Schematic band diagram of WSe$_2$ encapsulated by hBN, showing how hybridisation between the hBN and the $\Gamma$-point valence band of WSe$_2$ could potentially push the $\Gamma$-point valence band above the maximum at $K$, while such effects on the Q-point conduction band are likely to be too small to change the band edge location.}
\end{figure}
In Fig. \ref{fig:encaps_sketch} we sketch the band alignment of thick hBN and 2H-bilayer WSe$_2$, approximated from the band alignment of graphene and hBN\cite{Ogawa2019}, that of graphene and WSe$_2$\cite{Wilson2017}, and a scanning tunneling spectroscopy measurement of the quasiparticle band gap\cite{Park2016}. 
The valence band of hBN lies only a small difference in energy ($\sim 0.9$~eV) below that of WSe$_2$, and could hybridize strongly with the $\Gamma$ point valence band, while we expect effects on the K-point to be weak as the K-point states are localized on the inside of the WSe$_2$ layers. 
Using the resonant $\Gamma$-point valence band interlayer tunnelling strength for WSe$_2$ as an order-of-magnitude estimate of the likely strength of tunnelling resulting from hybridisation between the $\Gamma$ point states of WSe$_2$ and hBN, we can estimate the magnitude of the resulting upwards shift in energy of the WSe$_2$ $\Gamma$-point valence band states. 
This is found to be $\sim 100$~meV, which would be enough to change the order of the valence band valleys, shifting the valence band maximum to the $\Gamma$-point.
A similar but much smaller effect (we estimate the energy shift to be $\sim 10$~meV) could also be expected for the $Q$-point in the conduction band, which would reduce the splitting slightly between the Q and $K/K'$ valleys\cite{PhysRevResearch.3.023047}. 
Such effects of interlayer hybridisation should also be considered in other heterostructures, such as WSe$_2$-InSe\cite{Ubrig2020}.

\section{Discussion}\label{sec:discussion}

In this work, we have developed hybrid $\mathbf{k}\cdot\mathbf{p}$ tight-binding models to study the variation of conduction and valence band edge energies and moir\'e miniband structures in twisted P- and AP-WSe$_2$ bilayers caused by a number of factors: lattice reconstruction, piezoelectric charges, interlayer hybridisation and interlayer charge transfer. 

In marginally twisted AP-WSe$_2$ bilayers ($\theta_{AP}\lesssim 1.0^{\circ}$) the conduction and valence band edges form arrays of quantum dots for $K$- and $-K$-valley electrons and holes, respectively, formed by reconstruction-induced piezopotentials. Each quantum dot hosts localised states labeled by orbital momentum, that develop into flat moir\'e minibands upon account of the weak coupling between neighboring quantum dots, giving way to ordinary parabolic bands at larger twist angles ($1.0^{\circ}\lesssim \theta_{AP}\lesssim 2^{\circ}$). 

Here, we note that prevalence of a $K$-point band edge over $\Gamma$-valley states results from DFT calculations carried out for suspended samples. However, hBN encapsulation of the WSe$_2$ bilayers may influence the order of the $\Gamma$- and $K$-point state energies in the valence band. This is because $\Gamma$-valley states, formed by metal $d_{z^2}$ and chalcogen $p_z$ orbitals, will experience stronger hybridisation (repulsion) with hBN states, as compared to $K$-valley states, which consist of metal $d_{(x\pm iy)^2}$ orbitals.

For marginally twisted P-WSe$_2$ bilayers we found that the $K$-point conduction- and valence band edges and the $\Gamma$-point valence band edge lie in triangular domains with MX$'$ and XM$'$ registries, at extrema of the total piezo- and ferroelectric potentials, with the global valence band maximum located at the $K$-point. These large domains are physical realisations of WSe$_2$ bilayers belonging to the 3R polytype that lack inversion symmetry, giving rise to weak ferroelectricity \cite{Li2017}. 

As the MX$'$ and XM$'$ domains contract for increasing twist angles, the band edge states become laterally confined, forming a hexagonal lattice of localised states. The two sublattices, located at opposite laters, hybridise through interlayer tunnelling and form two distinct gapped graphene-like superlattices: one for the valence- and another for the conduction band edge.

Moreover, we have demonstrated the opportunity to control the indirect-to-direct band gap transition in 2H- and MX$'$/XM$'$-stacked WSe$_2$ bilayers via external stimuli such as uniform strain, pressure and out-of-plane electric field.

\begin{acknowledgements}
We thank R.\ Gorbachev, W.\ Yao, and C.\ Yelgel for fruitful discussions. We acknowledge support from the
European Graphene Flagship Core3 Project, ERC Synergy Grant Hetero2D,
EPSRC grants EP/S030719/1, EP/S019367/1, EP/P026850/1 and EP/N010345/1, and the
Lloyd Register Foundation Nanotechnology Grant.
D.\ R.-T.\ acknowledges funding from UNAM-DGAPA through its postdoctoral fellowship program. Computational resources were provided by the Computational Shared Facility of the University of Manchester, and the ARCHER2 UK National Supercomputing Service (https://www.archer2.ac.uk) through EPSRC Access to HPC project e672.
\end{acknowledgements}

\bibliography{references_list}

\appendix

\section{General approach for describing hybridisation in WSe$_2$ bilayers.}\label{app:GenApproach}

To derive an effective Hamiltonian describing interlayer hybridisation in twisted TMD homobilayers we first consider coupling between states of two aligned monolayers having a lateral shift $\bm{r}_0$ between their lattices (i.e. $\bm{R}_{i}^{t}=\bm{r}_0+\bm{R}_{i}^{b}$, with $i$ the unit cell index of a single monolayer) and assume that the bilayer crystal potential can be represented as a sum of those of constituent monolayers, $V_t(\bm{r},z)+V_b(\bm{r},z)$ (hereafter, indices $t$ and $b$ label the top and bottom layers). In such a system, the electronic states satisfy the following equation:
\begin{multline}\label{main_Eq}
\left[\frac{\hat{\bm{p}}^2}{2m_0}+\left(V_{t}(\bm{r},z)+\frac{E_0}{2}\right)+\left(V_{b}(\bm{r},z)+\frac{E_0}{2}\right)\right]\Psi\\
=E\Psi,
\end{multline}
where $\hat{\bm{p}}=-i\hbar\left(\bm{\nabla}_{\bm{r}},\partial_z\right)$ is the 3D momentum operator ($z$-axis is along normal to layers), $m_0$ is the free electron mass, and $E_0/2$ is a reference point for the potentials that we explicitly take into account to preserve gauge invariance of the equations derived below. As hybridisation of states in different parts of the Brillouin zone (valleys centered at the $\Gamma$  or $\mathcal{K}$ points) can be described independently, for each of the valleys we will use basis of Kohn-Luttinger functions \cite{Lutt_Kohn_1956} to expand an arbitrary state $\Psi$:
\begin{equation}\label{general_wf}
\Psi=\sum_{n_t,\bm{k}}C_{\bm{k},n_t}|\bm{k},n_t\rangle  + \sum_{n_b,\bm{k}}C_{\bm{k},n_b}|\bm{k},n_b\rangle.
\end{equation} 
Here, $|\bm{k},n_{t/b}\rangle = e^{i\bm{kr}}\psi_{\mathcal{K},n_{t,b}}(\bm{r})$ and $\psi_{\mathcal{K},n_{t/b}}$ are Bloch eigenfunctions of the $n_{t/b}$-th band at the $\mathcal{K}$-point of the TMD monolayer Brillouin zone (below we consider $\mathcal{K}=\Gamma$ in Appendix \ref{app:hamiltonianG} and $\mathcal{K}=K$ in Appendix \ref{app:hamiltonianK}), and $\bm{k}$ is the 2D wave vector measured from the given $\mathcal{K}$-point. The basis choice of Eq.\ (\ref{general_wf}) implies that the top layer crystal potential can be treated as a perturbation for bottom-layer states, and vice versa.  

Substituting the wave function (\ref{general_wf}) into Eq.\ (\ref{main_Eq}) we obtain a matrix equation for the column vector of expansion coefficients $C$:
\begin{equation}\label{gen_matrix_eq}
\hat{H}_0C = \left(E-E_0\right)\left(\hat{1}+\hat{T}\right)C. 
\end{equation}
Here, $\hat{1}$ is the unit matrix, and $\hat{T}$ is a matrix whose elements are given by the overlap intergrals between basis functions (\ref{general_wf}) of different layers, which are non-orthogonal. Note that, by definition, all diagonal elements of $\hat{T}$ are equal to zero.  The intralayer matrix elements of $\hat{H}_0$ read
\begin{equation}\label{diagonal_H0}
\begin{split}
\langle\bm{k},n_{t/b}|H_0|\bm{k}',n'_{t/b}\rangle\equiv
\langle\bm{k},n_{t/b}|\frac{\hat{\bm{p}}^2}{2m_0}+V_{t}+V_{b}|\bm{k}',n'_{t/b}\rangle \\
= \varepsilon_{n_{t/b}}(\bm{k})\delta_{\bm{k},\bm{k}'}\delta_{n_{t/b},n'_{t/b}} + \langle\bm{k},n_{t/b}|V_{b/t}|\bm{k}',n'_{t/b}\rangle\\
\cong  \varepsilon_{n_{t/b}}(\bm{k})\delta_{\bm{k},\bm{k}'}\delta_{n_{t/b},n'_{t/b}} + \langle\bm{k},n_{t/b}|V_{b/t}|\bm{k}',n_{t/b}\rangle
\end{split}
\end{equation}
where $\varepsilon_{n_{t/b}}(\bm{k})$ is the $n_{t/b}$-th band state energy in the top/bottom monolayer, measured from the vacuum level for $V_{t/b}$, i.e., $E_0/2$. For the interlayer matrix elements of $\hat{H}_0$ we have:
\begin{multline}\label{non_diagonal_H0}
\langle\bm{k},n_t|H_0|\bm{k}',n_b\rangle \\
= \langle\bm{k},n_t|\frac{\hat{\bm{p}}^2}{2m_0}+V_{t}+\frac{\hat{\bm{p}}^2}{2m_0}+V_{b}-\frac{\hat{\bm{p}}^2}{2m_0}|\bm{k}',n_b\rangle\\
=\delta_{\bm{k},\bm{k}'}\left[\varepsilon_{n_b}(\bm{k})+\varepsilon_{n_t}(\bm{k})\right]\langle\bm{k},n_t|\hat{T}|\bm{k},n_b\rangle\\
-\delta_{\bm{k},\bm{k}'}\langle\bm{k},n_t|\frac{\hat{\bm{p}}^2}{2m_0}|\bm{k},n_b\rangle.
\end{multline}

Note that conservation of crystal momentum in Eqs.\ (\ref{diagonal_H0}) and (\ref{non_diagonal_H0}) results from the alignment of constituent monolayers. 
The transformation $\widetilde{C}=\sqrt{\hat{1}+\hat{T}}C$ eliminates off diagonal elements in the normalization condition for the column vector $\widetilde{C}$ (i.e. $\widetilde{C}^\dagger\widetilde{C}=1$), and allows us to rewrite Eq.\ (\ref{gen_matrix_eq}) in standard form with a new Hamiltonian matrix:
 \begin{equation}\label{Eq:H_0_transformed}
\frac{1}{\sqrt{\hat{1}+\hat{T}}}\hat{H}_0\frac{1}{\sqrt{\hat{1}+\hat{T}}}\widetilde{C}=\left(E-E_0\right)\widetilde{C}.
\end{equation}
Below, we exclude the reference energy of the crystal potentials in Eq.\ (\ref{Eq:H_0_transformed}) by the energy shift $E-E_0\to E$. Since the matrix elements of $\hat{T}$ are much smaller than unity, we expand $1/\sqrt{\hat{1}+\hat{T}}$ in the previous equation up to second order in $\hat{T}$, and obtain the following equation:
\begin{equation}\label{Eq:gen_coef_eq}
\begin{split}
\sum_{\bm{k}'}\langle\bm{k},n_{t/b}\left|\hat{H}_0-\frac{1}{2}\left\{\hat{T},\hat{H}_0\right\}\right|\bm{k}',n_{t/b}\rangle \widetilde{C}_{\bm{k}',n_{t/b}}\\
+\sum_{\bm{k}',n_{b/t}}\langle\bm{k},n_{t/b}\left|\hat{H}_0-\frac{1}{2}\left\{\hat{T},\hat{H}_0\right\}\right|\bm{k}',n_{b/t}\rangle \widetilde{C}_{\bm{k}',n_{b/t}}\\
+\sum_{\bm{k}'}\langle\bm{k},n_{t/b}\left|\frac{3}{8}\left\{\hat{T}^2,\hat{H}_0\right\}+\frac{1}{4}\hat{T}\hat{H}_0\hat{T}\right|\bm{k}',n_{t/b}\rangle \widetilde{C}_{\bm{k}',n_{t/b}} \\
=E\widetilde{C}_{\bm{k},n_{t/b}},
\end{split}
\end{equation}
where $\left\{\hat{A},\hat{B}\right\}=\hat{A}\hat{B}+\hat{B}\hat{A}$ is the anti-commutator of operators $\hat{A}$ and $\hat{B}$. Equation (\ref{Eq:gen_coef_eq}) is the final result of this section, which we will exploit in the following sections to derive effective Hamiltonians describing hybridisation of top valence band states in $\Gamma$, as well as top valence and bottom conduction states at the $K$ points of two aligned monolayers.

\section{Hybridisation of top valence band states in $\Gamma$-point}\label{app:hamiltonianG}

\begin{table}
\caption{Character table for the relevant representations of point group $D_{3h}$. \label{tab:D3hgroup}}
\begin{tabular}{c|cccccc}
\hline\hline
        & E & 2$C_3$ & $3C_2$ & $\sigma_h$ & $2S_3$ & 3$\sigma_v$ \\
        \hline 
     A$'$ &  1 & 1 & 1 & 1 & 1 & 1 \\
     A$''$ & 1 & 1 & -1 & 1 & 1 & -1\\
     \hline\hline
\end{tabular}
\end{table}

Let us consider the top valence band states at the $\Gamma$ point of TMD monolayers, with energy $\varepsilon_{A'}$ and formed mostly by $d_{z^2}$ orbitals of metals \cite{kormanyos2015k}. As this state transforms according to the one-dimensional $A'$ representation of group $D_{3h}$ (see Table \ref{tab:D3hgroup}), the minimal model that describes the splitting of the states in bilayers comprises only terms with $\widetilde{C}_{A'_{t/b}}$  in Eq.\ (\ref{Eq:gen_coef_eq}), which would lead to a $2\times2$ effective Hamiltonian. However, interlayer coupling affects not only the splitting of the $A'$-states, but also changes the average energy of the split states, due to their hybridisation with other bands. To take into account the latter effect, we add to our model one more band in each layer, which is the closest in energy to the top valence band in monolayers and composed of selenium $p_z$ orbitals. These bands transform according to representation $A''$, and have energy $\varepsilon_{A''}$. Therefore, in a minimal effective Hamiltonian describing hybridisation of the top valence band states in $\Gamma$-point we leave only the two bands in each layer, so that system of equations (\ref{Eq:gen_coef_eq}) is reduced to
\begin{widetext}
\begin{equation}\label{Eq:Ham_4_bands}
\begin{pmatrix}
\varepsilon_{A'} + \Phi^{A'_t}(\bm{r}_0)-E & T^{A'_t,A'_b}(\bm{r}_0) & 0 & T^{A'_t,A''_b}(\bm{r}_0) \\
 T^{A'_t,A'_b}(\bm{r}_0) & \varepsilon_{A'}+\Phi^{A'_b}(\bm{r}_0)-E & T^{A'_b,A''_t}(-\bm{r}_0) & 0 \\
0 &  T^{A'_b,A''_t}(-\bm{r}_0) & \varepsilon_{A''}+\Phi^{A''_t}(\bm{r}_0)-E &  T^{A''_t,A''_b}(\bm{r}_0) \\
T^{A'_t,A''_b}(\bm{r}_0)&0 & T^{A''_t,A''_b}(\bm{r}_0) & \varepsilon_{A''}+\Phi^{A''_b}(\bm{r}_0)-E  
\end{pmatrix}
\begin{pmatrix}
\widetilde{C}_{A'_t} \\
\widetilde{C}_{A'_b} \\
\widetilde{C}_{A''_t}\\
\widetilde{C}_{A''_b} 
\end{pmatrix}=0,
\end{equation}
\end{widetext}  
where $T^{A'_t,A'_b}(\bm{r}_0)$, $T^{A''_t,A''_b}(\bm{r}_0)$, $T^{A'_t,A''_b}(\bm{r}_0)$, and $\Phi^{A'_{t/b}}(\bm{r}_0)$, $\Phi^{A''_{t/b}}(\bm{r}_0)$ are the overlap integrals characterizing the first- and  second order scattering processes, respectively, between corresponding Bloch amplitudes $u_{\Gamma,A'_{t/b}}(\bm{r},z)$, $u_{\Gamma,A''_{t/b}}(\bm{r},z)$, implicitly dependending on interlayer distance $d$. To find explicit $\bm{r}_0$-dependencies for terms in Eq.\ (\ref{Eq:Ham_4_bands}), we approximate the Bloch amplitudes by the lowest harmonics in their Fourier series in monolayer reciprocal vectors $\GG_j$ (same for two aligned layers):   
\begin{multline}\label{Eq:Bloch_amplitude_top}
u_{A'_{t}/A''_{t}}(\bm{r},z)\approx u_{A'_{t}/A''_{t}}^{(0)}(z)\\
+\sum_{j=1,2,3}\left[u_{A'_{t}/A''_{t}}^{(1)}(z)e^{i\bm{G}_j\cdot(\bm{r}-\bm{r}_0)}+c.c.\right],
\end{multline}
\begin{equation}\label{Eq:Bloch_amplitude_bottom}
u_{A'_{b}/A''_{b}}(\bm{r},z)\approx u_{A'_{b}/A''_{b}}^{(0)}(z)+\sum_{j=1,2,3}\left[u_{A'_{b}/A''_{b}}^{(1)}(z)e^{i\bm{G}_j\cdot\bm{r}}+c.c.\right].
\end{equation}
In this approximation, the resonant tunnelling between $A'$ or $A''$ bands reads
\begin{equation}\label{Eq:A'A'_tunneling}
\begin{split}
T^{A_t,A_b}(\bm{r}_0)=t_{A_tA_b}^{(0)}\varepsilon_{A} - \frac{\hbar^2(k_{z0}^2)_{A_tA_b}}{2m_0}\\ +2|t^{(1)}_{A_tA_b}|\left(\varepsilon_{A}-\frac{\hbar^2G^2}{2m_0}\right)\sum_{j=1,2,3}\cos(\bm{G}_j\cdot\bm{r}_0+\varphi_{A_tA_b}) \\
 -  2\frac{\hbar^2|(k_{z1}^2)_{A_tA_b}|}{2m_0}\sum_{j=1,2,3}\cos(\bm{G}_j\cdot\bm{r}_0+\widetilde{\varphi}_{A_tA_b}),
\end{split}
\end{equation}
where 
\begin{align}
    t^{(0)}_{A_t,A_b}=\int dzu_{A_t}^{(0)}u_{A_b}^{(0)},\\
    |t^{(1)}_{A_t,A_b}|e^{i\varphi_{A_tA_b}}=\int dzu_{A_t}^{(1)*}u_{A_b}^{(1)}, \\
(k_{z0}^2)_{A_t,A_b}=\int dz(\partial_zu_{A_t}^{(0)})(\partial_zu_{A_b}^{(0)})\\
 \left|(k_{z1}^2)_{A_t,A_b}\right|e^{i\widetilde{\varphi}_{A_tA_b}}=\int dz(\partial_zu_{A_t}^{(1)*})(\partial_zu_{A_b}^{(1)})
\end{align} 
are overlap integrals implicitly depending on interlayer distance ($A$ labels either $A'$ or $A''$ band). In Eqs.\ (\ref{Eq:Bloch_amplitude_top}), (\ref{Eq:Bloch_amplitude_bottom}) we use the fact that states with zero crystal momentum ($\Gamma$-point states) can be represented by a real-valued wave function leading to real values for $t^{(0)}_{A_tA_b}$ and $(k_{z0}^2)_{A_t,A_b}$. For P-bilayers, Eq.\ (\ref{Eq:A'A'_tunneling}) can be simplified due to $\sigma_h$ mirror symmetry in each monolayer. Indeed, as states in monolayers are either even or odd with respect to $\sigma_h$, one has relation $\int dz u_{A'_t}^{(1)*}(z)u_{A'_b}^{(1)}(z)=\int dz u_{A'_t}^{(1)*}(-z)u_{A'_b}^{(1)}(-z)=\int dz u_{A'_b}^{(1)*}(z)u_{A'_t}^{(1)}(z)$, leading to $\varphi_{A_tA_b}=0$. Similarly, $\widetilde{\varphi}_{A_tA_b}=0$. Thus, for P orientation, the matrix elements describing resonant tunneling in Eq.\ (\ref{Eq:Ham_4_bands}) are even functions of $\bm{r}_0$, 
 $T^{A'_t,A'_b}(\bm{r}_0)=T^{A'_t,A'_b}(-\bm{r}_0)$. By contrast, for AP-WSe$_2$ bilayers the phases of $u_{A'_t}^{(1)}(z)$ and $u_{A'_b}^{(1)}(z)$ are not equal to each other, allowing nonzero $\varphi_{A_tA_b}$ and $\widetilde{\varphi}_{A_tA_b}$ in Eq.\ (\ref{Eq:A'A'_tunneling}). 

The off-resonant interlayer coupling term is expressed as follows:
\begin{equation}\label{Eq:A'A''_tunneling}
\begin{split}
T^{A'_t,A''_b}(\bm{r}_0)=\\
t_{A'_tA''_b}^{(0)}\frac{\varepsilon_{A'}+\varepsilon_{A''}}{2} -\frac{\hbar^2(k_{z0}^2)_{A'_tA''_b}}{2m_0}\\ +2|t^{(1)}_{A'_tA''_b}|\left(\frac{\varepsilon_{A'} + \varepsilon_{A''}}{2}-\frac{\hbar^2G^2}{2m_0}\right)\sum_{j=1,2,3}\cos(\bm{G}_j\cdot\bm{r}_0+\varphi_{A'_tA''_b}) \\
 -  2\frac{\hbar^2|(k_{z1}^2)_{A'_tA''_b}|}{2m_0}\sum_{j=1,2,3}\cos(\bm{G}_j\cdot\bm{r}_0+\widetilde{\varphi}_{A'_tA''_b}),
\end{split}
\end{equation}
where 
\begin{align}
    t^{(0)}_{A'_t,A''_b}=\int dzu_{A'_t}^{(0)}u_{A''_b}^{(0)},\\
    |t^{(1)}_{A'_tA''_b}|e^{i\varphi_{A'_tA''_b}}=\int dzu_{A'_t}^{(1)*}u_{A''_b}^{(1)}, \\
    (k_{z0}^2)_{A'_t,A''_b}=\int dz(\partial_zu_{A'_t}^{(0)})(\partial_zu_{A''_b}^{(0)}), \\
   |(k_{z1}^2)_{A'_t,A''_b}|e^{i\widetilde{\varphi}_{A'_tA''_b}}=\int dz(\partial_zu_{A'_t}^{(1)*})(\partial_zu_{A''_b}^{(1)}) 
\end{align}
are interband-interlayer overlap integrals. For P-bilayers, $\sigma_h$ symmetry relates the off-resonant interlayer matrix elements
 \begin{equation}\label{Eq:relation_A'_A''}
     T^{A'_t,A''_b}(\bm{r}_0)=-T^{A'_b,A''_t}(-\bm{r}_0).
 \end{equation}

The second-order contribution to the diagonal elements of Eq.\ (\ref{Eq:Ham_4_bands}) for $A'$ states is expressed as
\begin{widetext}
\begin{equation}\label{Eq:A't_poten_energy}
\begin{split}
    &\Phi^{A'_t}(\bm{r}_0)=-\varepsilon_{A'}\left[t^{(0)}_{A'_tA'_b}+2|t^{(1)}_{A'_tA'_b}|\sum_{j=1}^3\cos\left(\bm{G}_j\cdot\bm{r}_0+\varphi_{A'_tA'_b}\right) \right]^2 +\widetilde{V}^{(0)}_b + 4|\widetilde{V}^{(1)}_b|\sum_{j=1}^3\cos\left(\bm{G}_j\cdot\bm{r}_0+\varphi_{V_b}\right)\\
    &+ \left[t^{(0)}_{A'_tA'_b}+2|t^{(1)}_{A'_tA'_b}|\sum_{j=1}^3\cos\left(\bm{G}_j\cdot\bm{r}_0+\varphi_{A'_tA'_b}\right) \right]\left[\frac{\hbar^2(k_{z0}^2)_{A'_t,A'_b}}{2m_0}+2\frac{\hbar^2|(k_{z1}^2)_{A'_t,A'_b}|}{2m_0}\sum_{j=1}^3\cos\left(\bm{G}_j\cdot\bm{r}_0+\widetilde{\varphi}_{A'_tA'_b}\right) \right] \\
    &-\frac14\left(\varepsilon_{A'}+3\varepsilon_{A''}\right)\left[t^{(0)}_{A'_tA''_b}+2|t^{(1)}_{A'_tA''_b}|\sum_{j=1}^3\cos\left(\bm{G}_j\cdot\bm{r}_0+\varphi_{A'_tA''_b}\right) \right]^2\\
    &+ \left[t^{(0)}_{A'_tA''_b}+2|t^{(1)}_{A'_tA''_b}|\sum_{j=1}^3\cos\left(\bm{G}_j\cdot\bm{r}_0+\varphi_{A'_tA''_b}\right) \right]\left[\frac{\hbar^2(k_{z0}^2)_{A'_t,A''_b}}{2m_0}+2\frac{\hbar^2|(k_{z1}^2)_{A'_t,A''_b}|}{2m_0}\sum_{j=1}^3\cos\left(\bm{G}_j\cdot\bm{r}_0+\widetilde{\varphi}_{A'_tA''_b}\right) \right],
\end{split}
\end{equation}

\begin{equation}\label{Eq:A'b_poten_energy}
\begin{split}
    &\Phi^{A'_b}(\bm{r}_0)=-\varepsilon_{A'}\left[t^{(0)}_{A'_tA'_b}+2|t^{(1)}_{A'_tA'_b}|\sum_{j=1}^3\cos\left(\bm{G}_j\cdot\bm{r}_0+\varphi_{A'_tA'_b}\right) \right]^2 +\widetilde{V}^{(0)}_t + 4|\widetilde{V}^{(1)}_t|\sum_{j=1,2,3}\cos\left(\bm{G}_j\cdot\bm{r}_0-\varphi_{V_t}\right)\\
    &+ \left[t^{(0)}_{A'_tA'_b}+2|t^{(1)}_{A'_tA'_b}|\sum_{j=1}^3\cos\left(\bm{G}_j\cdot\bm{r}_0+\varphi_{A'_tA'_b}\right) \right]\left[\frac{\hbar^2(k_{z0}^2)_{A'_t,A'_b}}{2m_0}+2\frac{\hbar^2|(k_{z1}^2)_{A'_t,A'_b}|}{2m_0}\sum_{j=1,2,3}\cos\left(\bm{G}_j\cdot\bm{r}_0+\widetilde{\varphi}_{A'_tA'_b}\right) \right] \\
    &-\frac14\left(\varepsilon_{A'}+3\varepsilon_{A''}\right)\left[t^{(0)}_{A'_tA''_b}+2|t^{(1)}_{A'_tA''_b}|\sum_{j=1,2,3}\cos\left(\bm{G}_j\cdot\bm{r}_0-\varphi_{A'_tA''_b}\right) \right]^2\\
    &+ \left[t^{(0)}_{A'_tA''_b}+2|t^{(1)}_{A'_tA''_b}|\sum_{j=1}^3\cos\left(\bm{G}_j\cdot\bm{r}_0-\varphi_{A'_tA''_b}\right) \right]\left[\frac{\hbar^2(k_{z0}^2)_{A'_t,A''_b}}{2m_0}+2\frac{\hbar^2|(k_{z1}^2)_{A'_t,A''_b}|}{2m_0}\sum_{j=1,2,3}\cos\left(\bm{G}_j\cdot\bm{r}_0-\widetilde{\varphi}_{A'_tA''_b}\right) \right],
\end{split}
\end{equation}
where \mbox{$\widetilde{V}_{t/b}^{(0)}=\int V_{t/b}^{(0)}\left[\left(u_{A'_{b/t}}^{(0)}\right)^2+2u_{A'_{b/t}}^{(1)}u_{A'_{b/t}}^{(1)*}\right]$}, $|\widetilde{V}_{t/b}^{(1)}|e^{i\varphi_{V_{t/b}}}=\int V_{t/b}^{(1)}\left(u_{A'_{b/t}}^{(1)}\right)^2$, and we exploited the lowest harmonics of the Fourier series for the monolayer potentials
\begin{equation}
    V_{t}(\bm{r},z)=V_t^{(0)} + \sum_{j=1}^3\left\{V_{t}^{(1)}e^{i\bm{G}_j(\bm{r}-\bm{r}_0)}+c.c.\right\},\quad
    V_{b}(\bm{r},z)=V_t^{(0)} + \sum_{j=1}^3\left\{V_{t}^{(1)}e^{i\bm{G}_j\bm{r}}+c.c.\right\}.
\end{equation}
For $A''$ subbands, the terms $\Phi^{A''_{t/b}}_{\bm{k},\bm{G}}(\bm{r}_0)$ can be obtained from Eqs.\ (\ref{Eq:A't_poten_energy}) and (\ref{Eq:A'b_poten_energy}) by exchanging $A'\leftrightarrow A''$ in all terms.

Having established the explicit form for the matrix elements of Hamiltonian (\ref{Eq:Ham_4_bands}), we now exclude all but the lowest-energy $A''$ states to obtain a minimal effective Hamiltonian describing hybridisation of top valence states at the $\Gamma$ point:
\begin{equation}\label{Eq:Ham_2_bands_structure}
\begin{pmatrix}
\varepsilon_{A'}+\Phi^{A'_t}(\bm{r}_0)+\frac{\left[T^{A'_t,A''_b}(\bm{r}_0)\right]^2}{\varepsilon_{A'}-\varepsilon_{A''}}-E & T^{A'_t,A'_b}(\bm{r}_0)   \\
 T^{A'_t,A'_b}(\bm{r}_0) & \varepsilon_{A'}+\Phi^{A'_b}(\bm{r}_0)+\frac{\left[T^{A'_b,A''_t}(-\bm{r}_0)\right]^2}{\varepsilon_{A'}-\varepsilon_{A''}}-E  
\end{pmatrix}
\begin{pmatrix}
\widetilde{C}_{A'_t} \\
\widetilde{C}_{A'_b} 
\end{pmatrix}=0.
\end{equation}
\end{widetext}

\begin{figure}
	\includegraphics[width = 0.9\linewidth]{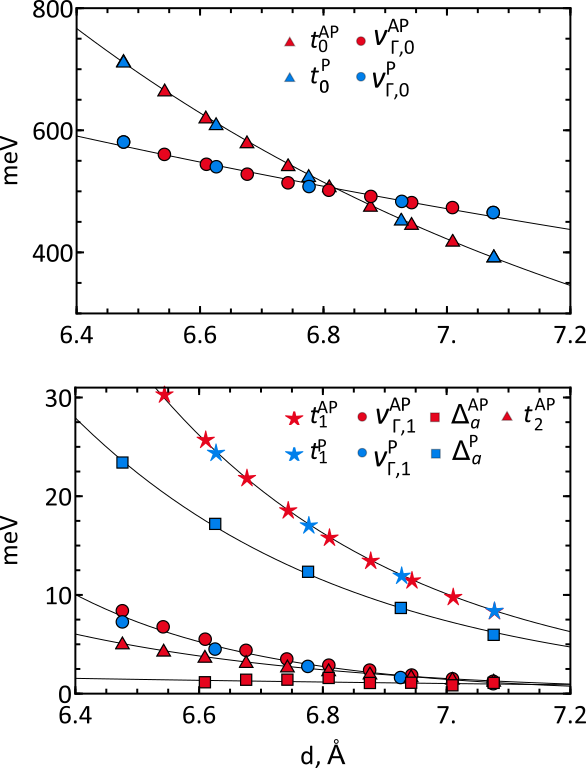}
	\caption{
		\label{Fig:App_A1}
	$z$-dependences for $v^{P/AP}_{\Gamma;0,1}$, $t^{P/AP}_{0,1}$, $t_2^{AP}$, and $\Delta^{P/AP}_a$ extracted from DFT. Top panel shows that $v_{\Gamma,0}^{AP}=v_{\Gamma,0}^P$, $t_{0}^{AP}=t_0^P$. Bottom panel demonstrates validity of the following relations: $v_{\Gamma,1}^{AP}=v_{\Gamma,1}^P$, $t_{1}^{AP}=t_1^P$ and $\Delta^{AP}_a= 0$. $H^{P/AP}_{\Gamma}$, consisting of the greatest terms with $t^{P/AP}_{0,1}$, $v_{\Gamma,0}$ and $\Delta^{P/AP}_a$, determines major effects in the hybridisation model (\ref{Eq:AppPhenom_Ham_Gamma}), while the other terms $v_{\Gamma,1}^{P/AP}$, $t_2^{P/AP}$, gathered in $\delta H^{P/AP}_{\Gamma}$, result in no more than $\approx 10\%$ amendments (compare dashed and solid lines in Figs. \ref{Fig:enrg_vs_conf_AP} and \ref{Fig:enrg_vs_conf_P}).  
	}
\end{figure}

\begin{figure}[!t] 
   \includegraphics[width = 0.9\linewidth]{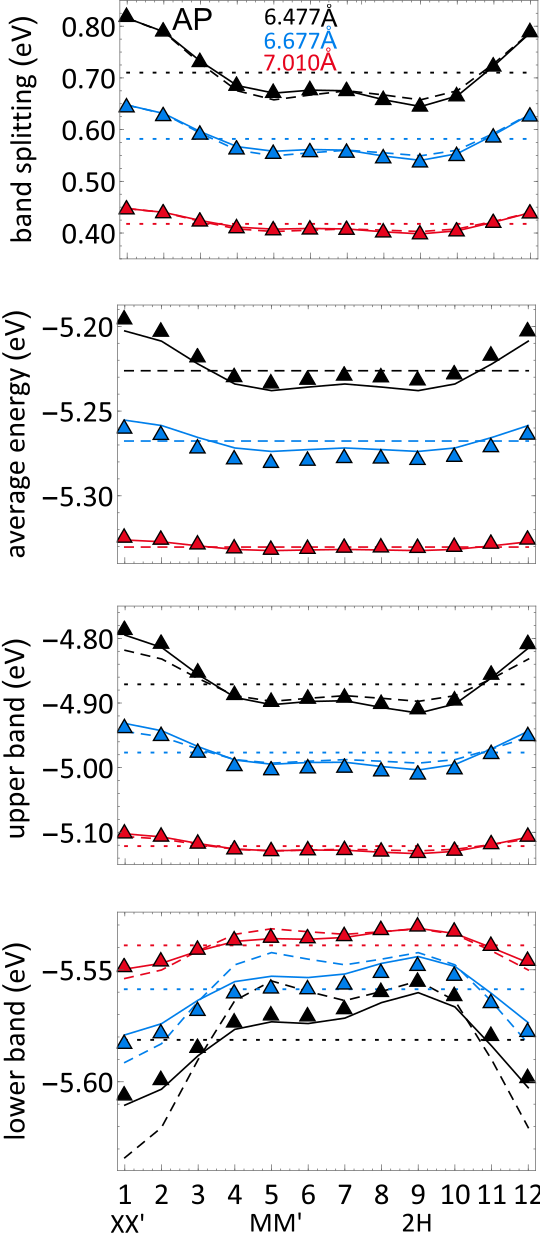}
	\caption{
		\label{Fig:enrg_vs_conf_AP}
DFT-computed values (triangles) for splitting, average, and individual energies of the two top-most valence band states in $\Gamma$-point of AP-WSe$_2$ bilayers along the stacking configuration path in Fig. \ref{fig:configurations}(b) {\it versus} corresponding values obtained with $H^{AP}_{\Gamma}+\delta H^{AP}_{\Gamma}$ (solid) and  $H^{AP}_{\Gamma}$ (dashed). Dotted straight lines show results for the energies calculated with $H^{AP}_{\Gamma}$ at $t^{AP}_1=0$, (i.e. with account of only the largest interlayer-distance-dependent terms), emphasizing major role of interlayer distance variation on position of band edge in $\Gamma$-point of the bilayers. 
	}
\end{figure}

\begin{figure}[t!]
	\includegraphics[width = 0.9\linewidth]{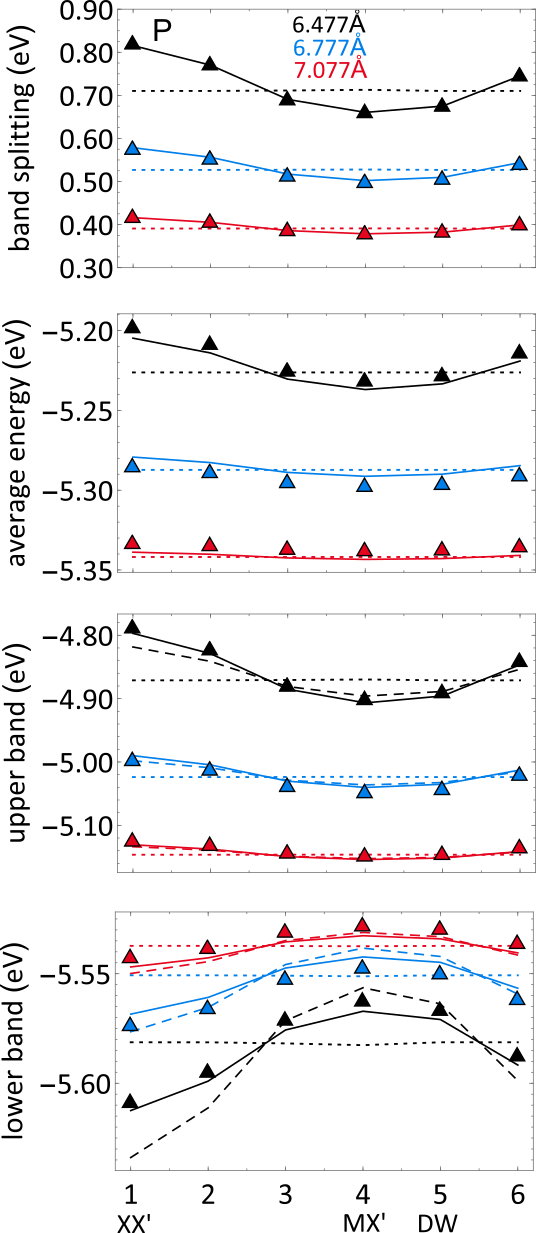}
	\caption{
		\label{Fig:enrg_vs_conf_P}
	DFT-computed values (triangles) for splitting, average, and individual energies of the two top-most valence band states in $\Gamma$-point of P-WSe$_2$ bilayers along the stacking configuration path in Fig. \ref{fig:configurations}(b) {\it versus} corresponding values obtained with $H^{P}_{\Gamma}+\delta H^{P}_{\Gamma}$ (solid) and  $H^{P}_{\Gamma}$ (dashed).  Dotted straight lines show results for the energies calculated with $H^{AP}_{\Gamma}$ at $t^{AP}_1=0$, (i.e. with account of only the largest interlayer-distance-dependent terms), emphasizing major role of interlayer distance variation on position of band edge in $\Gamma$-point of the bilayers.  
	}
\end{figure}

The Hamiltonian (\ref{Eq:Ham_2_bands_structure}) comprises quite a few microscopic parameters that are impossible to extract independently using DFT-computed band structures for aligned P- and AP-vilayers. Therefore, below we will keep only its structure using a simplified expression
\begin{equation}\label{Eq:AppPhenom_Ham_Gamma}
\begin{split}
    H_{\Gamma}^{P/AP}&+\delta H_\Gamma^{P/AP} =\\ &\begin{pmatrix}
    \varepsilon_\Gamma^{P/AP} - \tfrac{S_\Gamma^{P/AP}}{2} & T_\Gamma^{P/AP}\\
    T_\Gamma^{P/AP} & \varepsilon_\Gamma^{P/AP} + \tfrac{S_\Gamma^{P/AP}}{2}
    \end{pmatrix}\\
    &+\begin{pmatrix}
    \delta\varepsilon_\Gamma^{P/AP} & \delta T_\Gamma^{P/AP}\\
    \delta T_\Gamma^{P/AP} & \delta\varepsilon_\Gamma^{P/AP}
    \end{pmatrix},
\end{split}
\end{equation}
with matrix elements defined in Eqs.\ (\ref{Eq:potential_jump}) and (\ref{Eq:H0_matrix_elements}) of the main text. To quantify the parameters in the model (\ref{Eq:AppPhenom_Ham_Gamma}), we calculate the band energies within the framework of DFT for several lateral offsets identified in Fig.\ \ref{fig:configurations}, and multiple interlayer distances between the adhesion energy minima of the lowest and highest configurations (see Fig.\ \ref{vdWen}). To fit the interlayer distance dependencies of $T_{\Gamma}$, $\delta T_{\Gamma}$ $\varepsilon_{\Gamma}$, $\delta\varepsilon_{\Gamma}$ and $S^{P/AP}_{\Gamma}$, we calculate the mean, $(E^{P/AP}_{+}+E^{P/AP}_{-})/2$, and the difference, $(E^{P/AP}_{+}-E^{P/AP}_{-})$, of the energy eigenvalues of Hamiltonian (\ref{Eq:AppPhenom_Ham_Gamma}), \begin{multline}\label{Eq:App_bands}
    E_{\pm}^{P/AP}(\bm{r}_0,z)=\varepsilon_{\Gamma}^{P/AP}(\bm{r}_0,z)\\
    \pm\sqrt{\left(T_{\Gamma}^{P/AP}(\bm{r}_0,z)\right)^2+\left(\frac{S^{P/AP}_{\Gamma}(\bm{r}_0,z)}{2}\right)^2}.
\end{multline}
The analysis of the $z$-dependencies of different terms in the matrix elements of (\ref{Eq:AppPhenom_Ham_Gamma}) (see Fig. \ref{Fig:App_A1}) allows us to fit them by exponential functions, with the parameters given in Table \ref{tab_fit_splitting}, and leads to the following relations: $v_{\Gamma,0}^{P}=v_{\Gamma,0}^{AP}$, $v_{\Gamma,1}^{P}=v_{\Gamma,1}^{AP}$, $t_0^{P}=t_0^{AP}$, $t_1^{P}=t_1^{AP}$, and $S^{AP}_{\Gamma}=\Delta_{a}^{AP}f_a(\bm{r}_0)=0$. Moreover, since the values of $t^{P/AP}_{0,1}$ and $v_{\Gamma,0}$ in $H_{\Gamma}^{P/AP}$ are much larger than the others gathered in $\delta H_{\Gamma}^{P/AP}$, the latter can be used as a perturbation to the former, which characterises the main features of the hybridisation model at the $\Gamma$ point. In Figs.\ \ref{Fig:enrg_vs_conf_AP} and \ref{Fig:enrg_vs_conf_P} we compare the DFT-computed (triangles) energies for splitting, average and individual energies of the hybridised $A'$-states in WSe$_2$ bilayers with those calculated with the help of $H_{\Gamma}$ (dashed) and $H^{P/AP}_{\Gamma}+\delta H^{P/AP}_{\Gamma}$ (solid). The figures demonstrate that the $\Gamma$-point state energy for the top valence band is mainly determined by the optimal interlayer distances of corresponding stacking configurations rather than $\bm{r}_0$-dependencies of matrix elements at a fixed distance.

\section{Derivation of the $K$-point hybridisation Hamiltonians}\label{app:hamiltonianK}
Applying the formalism of Appendix \ref{app:GenApproach} to the valence-band states of spin projection $s$ at the $\tau K$ point, we may write the effective three-band Hamiltonian
\begin{widetext}
\begin{equation}\label{eq:K3band}
{
	H_{K}^{\tau,s} = \begin{pmatrix}
	\varepsilon_{E'}+\varepsilon_{E',AP}^{\tau,s} & T_{E',E'} & 0 & T_{E'',E'} & 0 & T_{A',E'}\\
	T_{E',E'}^* & \varepsilon_{E'} - \varepsilon_{E',AP}^{\tau,s}  & T_{E',E''}^* & 0 & T_{E',A'}^* & 0 \\
	0         & T_{E',E''} & \varepsilon_{E''}+\varepsilon_{E'',AP}^{\tau,s} & T_{E'',E''} & 0 & T_{A',E''} \\
	T_{E'',E'}^*& 0 & T_{E'',E''}^* & \varepsilon_{E''}-\varepsilon_{E'',AP}^{\tau,s} & T_{E'',A'}^* & 0\\
	0         & T_{E',A'} & 0 & T_{E'',A'} & \varepsilon_{A'}+\varepsilon_{A',AP}^{\tau,s} & T_{A',A'}\\
	T_{A',E'}^* & 0 & T_{A',E''}^* & 0 & T_{A',A'}^* & \varepsilon_{A'} - \varepsilon_{A',AP}^{\tau,s} 
	\end{pmatrix},
}
\end{equation}
\end{widetext}
for bands VB($K$), VB-1($K$) and VB-2($K$) of the top and bottom layers, which transform like the irreducible representations $E'$, $E''$ and $A'$ of group $C_{3h}$, respectively (Table \ref{tableC3h}).
\begin{table}
	\caption{
		{Character table for the point group $C_{3h}$, describing the symmetry of the $K$ point bands. Here, $\varepsilon=e^{i2\pi/3}$. \label{tableC3h}
		}
	}
	\begin{tabular}{P{1.55cm} | P{1.55cm} P{1.55cm} P{1.55cm} P{1.55cm}}
		\hline\hline
		\, & $E$ & $2C_3$ & $\sigma_h$ & $2S_3$\\
		\hline
		$A'$ & 1 & 1 & 1 & 1\\
		\hline
		$A''$ & 1 & 1 & -1 & -1\\
		\hline
		\multirow{2}{*}{$E'$} & 1 & $\varepsilon$ & 1 & $\varepsilon$\\
		& 1 & $\varepsilon^*$ & 1 & $\varepsilon^*$\\
		\hline
		\multirow{2}{*}{$E''$} & 1 & $\varepsilon$ & 1 & $\varepsilon$\\
		& 1 & $\varepsilon^*$ & 1 & $\varepsilon^*$\\
		\hline
		\hline
	\end{tabular}
\end{table}

Here, we consider only interlayer hybridisation terms ($\alpha=E',E'',A'$)
\begin{equation}\label{eq:TBand}
	T_{\alpha_t,\alpha_b} = \langle \tau_t \mathbf{K}_t, \alpha_t |\left[\left(\varepsilon_{\alpha_b}+\varepsilon_{\alpha_t}\right)\hat{T} - \frac{\hat{\mathbf{p}}^2}{2m_0} \right]| \tau_b \mathbf{K}_b, \alpha_b \rangle,
\end{equation}
up to first order in the interlayer overlap integrals, and ignore the potential scattering terms appearing in Eq.\ \eqref{diagonal_H0}. For $P$ type bilayers the same-band spin-$s$ levels in both layers coincide; for $AP$ structures these levels are separated by the band's spin-orbit splitting. This is the origin of the terms
\begin{equation}
	\varepsilon_{\alpha,AP}^{\tau,s}=\left\{ \begin{array}{ccc}
	0 &,& P\,\text{stacking}\\
	\tau s \frac{\Delta_{\alpha}^{SO}}{2} &,& AP\,\text{stacking}
	\end{array} \right. .
\end{equation}
in Eq.\ \eqref{eq:K3band}.

\subsection{$P$ stacking}
\begin{figure}[t!]
\begin{center}
\includegraphics[width=0.9\columnwidth]{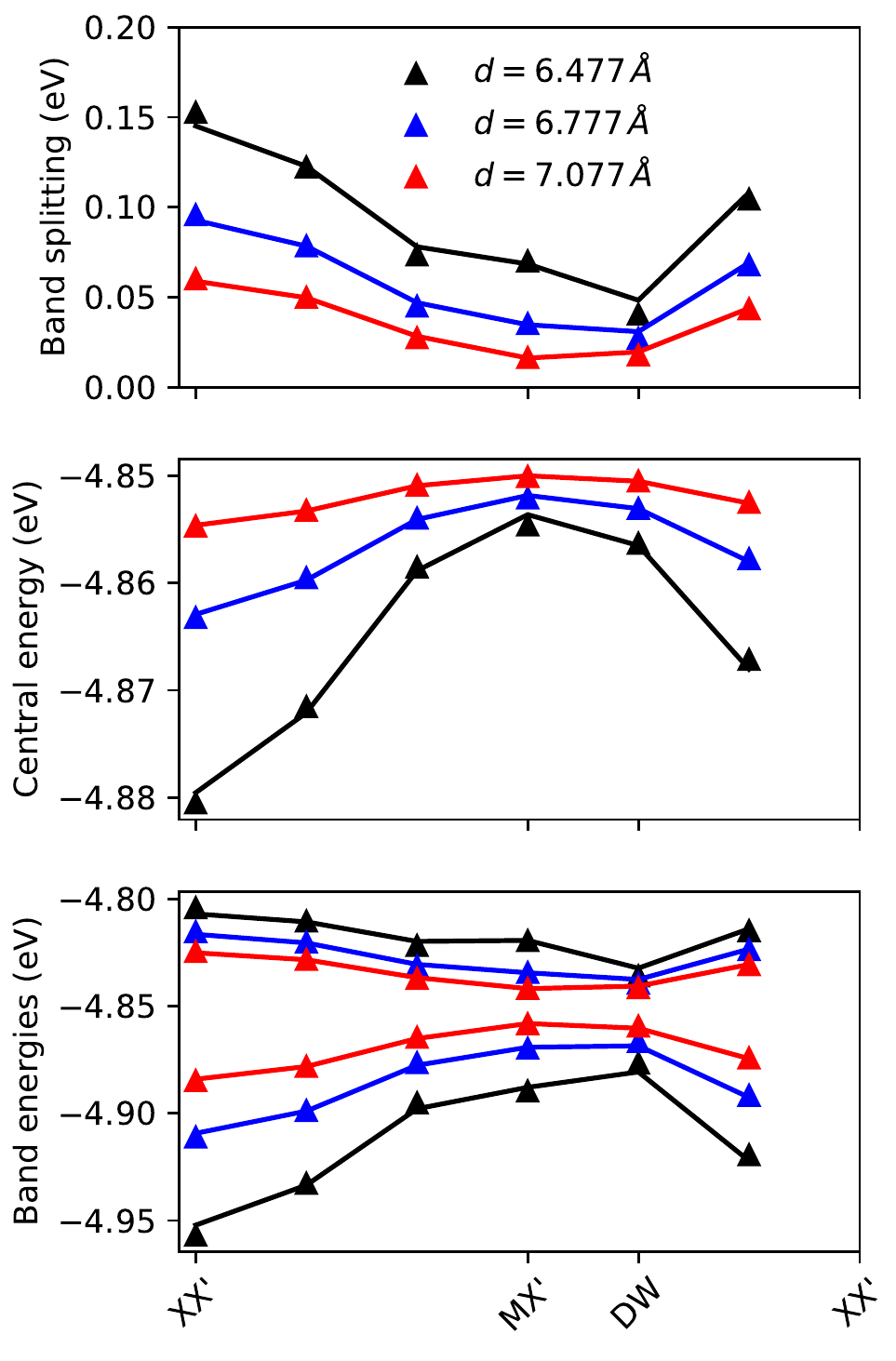}
\caption{Comparison between the DFT results and the model \eqref{eq:H2bandPFinal} for (top) the band splitting, (middle) central energy and (bottom) band energies as a function of stacking configuration, using the parameters of Table \ref{tab:VBKPFit}.}
\label{fig:ThVsDFTVBKP}
\end{center}
\end{figure}
Next, we do a L\"owdin transformation to project out bands VB-1($K$) and VB-2($K$) of both layers up to second order in perturbation theory, keeping only the necessary second order terms to correctly fit the band splittings at the $K$ point predicted by DFT calculations. In the case of $P$ stacking, this means ignoring the VB-2($K$) band altogether. Due to spin-valley locking in TMDs, the band edge at valley $\tau$ belongs to the spin $s=-\tau$ band, described by the Hamiltonian
\begin{equation}\label{eq:H2bandPFinal}
	H_{P,K}^\tau=\begin{pmatrix}
	\varepsilon_{\rm VB}^P(\mathbf{r}_0)  - \tfrac{\Delta^P(\mathbf{r}_0)}{2} & T_{\rm VB}^{P}(\mathbf{r}_0)\\ \\
	T_{\rm VB}^{P*}(\mathbf{r}_0) &  \varepsilon_{\rm VB}^P(\mathbf{r}_0)  + \tfrac{\Delta^P(\mathbf{r}_0)}{2}
	\end{pmatrix},
\end{equation}
where we have abbreviated $T_{\rm VB}\equiv T_{E',E'}$ and defined
\begin{equation}\label{eq:definitionsKP}
\begin{split}
	\varepsilon_{\rm VB}^P(\mathbf{r}_0) =& \varepsilon_{E'} +  \frac{|T_{E'',E'}^P(\mathbf{r}_0)|^2 + |T_{E',E''}^P(\mathbf{r}_0)|^2}{4(\varepsilon_{E'}-\varepsilon_{E''})},\\
	\Delta^P(\mathbf{r}_0) =&  \frac{|T_{E'',E'}^P(\mathbf{r}_0)|^2 - |T_{E',E''}^P(\mathbf{r}_0)|^2}{2(\varepsilon_{E'}-\varepsilon_{E''})}.
\end{split}
\end{equation}
Note that on symmetry grounds we have obtained an interlayer splitting term, which we shall use to account for the interlayer bias $\Delta_P(\bm{r}_0)$ found in our \emph{ab initio} calculations.

The hopping matrix elements $T_{\alpha,\beta}^P$ between bottom-layer band $\beta$ and top-layer band $\alpha$ are obtained by Fourier expanding the corresponding Bloch functions at momentum $\tau \mathbf{K}$ in the in-plane coordinates, and keeping only those Bragg vectors $\mathbf{G}$ such that $\tau \mathbf{K}+\mathbf{G}=\tau C_3^{\mu}\mathbf{K}$, with $\mu=1,2$. This approximation gives
\begin{equation}\label{eq:TabGeneral}
	T_{\alpha,\beta}(\mathbf{r}_0)=\sum_{\mu=0}^2e^{i\tau C_3^\mu\mathbf{K}\cdot\mathbf{r}_0}t_{\alpha,\beta}(\tau C_3^\mu\mathbf{K}),
\end{equation}
where $t_{\alpha,\beta}(\mathbf{q})$ has the form
\begin{equation}
\begin{split}
	t_{\alpha,\beta}(\mathbf{q}) =& t_{\alpha,\beta}^{(1)}\int dz\,u_{\alpha,t}^*(\mathbf{q},z)u_{\beta,b}(\mathbf{q},z)\\
	&+ t_{\alpha,\beta}^{(2)}\int dz\,\partial_z u_{\alpha,t}^*(\mathbf{q},z) \partial_z u_{\beta,b}(\mathbf{q},z).
\end{split}
\end{equation}
To relate the three coefficients $t_{\alpha,\beta}(\tau C_3^\mu \mathbf{K})$ in \eqref{eq:TabGeneral}, we use the following symmetry property of the Bloch functions at the $\tau \mathbf{K}$ point and of their Fourier coefficients:
\begin{equation}
	u_{\alpha,t/b}(\tau C_3\mathbf{K}) = \phi_{\alpha,\tau}u_{\alpha,t/b}(\tau \mathbf{K}),
\end{equation}
where (Table \ref{tableC3h})
\begin{equation}\label{eq:symrulesK}
	\phi_{E',\tau} = e^{-i\tfrac{2\pi}{3}\tau},\,\phi_{E'',\tau} = e^{i\tfrac{2\pi}{3}\tau},\,\text{and}\,\,\phi_{A',\tau} = 1.
\end{equation}
This immediately gives
\begin{equation}\label{eq:TabFinal}
	T_{\alpha,\beta}(\mathbf{r}_0)=t_{\alpha,\beta} \sum_{\mu=0}^2e^{i\tau C_3^\mu\mathbf{K}\cdot\mathbf{r}_0}\phi_{\alpha,\tau\mu}^*\phi_{\beta,\tau\mu},
\end{equation}
with $t_{\alpha,\beta} \equiv t_{\alpha,\beta}(\tau \mathbf{K})$ for short. The hopping terms relevant to Eq.\ \eqref{eq:H2bandPFinal} are
\begin{equation}
\begin{split}
	T_{\rm VB}^P(\mathbf{r}_0) =& t_{\rm VB}^P\sum_{\mu=0}^2e^{iC_3^\mu \tau\mathbf{K}\cdot\mathbf{r}_0}\\
	T_{E',E''}^P(\mathbf{r}_0) =& t_{E',E''}^P\sum_{\mu=0}^2e^{iC_3^\mu \tau\mathbf{K}\cdot\mathbf{r}_0}e^{-i\tfrac{2\pi}{3}\tau\mu}\\
	T_{E'',E'}^P(\mathbf{r}_0) =& t_{E'',E'}^P\sum_{\mu=0}^2e^{iC_3^\mu \tau\mathbf{K}\cdot\mathbf{r}_0}e^{i\tfrac{2\pi}{3}\tau\mu},
\end{split}
\end{equation}
where inversion symmetry (simultaneous layer exchange and $\mathbf{r}_0 \rightarrow -\mathbf{r}_0$) requires that $t_{E',E''}=t_{E'',E'}$. Note, however, that $T_{E',E''}^P$ and $T_{E'',E'}^P$ depend differently on $\mathbf{r}_0$. Substitution into \eqref{eq:definitionsKP} finally gives
\begin{equation}\label{eq:finalformsKP}
\begin{split}
	\varepsilon_{\rm VB}^P(\mathbf{r}_0) =& \varepsilon_{E'} - v_{{\rm VB},0}^P - v_{{\rm VB},1}^{P}\sum_{j=1}^3\cos{\left(\mathbf{G}_j\cdot\mathbf{r}_0 \right)},\\
	\Delta^P(\mathbf{r}_0) =& \Delta_a^P\sum_{j=1}^3\sin{\left(\mathbf{G}_j\cdot\mathbf{r}_0 \right)}.
\end{split}
\end{equation}
where the expression obtained for $\Delta^P(\rr_0)$ matches that of Eq.\ \eqref{Eq:potential_jump} for the ferroelectric potential energy difference between the layers.

Each parameter appearing in Eq.\ \eqref{eq:finalformsKP} was fitted to DFT data for different interlayer distances, and interpolated as $P(d)=Ae^{-q(d-d_0)}$. The results are shown in Table \ref{tab:VBKPFit}, and a comparison between the DFT band structures and the fitted model \eqref{eq:H2bandPFinal} is shown in Fig.\ \ref{fig:ThVsDFTVBKP}.

\begin{table}[h!]
\caption{Fitting parameters for the exponential $d$ dependence of the coefficients entering Eq.\ \eqref{eq:H2bandPFinal}. The interpolation formula used was $P(d)=Ae^{-q z}$.}
\begin{center}
\begin{tabular}{P{2.0cm} Q{1.3cm}@{.}R{1.7cm} Q{1.3cm}@{.}R{1.7cm}}
\hline\hline
 \, & \multicolumn{2}{c}{$A\,({\rm meV})$} & \multicolumn{2}{c}{$q\,({\rm \AA^{-1}})$} \\
 \hline\hline
 $v_{{\rm VB},0}^P$  &    10&8       &  1&8  \\
 $v_{{\rm VB},1}^{P}$  &  3&0      &  2&9  \\
 $|t_{{\rm VB}}^{P}|$  & 17&2     & 1&5  \\
 $\Delta_P$      & 7&8  & 2&3  \\
  \hline\hline
\end{tabular}
\end{center}
\label{tab:VBKPFit}
\end{table}

\subsection{$AP$ stacking}
\begin{figure}[b!]
\begin{center}
\includegraphics[width=0.9\columnwidth]{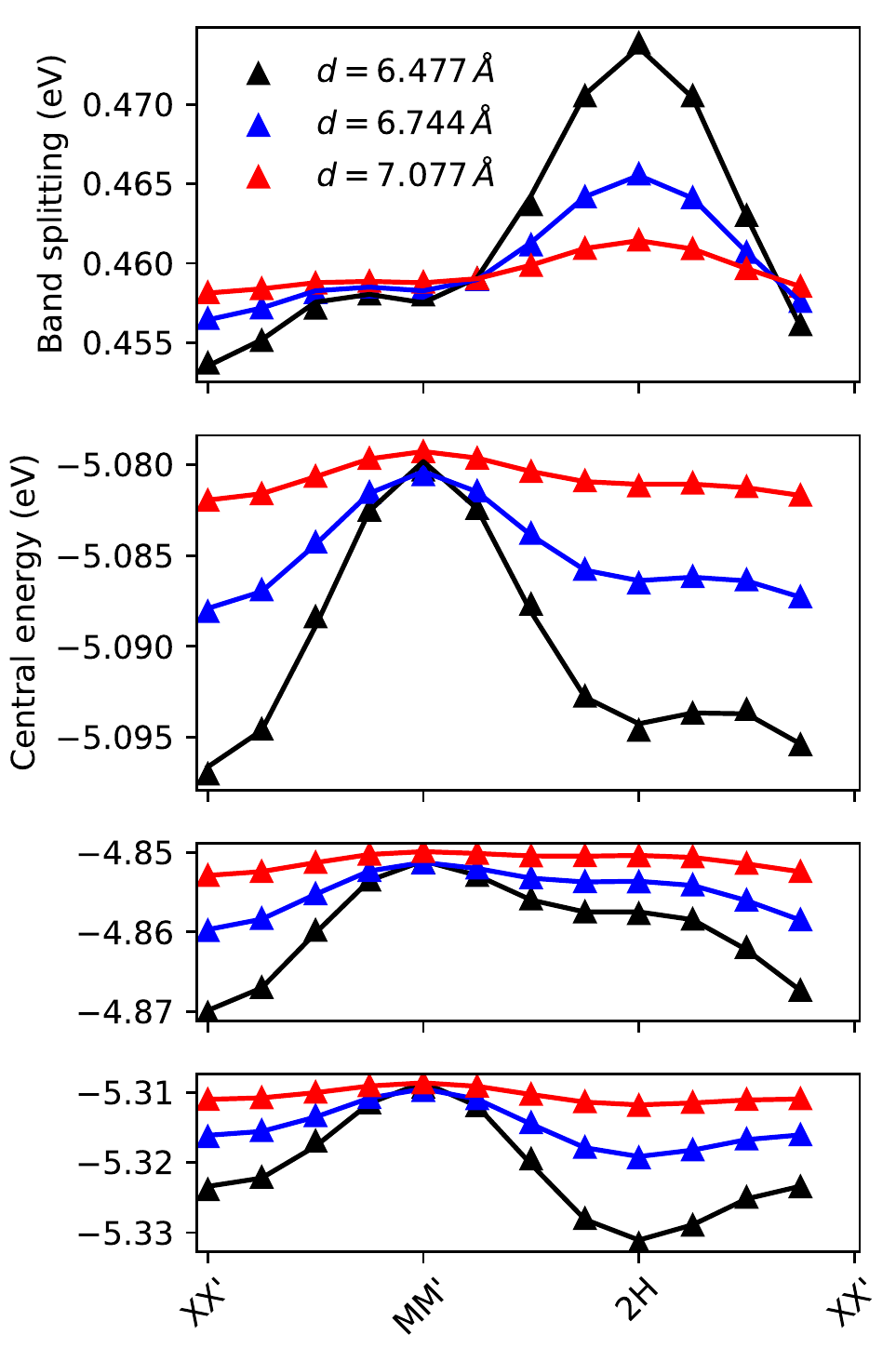}
\caption{Comparison between the DFT results and the model \eqref{eq:H2bandAPFinal} for (top) the band splitting, (middle) central energy and (bottom) band energies as a function of stacking configuration, using the parameters of Table \ref{tab:VBKAPFit}.}
\label{fig:VBKAPFit}
\end{center}
\end{figure}
The same analysis for $AP$ stacking, this time including also band $A'$, gives the Hamiltonian
\begin{equation}\label{eq:H2bandAPFinal}
	H_{AP,K}^\tau=\begin{pmatrix}
	\varepsilon_{{\rm VB}}^{AP}(\mathbf{r}_0) +\tau s \frac{\Delta_{{\rm VB}}^{SO}(\mathbf{r}_0)}{2}   & T_{{\rm VB}}^{AP }(\mathbf{r}_0)\\ \\
	T_{\rm VB}^{AP*}(\mathbf{r}_0) &  \varepsilon_{\rm VB}^{AP}(\mathbf{r}_0) - \tau s \frac{\Delta_{\rm VB}^{SO}(\mathbf{r}_0)}{2}
	\end{pmatrix},
\end{equation}
where
\begin{equation}\label{eq:superfluous}
\begin{split}
	\varepsilon_{\rm VB}^{AP}(\mathbf{r}_0) =& \varepsilon_{E'} + \frac{1}{2}\sum_{\alpha=E'',A'}^2\frac{(\varepsilon_{E'}-\varepsilon_{\alpha})|T_{\alpha,E'}^{AP}(\mathbf{r}_0)|^2}{(\varepsilon_{E'}-\varepsilon_{\alpha})^2 - \frac{(\Delta^{SO}_{ E'}+\Delta^{SO}_{ \alpha})^2}{4}},\\
	\Delta^{SO}_{\rm VB}(\mathbf{r}_0) =&  \Delta^{SO}_{\rm VB}+\frac{1}{2}\sum_{\alpha=E'',A'}^2\frac{(\Delta^{SO}_{ E'}+\Delta^{SO}_{ \alpha})|T_{\alpha,E'}^{AP}(\mathbf{r}_0)|^2}{(\varepsilon_{E'}-\varepsilon_{\alpha})^2 - \frac{(\Delta^{SO}_{ E'}+\Delta^{SO}_{ \alpha})^2}{4}}.
\end{split}
\end{equation}
The corresponding hopping terms are
\begin{equation}\label{eq:defsKAP}
\begin{split}
	T_{{\rm VB}}^{AP}(\mathbf{r}_0) =& t_{{\rm VB}}^{AP}\sum_{\mu=0}^2e^{iC_3^\mu \tau\mathbf{K}\cdot\mathbf{r}_0}e^{i\tfrac{2\pi}{3}\tau\mu},\\
	T_{E',E''}^{AP}(\mathbf{r}_0) =& t_{E',E''}\sum_{\mu=0}^2e^{iC_3^\mu \tau\mathbf{K}\cdot\mathbf{r}_0},\\
	T_{E'',E'}^{AP}(\mathbf{r}_0) =& t_{E'',E'}\sum_{\mu=0}^2e^{iC_3^\mu \tau\mathbf{K}\cdot\mathbf{r}_0},\\
	T_{E',A'}^{AP}(\mathbf{r}_0) =& t_{E',A'}\sum_{\mu=0}^2e^{iC_3^\mu \tau\mathbf{K}\cdot\mathbf{r}_0}e^{-i\tfrac{2\pi}{3}\tau\mu},\\
	T_{A',E'}^{AP}(\mathbf{r}_0) =& t_{A',E'}\sum_{\mu=0}^2e^{iC_3^\mu \tau\mathbf{K}\cdot\mathbf{r}_0}e^{-i\tfrac{2\pi}{3}\tau\mu}.
\end{split}
\end{equation}
Contrary to the case of $P$ stacking, the matrix elements $T_{E',E''}^{AP}$ and $T_{E',A'}^{AP}$ have the same $\mathbf{r}_0$ dependence as  $T_{E'',E'}^{AP}$ and $T_{A',E'}^{AP}$, respectively. However, they appear with different denominators in Eq.\ \eqref{eq:superfluous} due to spin-orbit coupling. Substituting \eqref{eq:defsKAP}  into \eqref{eq:superfluous} gives
\begin{equation}
\begin{split}
	\varepsilon_{{\rm VB}}^{AP}(\mathbf{r}_0) = &  \varepsilon_{E'} - v_{{\rm VB},0}^{AP} - v_{{\rm VB},1}^{AP}\sum_{j=1}^3\cos{\left(\mathbf{G}_j\cdot\mathbf{r}_0\right)}\\
	& - v_{{\rm VB},2}^{AP}\sum_{j=1}^3\sin{\left(\mathbf{G}_j\cdot\mathbf{r}_0\right)},\\
	\Delta_{{\rm VB}}^{SO}(\mathbf{r}_0) =&  \Delta_{{\rm VB}}^{SO}+\tilde{\Delta}_{{\rm VB},0}^{SO}+\tilde{\Delta}_{{\rm VB},1}^{SO}\sum_{j=1}^3\cos{\left(\mathbf{G}_j\cdot\mathbf{r}_0 \right)}\\ &+\tilde{\Delta}_{{\rm VB},2}^{SO}\sum_{j=1}^3\sin{\left(\mathbf{G}_j\cdot\mathbf{r}_0 \right)},
\end{split}
\end{equation}
where inversion symmetry dictates that $|T_{E',\alpha}^{AP}|^2=|T_{\alpha,E'}^{AP}|^2$.

Fitting the model parameters to the DFT data reveals that $v_{{\rm VB},2}^{AP}\approx -v_{{\rm VB},1}^{AP}$, that $\tilde{\Delta}_{{\rm VB},2}^{SO}\approx\tilde{\Delta}_{{\rm VB},1}^{SO}$ and $\tilde{\Delta}_{{\rm VB},0}^{SO}$ is negligible by comparison to all other SO terms  (Table \ref{tab:KpointAll}), and that the constant energy shift $v_{{\rm VB},0}^{AP} \approx v_{{\rm VB},0}^{P}$ (see Tables \ref{tab:VBKPFit} and \ref{tab:VBKAPFit}). The results are shown in Fig.\ \ref{fig:VBKAPFit}.

\begin{table}[th!]
\caption{Fitting parameters for the exponential $d$ dependence of the coefficients entering Eq.\ \eqref{eq:H2bandAPFinal}. The interpolation formula used was $P(d)=Ae^{-q z}$.}
\begin{center}
\begin{tabular}{P{2.0cm} Q{1.3cm}@{.}R{1.7cm} Q{1.3cm}@{.}R{1.7cm}}
\hline\hline
  & \multicolumn{2}{c}{$A\,({\rm meV})$} & \multicolumn{2}{c}{$q\,({\rm \AA}^{-1})$}\\
 \hline\hline
 $v_{{\rm VB},0}^{AP}$               &  10&5     &   1&6    \\
 $v_{{\rm VB},1}^{AP}$               &  1&1     &   2&8    \\
 $v_{{\rm VB},2}^{AP}$               &  -1&3    &   3&3    \\
 $|t_{{\rm VB}}^{AP}|$             &  9&8        &   1&5    \\
 $\tilde{\Delta}_{{\rm VB},1}^{SO}$   & -0&9     &   2&9    \\
 $\tilde{\Delta}_{{\rm VB},2}^{SO}$   & -0&8     &   2&9    \\
  \hline\hline
\end{tabular}
\end{center}
\label{tab:VBKAPFit}
\end{table}%

\subsection{Hybridisation models for the conduction bands}
The effective models for the bilayer conduction bands in $P$- and $AP$-type bilayers can be constructed in analogy with Eqs.\ \eqref{eq:H2bandPFinal} and \eqref{eq:H2bandAPFinal}. The symmetry rules for the matrix elements $T_{\alpha,\beta}$, where $\alpha,\beta = A',E'',E'$ for CB($K$), CB+1$(K)$ and CB+2($K$), respectively, are given by
\begin{equation}
	\phi_{A',\tau} = 1,\,\phi_{E'',\tau} = e^{-i\tfrac{2\pi}{3}\tau},\,\text{and}\,\,\phi_{E',\tau} = e^{i\tfrac{2\pi}{3}\tau}.
\end{equation}
The corresponding minimal Hamiltonians for $P$ and $AP$ stacking are
\begin{subequations}
\begin{equation}\label{eq:CBKP}
    H_{P,\tau\mathbf{K}}^{(CB)}=\begin{pmatrix}
    \varepsilon_{\rm CB}^{P}(\mathbf{r}_0)-\tfrac{\Delta^P(\mathbf{r}_0)}{2} & T_{\rm CB}(\mathbf{r}_0)\\
    T_{\rm CB}^{*}(\mathbf{r}_0) & \varepsilon_{\rm CB}^P(\mathbf{r}_0) + \tfrac{\Delta^P(\mathbf{r}_0)}{2}
    \end{pmatrix},
\end{equation}
\begin{equation}\label{eq:CBKAP}
    H_{AP,\tau\mathbf{K}}^{(CB)}=\begin{pmatrix}
    \varepsilon_{\rm CB}^{AP}(\mathbf{r}_0) - \tfrac{\tau s \Delta_{\rm CB}^{SO}}{2}(\mathbf{r}_0) & T_{\rm CB}(\mathbf{r}_0)\\
    T_{\rm CB}^{*}(\mathbf{r}_0) & \varepsilon_{\rm CB}^{AP}(\mathbf{r}_0) +  \tfrac{\tau s\Delta_{\rm CB}^{SO}(\mathbf{r}_0)}{2}
    \end{pmatrix},
\end{equation}
\end{subequations}
with the definitions
\begin{subequations}
\begin{equation}
    \varepsilon_{\rm CB}^P(\mathbf{r}_0)=\varepsilon_{A'}-v_{{\rm CB},0}^P-v_{{\rm CB},1}^{P}\sum_{j=1}^3\cos{\left(\mathbf{G}_j\cdot\mathbf{r}_0 \right)},
\end{equation}
\begin{equation}
\begin{split}
    \varepsilon_{\rm CB}^{AP}(\mathbf{r}_0)=\varepsilon_{A'}-v_{{\rm CB},0}^{AP}&-v_{{\rm CB},1}^{AP}\sum_{j=1}^3\cos{\left(\mathbf{G}_j\cdot\mathbf{r}_0 \right)}\\ &-v_{{\rm CB},2}^{AP}\sum_{j=1}^3\sin{\left(\mathbf{G}_j\cdot\mathbf{r}_0 \right)},
\end{split}
\end{equation}
\begin{equation}
\begin{split}
    \Delta_{\rm CB}^{SO}(\mathbf{r}_0)=&\Delta_{\rm CB}^{SO} +\tilde{\Delta}_{{\rm CB},1}^{SO}\sum_{j=1}^3\cos{\left(\mathbf{G}_j\cdot\mathbf{r}_0 \right)}\\
    &+ \tilde{\Delta}_{{\rm CB},2}^{SO\,'}\sum_{j=1}^3\sin{\left(\mathbf{G}_j\cdot\mathbf{r}_0 \right)},
\end{split}
\end{equation}
\begin{equation}
    T_{\rm CB}(\mathbf{r}_0)=t_{\rm CB}\sum_{\mu=0}^2e^{i\tau C_3^\mu \mathbf{K}\cdot \mathbf{r}_0}.
\end{equation}
\end{subequations}

Our DFT calculations show that (i) $|\tilde{\Delta}_{{\rm CB},2}^{SO}| \ll |\tilde{\Delta}_{{\rm CB},1}|,|\Delta_{\rm CB}^{SO}|$, and thus can be safely neglected, whereas $\tilde{\Delta}_{{\rm CB},1}^{SO}\approx\tilde{\Delta}_{{\rm VB},1}^{SO}$; (iii) as in the valence band case, the constant energy shifts $v_{{\rm CB},0}^P\approx v_{{\rm CB},0}^{AP}$, and in fact they coincide with $v_{{\rm VB},0}^{P/AP}$ to within $\sim 1\,{\rm meV}$; moreover, (iii) the conduction-band hopping parameters $t_{\rm CB}$ coincide for $P$ and $AP$ stacking (see Tables \ref{tab:VBKPFit}-\ref{tab:CBKAPFit}), and (iv) $v_{{\rm CB},2}^{AP}\approx v_{{\rm CB},2}^{P}$. Importantly, the DFT results also show that the same splitting $\Delta^P(\rr_0)$ appearing for the $K$-point valence bands, as well as for both the conduction and valence bands at the $\Gamma$ point, is also present for the $K$-point conduction bands, which is consistent with a purely electrostatic interlayer bias acting on the metallic orbitals.

\begin{table}[h!]
\caption{Fitting parameters for the exponential $d$ dependence of the coefficients entering Eq.\ \eqref{eq:CBKP}. The interpolation formula used was $P(d)=Ae^{-q z}$.}
\begin{center}
\begin{tabular}{P{2.0cm} Q{1.3cm}@{.}R{1.7cm} Q{1.3cm}@{.}R{1.7cm}}
\hline\hline
  & \multicolumn{2}{c}{$A\,({\rm meV})$} & \multicolumn{2}{c}{$q\,({\rm \AA^{-1}})$} \\
 \hline\hline
 $v_{{\rm CB},0}^P$       & 11&2      &  1&8  \\
 $v_{{\rm CB},1}^{P}$       &  1&5        &  2&7  \\
 $|t_{\rm CB}|$  &  3&7           &  1&5  \\
  \hline\hline
\end{tabular}
\end{center}
\label{tab:CBKPFit}
\end{table}

\begin{table}[h!]
\caption{Fitting parameters for the exponential $d$ dependence of the coefficients entering Eq.\ \eqref{eq:CBKAP}. The interpolation formula used was $P(d)=Ae^{-q z}$.}
\begin{center}
\begin{tabular}{P{2.0cm} Q{1.3cm}@{.}R{1.7cm} Q{1.3cm}@{.}R{1.7cm}}
\hline\hline
  & \multicolumn{2}{c}{$A\,({\rm meV})$} & \multicolumn{2}{c}{$q\,({\rm \AA^{-1}})$} \\
 \hline\hline
$v_{{\rm CB},0}^{AP}$                 & 11&2      &  1&3  \\
 $v_{{\rm CB},1}^{AP}$                 &  0&7  &  2&9  \\
 $v_{{\rm CB},2}^{AP}$                 &  0&4    &  2&3  \\
$|t_{\rm CB}^{AP}|$            &  3&7       &  1&0  \\
 $\tilde{\Delta}_{{\rm CB},1}^{SO}$    & -1&6     &  2&0  \\
  $\tilde{\Delta}_{{\rm CB},2}^{SO'}$    & 0&1     &  1&3  \\
  \hline\hline
\end{tabular}
\end{center}
\label{tab:CBKAPFit}
\end{table}

\subsection{Summary}
Put together, our symmetry analysis and DFT results show that the effective Hamiltonian for the top two valence subbands at the $K$ point is given by
\begin{equation}
    H_{VB,K}=\begin{pmatrix}
    \varepsilon_{\rm VB}(\rr_0,d) - \tfrac{S_{\rm VB}^{P/AP}(\rr_0)}{2} & T_{{\rm VB},\tau}^{P/AP}(\rr_0,d)\\
    T_{{\rm VB},\tau}^{P/AP*}(\rr_0,d) & \varepsilon_{\rm VB}(\rr_0,d) + \tfrac{S_{VB}^{P/AP}(\rr_0)}{2}
    \end{pmatrix},
\end{equation}
with the definitions
\begin{subequations}
\begin{equation}
\begin{split}
    \varepsilon_{\rm VB}^P(\rr_0,d)=&\varepsilon_{E'}-v_0-v_{{\rm VB},1}^P\sum_{j=1}^3\cos{\left(\GG_j\cdot\rr_0 \right)},\\
    \varepsilon_{\rm VB}^{AP}(\rr_0,d)=&\varepsilon_{E'}-v_0-v_{\rm VB}^{AP}\sqrt{2}\sum_{j=1}^3\cos{\left(\GG_j\cdot\rr_0 + \tfrac{\pi}{4} \right)},
\end{split}
\end{equation}
\begin{equation}
    S_{\rm VB}^{P}(\rr_0,d) = \Delta^P(\rr_0,d);\quad \Delta_{\rm VB}^{AP}(\rr_0,d) = -\tau s \Delta^{SO}_{\rm VB},
\end{equation}
\begin{equation}\label{eq:TappVB}
\begin{split}
    T_{\rm VB}^P(\rr_0,d)=&t_{\rm VB}^P(d)\sum_{\mu=0}^2e^{i\tau C_3^\mu \KK\cdot\rr_0},\\
    T_{\rm VB}^{AP}(\rr_0,d) =&t_{\rm VB}^{AP}(d)\sum_{\mu=0}^2e^{i\tau C_3^\mu \KK\cdot\rr_0}e^{i\tfrac{2\pi}{3}\tau\mu},
\end{split}
\end{equation}
\end{subequations}
where we have used the fact that $v_{{\rm VB},2}^{AP}\approx -v_{{\rm VB},1}^{AP}$ and the identity $\cos{(x\pm \pi/4)}=(\cos{(x)} \mp \sin{(x)})/\sqrt{2}$ to simplify $\varepsilon_{\rm VB}^{AP}(\rr_0,d)$.

Similarly, the effective Hamiltonian for the bottom two conduction subbands takes the form
\begin{equation}
    H_{CB,K}=\begin{pmatrix}
    \varepsilon_{\rm CB}(\rr_0,d) - \tfrac{S_{\rm CB}^{P/AP}(\rr_0)}{2} & T_{{\rm CB},\tau}(\rr_0,d)\\
    T_{{\rm CB},\tau}^{*}(\rr_0,d) & \varepsilon_{\rm CB}(\rr_0,d) + \tfrac{S_{\rm CB}^{P/AP}(\rr_0)}{2}
    \end{pmatrix},
\end{equation}
with the definitions
\begin{subequations}
\begin{equation}
\begin{split}
    \varepsilon_{\rm CB}^P(\rr_0,d)=&\varepsilon_{\rm CB}-v_0-v_{{\rm CB},1}^{P}\sum_{j=1}^3\cos{\left(\GG_j\cdot\rr_0 \right)},\\
    \varepsilon_{\rm CB}^{AP}(\rr_0,d)=&\varepsilon_{\rm CB}-v_0-v_{{\rm CB},1}^{AP}\sqrt{2}\sum_{j=1}^3\cos{\left(\GG_j\cdot\rr_0 - \tfrac{\pi}{4} \right)},
\end{split}
\end{equation}
\begin{equation}
    S_{\rm CB}^{P}(\rr_0,d) = \Delta_P(\rr_0,d);\quad S_{\rm CB}^{AP}(\rr_0,d) = \tau s \Delta_{\rm CB}^{SO},
\end{equation}
\begin{equation}\label{eq:TappCB}
    T_{\rm CB}(\rr_0,d)= t_{\rm CB}(d)\sum_{\mu=0}^2e^{i\tau C_3^\mu \KK\cdot\rr_0},
\end{equation}
\end{subequations}
using $v_{{\rm CB},2}^{AP} \approx v_{{\rm CB},1}^{AP}$. It is then convenient to rescale $\sqrt{2}v_{{\rm CB},1}^{AP}\rightarrow v_{{\rm CB},1}^{AP}$, and rewrite
\begin{equation}
\begin{split}
    \sum_{\mu=0}^2e^{i\tau C_3^\mu \KK\cdot\rr_0} &= e^{i\tau\frac{4\pi x_0}{3a}} + 2e^{-i\tau\frac{2\pi x_0}{3a}}\cos\left(\tfrac{2\pi y_0}{a\sqrt{3}}\right),\\
    \sum_{\mu=0}^2e^{i\tau C_3^\mu \KK\cdot\rr_0}e^{i\tfrac{2\pi}{3}\mu\tau} &= e^{i\tau\frac{4\pi x_0}{3a}} + 2e^{-i\tau\frac{2\pi x_0}{3a}}\cos\left(\tfrac{2\pi y_0}{a\sqrt{3}}+\tfrac{2\pi}{3}\right),
\end{split}
\end{equation}
where we have used $\KK=\tfrac{4\pi}{3a}\hat{\mathbf{x}}$. Finally, writing all sums of sines and cosines in terms of $f_s(\rr_0)$ and $f_a(\rr_0)$ [see Eq.\ \eqref{Eq:Adhesion_energy}] gives the Hamiltonian shown in Sec.\ \ref{sec:Khyb}.

\section{DFT calculations, and configurations used for model parametrization}
\label{app:DFT_configs}
\begin{figure}
    \centering
    \includegraphics[width = 0.8\linewidth]{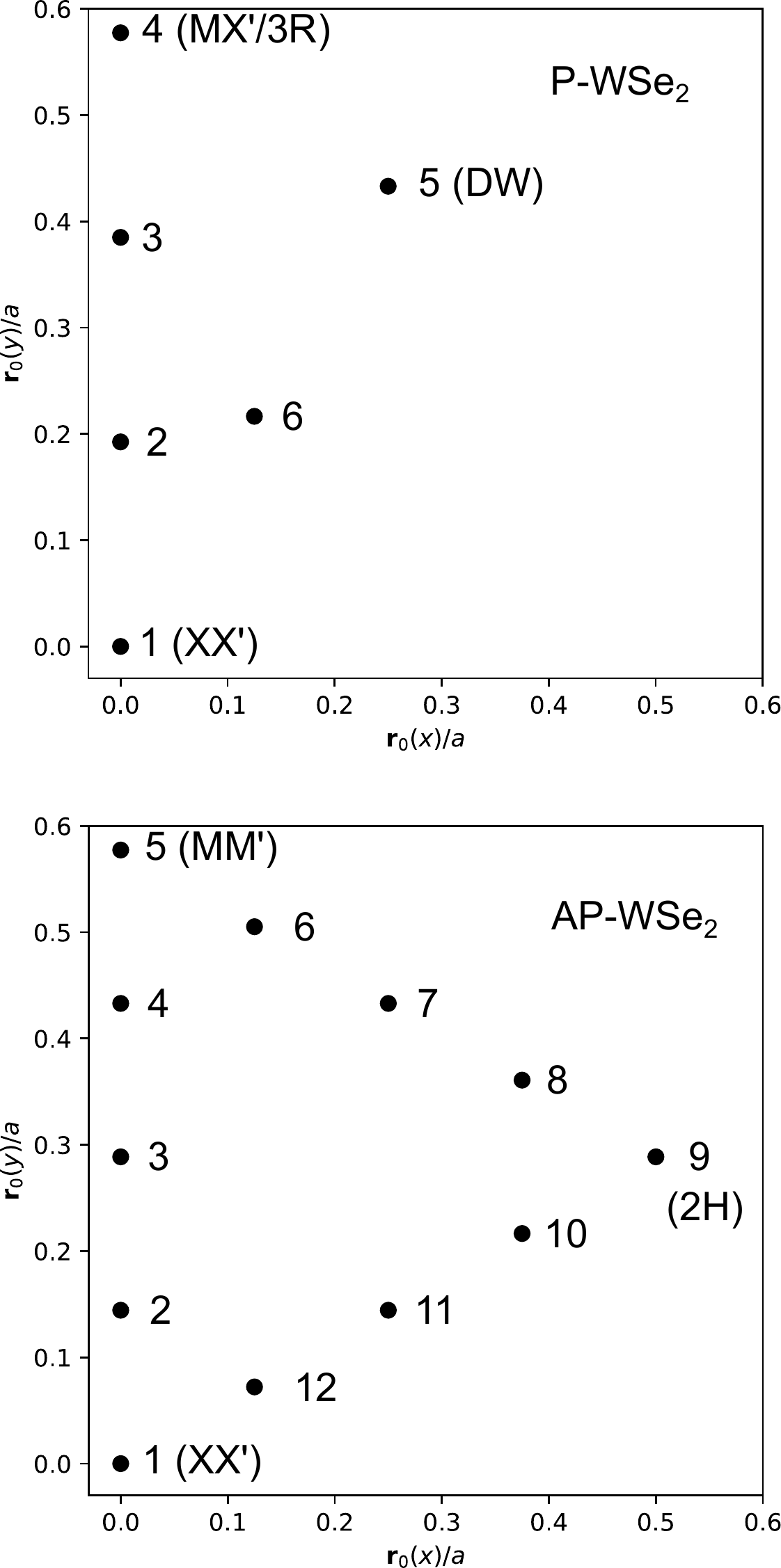}
    \caption{Schematic of in-plane shifts, $\mathbf{r}_0$, in configuration space, used in the DFT parametrizations of the models described in appendices \ref{app:hamiltonianG} and \ref{app:hamiltonianK} in units of the lattice constant $a=3.282$~\AA, for  P-stacking and AP-stacking in the upper and lower panel, respectively. The points are numbered to identify their progression around the paths.}
    \label{fig:configurations}
\end{figure}

These calculations are carried out using the plane-wave based \textsc{VASP}
code \cite{VASP} with spin-orbit coupling taken into account using projector augmented
wave (PAW) pseudopotentials. We approximated the exchange correlation functional using the generalised gradient approximation (GGA) of Perdew, Burke and Ernzerhof \cite{PBE}. 
The cutoff energy for the plane-waves is set to 600 eV with the in-plane Brillouin zone sampled by a
$12 \times 12$ grid. Above, we introduced the effective Hamiltonians for the $\Gamma$- and K-point top valence band states in WSe$_2$ bilayers with phenomenological parameters fitted to energy bands calculated with DFT for 6 (P-stacked bilayers) and 12 (AP-stacked) in-plane shifts $\mathbf{r}_0$, repeated for 6(11) interlayer distances spanning $d=6.477$~\AA~to~$d=7.077$~\AA for P-(AP-)stacking. 
The paths in configuration space traced by the in-plane shifts are plotted in Fig. \ref{fig:configurations}. From the point of view of the band energies, the paths form the boundary of the irreducible portion of in-plane shifts. To minimize the interaction between them, repeated images of bilayers are placed 30~\AA~apart, with P-stacked bilayers using supercells containing two $\sigma_h$-reflected images of the bilayers to maintain periodicity along $z$ taking into account the potential drop across a P-stacked bilayer due to interlayer charge transfer, as set out in the main text. The structure parameters for the monolayer are taken from experimental measurements of the bulk crystal\cite{schutte1987crystal}.

\section{Piezo-electric potential for WSe$_2$ bilayers encapsulated in hBN}\label{app_electrostatics}

To calculate electric potential, $\varphi(\bm{r},z)$, created by piezocharges in twisted AP/P-WSe$_2$ bilayers we expand the piezocharge densities in Fourier series over superlattice reciprocal vectors $\rho_{t/b}=\delta(z-z_{t/b})\sum\rho_n^{t/b}e^{i\bm{g}_n\bm{r}}$ and take into account polarisation charges, $\rho^{t/b}_{ind}=\alpha^{t/b}_{2D}\delta\left(z-z_{t/b}\right)\bm{\nabla}^2_{\bm{r}}\varphi(\bm{r},z_{t/b})$, which are induced by the piezocharges ($\alpha^{t/b}_{2D}=d_0\left(\varepsilon^{t/b}_{||}-1\right)/4\pi$ is the in-plane 2D polarisability of the top/bottom monolayer expressed via in-plane dielectric permittivity of bulk WSe$_2$ crystals \cite{laturia2018dielectric}. 

\begin{figure}
	\includegraphics[width=0.7\linewidth]{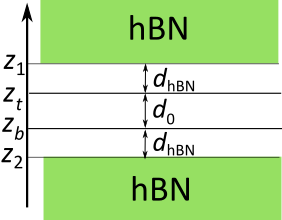}
	\caption{\label{Fig:APPelectrostatics} Model used to calculate piezo-electric potential of WSe$_2$ bilayer encapsulated in hBN.}
\end{figure}

The total charge densities in each layer are sum of the piezo- and polarisation charge densities. To find piezo-electric potential created by the total charge densities, $\rho^{t/b}_{tot}=\rho^{t/b}_{piezo}+\rho^{t/b}_{ind}$ we solve the Poisson equation,  
\begin{multline}\label{Poisson_eq}
\left[\epsilon_{\perp}\partial^2_{zz}+\epsilon_{||}\bm{\nabla}^2_{\bm{r}}\right]\varphi=0,\, z>z_1,\, z<z_2 \\
\left[\partial^2_{zz}+\bm{\nabla}^2_{\bm{r}}\right]\varphi=-4\pi\left(\rho^{t}_{tot}+\rho^{b}_{tot}\right),\, z_2\leq z\leq z_1,
\end{multline}
by expanding the potential in Fourier series over the superlattice reciprocal vectors, $\varphi(\bm{r},z)=\sum_n\widetilde{\varphi}_n(z)e^{i\bm{g}_n\bm{r}}$. In Eq. (\ref{Poisson_eq}) $\epsilon_{\perp}=3.76$, $\epsilon_{||}=6.93$ are in-layer and out-of-layer dielectric permittivities of bulk hBN crystals (see Fig. \ref{Fig:APPelectrostatics}). Solving the Poisson equation (\ref{Poisson_eq}) in each region with corresponding boundary conditions at interfaces we find amplitudes of the potential harmonics in the top and bottom layers:
\begin{multline}\label{potential_t}
\varphi_n(z_{t})=\frac{2\pi}{g_n}\left[\cosh (g_nd_{{\rm hBN}})+\sqrt{\epsilon_{||}\epsilon_{\perp}}\sinh (g_nd_{\rm hBN})\right] \\
\times\frac{\rho_n^{t}(F_1^be^{g_nd_0}+F_2^b)+\rho_n^{b}\left(F_1^b+F_2^be^{g_nd_0}\right)}{F_1^tF_1^b-F_2^tF_2^b}, 
\end{multline}
\begin{multline}\label{potential_b}
\varphi_n(z_{b})=\frac{2\pi}{g_n}\left[\cosh (g_nd_{\rm hBN})+\sqrt{\epsilon_{||}\epsilon_{\perp}}\sinh (g_nd_{\rm hBN})\right] \\
\times\frac{\rho_n^{t}(F_1^t+F_2^te^{g_nd_0})+\rho_n^{b}\left(F_1^te^{g_nd_0}+F_2^t\right)}{F_1^tF_1^b-F_2^tF_2^b},
\end{multline}
where the Fourier amplitudes of piezo-charge density read as 
\begin{equation}\label{piezocharge_harmonics}
\rho_n^{t/b}=e_{11}^{t/b}\left[2g_{nx}g_{ny}u_{nx}^{t/b}+(g_{nx}^2-g_{ny}^2)u_{ny}^{t/b}\right],
\end{equation}
with $e_{11}=2.03\times10^{-10}$C/m \cite{rostami2018piezoelectricity}, and 
\begin{multline}\label{Eq:App_electrostatics_F1}
    F_{1}^{t/b}=e^{g_n(d_0+d_{{\rm hBN}})} \frac{1+\sqrt{\epsilon_{||}\epsilon_{\perp}}}{2}\\
    + 2\pi\alpha_{2D}^{t/b} g_n e^{g_nd_0}\left[\cosh (g_nd_{\rm hBN})+\sqrt{\epsilon_{||}\epsilon_{\perp}}\sinh (g_nd_{\rm hBN})\right]
\end{multline}

\begin{multline}\label{Eq:App_electrostatics_F2}
    F_{2}^{t/b}=e^{-g_nd_{{\rm hBN}}} \frac{1-\sqrt{\epsilon_{||}\epsilon_{\perp}}}{2}\\
    - 2\pi\alpha_{2D}^{t/b} g_n \left[\cosh (g_nd_{\rm hBN})+\sqrt{\epsilon_{||}\epsilon_{\perp}}\sinh (g_nd_{\rm hBN})\right].
\end{multline}
Here, $d_{hBN}=z_1-z_t=z_b-z_2=(6.71+3.35)/2$\AA is distance between WSe$_2$ bilayer and hBN. In main manuscript we used the shorthand for potential on top and bottom layers as follows: $\varphi_{t/b}(\bm{r})=\varphi(\bm{r},z_{t/b})$. The total charge density in the top layer, $\rho^{t}_{tot}=\sum_n\left(\rho_n^{t}-4\pi g_n^2\alpha_{2D}^{t}\varphi_n^{t}\right)e^{i\bm{g}_n\bm{r}}$. 

\section{Fourier components of the interlayer distance variation and piezoelectric potential}\label{sec:minibands}
Below, we report the first Fourier components necessary to reconstruct the interlayer distance modulation [Eq.\ \eqref{Eq:z_local_P}] and piezoelectric potential [Eq.\ \eqref{eq:piezoharmonics}] for twisted P and AP WSe${}_2$ bilayers, and the in-plane relaxation field [Eq.\ \eqref{eq:ufield}] for P bilayers. In all cases, the monolayer basis Bragg vectors are defined as
\begin{equation}
\begin{split}
    \mathbf{G}_1 =& \frac{4\pi}{\sqrt{3}a}\left(\frac{\sqrt{3}}{2}\hat{\mathbf{x}} + \frac{1}{2}\hat{\mathbf{y}} \right),\\
    \mathbf{G}_2 =& \frac{4\pi}{\sqrt{3}a}\left(-\frac{\sqrt{3}}{2}\hat{\mathbf{x}} + \frac{1}{2}\hat{\mathbf{y}} \right).
\end{split}
\end{equation}
The basis moir\'e Bragg vectors for twist angle $\theta\ll 1$ (in radians) are then defined as
\begin{equation}
\begin{split}
    \gG_1 \equiv \theta \left( G_{1,y}\hat{\mathbf{x}} - G_{1,x}\hat{\mathbf{y}} \right) \approx \left(1 - \mathcal{R}_\theta\right) \GG_1,\\
    \gG_2 \equiv \theta \left( G_{2,y}\hat{\mathbf{x}} - G_{2,x}\hat{\mathbf{y}} \right) \approx \left(1 - \mathcal{R}_\theta\right) \GG_2.
\end{split}
\end{equation}
Note that $\GG_2 = C_3 \GG_1$ and $\gG_2 = C_3 \gG_2$, and we can define additional vectors $\GG_3 = C_3^2\GG_1$ and $\gG_3=C_3^2\gG_1$. Together, $\GG_1,\,\GG_2,\,\GG_3$ ($\gG_1,\,\gG_2,\,\gG_3$) form the first star of (moir\'e) Bragg vectors. 

Tables \ref{tab:dSparamsP}--\ref{tab:ufieldP} list the expansion coefficients associated with the $\gG_n$ moir\'e vectors. These are constructed from the basis moir\'e vectors ($\gG_1$ and $\gG_2$) as follows:
\begin{equation}
    \gG_{3n+2} = C_3 \gG_{3n+1},\quad \gG_{3n+3} = C_3^2 \gG_{3n+1},\, n = 0,1,2,\ldots
\end{equation}
and the reference vectors $\gG_1,\,\gG_4,\ldots,\,\gG_{3n+1},\ldots$ are listed in Table \ref{tab:refstarvectors} in the form
\begin{equation}\label{eq:refstarvectors}
    \gG_{3n+1} = m_1^{(3n+1)} \gG_1 + m_2^{(3n+1)} \gG_2.
\end{equation}

\begin{table}[h!]
\caption{Coefficients of the reference vectors $\gG_{3n+1}$, as defined in Eq.\ \eqref{eq:refstarvectors}.}
\begin{center}
\begin{tabular}{P{2cm} P{3cm} P{3cm}}
\hline\hline
  & $m_1$ & $m_2$ \\
 \hline\hline
$\gG_1$&1&0\\
$\gG_4$&1&-1\\
$\gG_{7}$&2&0\\
$\gG_{10}$&3&2\\
$\gG_{13}$&3&1\\
$\gG_{16}$&3&0\\
$\gG_{19}$&4&2\\
$\gG_{22}$&4&1\\
$\gG_{25}$&3&-1\\
$\gG_{28}$&4&0\\
$\gG_{31}$&5&3\\
$\gG_{34}$&5&2\\
$\gG_{37}$&5&4\\
$\gG_{40}$&5&1\\
$\gG_{43}$&5&0\\
$\gG_{46}$&6&3\\
$\gG_{49}$&6&4\\
$\gG_{52}$&6&2\\
$\gG_{55}$&6&1\\
$\gG_{58}$&6&5\\
\hline\hline
\end{tabular}
\end{center}
\label{tab:refstarvectors}
\end{table}

\begin{table*}[h!]
\caption{Coefficients of the moir\'e harmonics expansion Eq.\ \eqref{Eq:z_local_P} for the spatial variation of the interlayer distance in $P$ stacked bilayers with different twist angles. All values are given in units of $10^{-2}\,{\rm \AA}$.}
\begin{center}
\begin{tabular}{P{2.7cm} Q{0.46cm}@{.}R{1.2cm} Q{0.46cm}@{.}R{1.2cm} Q{0.46cm}@{.}R{1.2cm} Q{0.46cm}@{.}R{1.2cm} Q{0.46cm}@{.}R{1.2cm} Q{0.46cm}@{.}R{1.2cm} Q{0.46cm}@{.}R{1.2cm} Q{0.46cm}@{.}R{1.2cm} }
\hline\hline
  \, & \multicolumn{4}{c}{$\theta_P=1.4^\circ$} & \multicolumn{4}{c}{$\theta_P=2.0^\circ$} & \multicolumn{4}{c}{$\theta_P=3.0^\circ$} & \multicolumn{4}{c}{$\theta_P=4.0^\circ$} \\
\, & \multicolumn{2}{c}{$z_n^s$} & \multicolumn{2}{c}{$z_n^a$} & \multicolumn{2}{c}{$z_n^s$} & \multicolumn{2}{c}{$z_n^a$} & \multicolumn{2}{c}{$z_n^s$} & \multicolumn{2}{c}{$z_n^a$} & \multicolumn{2}{c}{$z_n^s$} & \multicolumn{2}{c}{$z_n^a$}\\
 \hline\hline
 $n=0$ &-25&58697 &  0&00000 & -20&55009 &  0&00000 & -12&39398 &  0&00000 & -7&55040 &  0&00000\\
 $n=1,\,2,\,3$ & 5&94496 &  0&00000 &  7&99818 &  0&00000 &  10&56258 &  0&00000 &  11&81364 &  0&00000\\
 $n=4,\,5,\,6$ & 2&21727 &  0&00000 &  2&48982 &  0&00000 &  1&93853 &  0&00000 &  1&23485 &  0&00000\\
 $n=7,\,8,\,9$ & 3&34285 &  0&00000 &  3&19196 &  0&00000 &  2&12191 &  0&00000 &  1&22791 &  0&00000\\
 $n=10,\,11,\,12$ & 0&93257 &  0&00000 &  0&70117 &  0&00000 &  0&15368 &  0&00000 & -0&01490 &  0&00000\\
 $n=13,\,14,\,15$ & 0&94016 &  0&00000 &  0&71081 &  0&00000 &  0&14847 &  0&00000 & -0&02342 &  0&00000\\
 $n=16,\,17,\,18$ & 1&38383 &  0&00000 &  0&84737 &  0&00000 &  0&19193 &  0&00000 &  0&02006 &  0&00000\\
 $n=19,\,20,\,21$ & 0&33587 &  0&00000 &  0&11963 &  0&00000 & -0&03378 &  0&00000 & -0&03073 &  0&00000\\
 $n=22,\,23,\,24$ & 0&34418 &  0&00000 &  0&12695 &  0&00000 & -0&02191 &  0&00000 & -0&02212 &  0&00000\\
 $n=25,\,26,\,27$ & 0&34708 &  0&00000 &  0&13040 &  0&00000 & -0&02701 &  0&00000 & -0&02360 &  0&00000\\
 $n=28,\,29,\,30$ & 0&50022 &  0&00000 &  0&15393 &  0&00000 & -0&00077 &  0&00000 &  0&00047 &  0&00000\\
 $n=31,\,32,\,33$ & 0&10877 &  0&00000 & -0&01108 &  0&00000 & -0&01406 &  0&00000 & -0&00942 &  0&00000\\
 $n=34,\,35,\,36$ & 0&10720 &  0&00000 & -0&02050 &  0&00000 & -0&01697 &  0&00000 &  0&00252 &  0&00000\\
 $n=37,\,38,\,39$ & 0&12186 &  0&00000 & -0&00057 &  0&00000 & -0&00550 &  0&00000 &  0&00210 &  0&00000\\
 $n=40,\,41,\,42$ & 0&10738 &  0&00000 &  0&00533 &  0&00000 & -0&01713 &  0&00000 &  0&00735 &  0&00000\\
 $n=43,\,44,\,45$ & 0&16345 &  0&00000 &  0&01498 &  0&00000 & -0&00487 &  0&00000 &  0&00507 &  0&00000\\
 $n=46,\,47,\,48$ & 0&01585 &  0&00000 & -0&01386 &  0&00000 & -0&00050 &  0&00000 & -0&00480 &  0&00000\\
 $n=49,\,50,\,51$ & 0&01925 &  0&00000 & -0&01096 &  0&00000 & -0&01046 &  0&00000 & -0&00874 &  0&00000\\
 $n=52,\,53,\,54$ & 0&03639 &  0&00000 & -0&00787 &  0&00000 & -0&00559 &  0&00000 &  0&00228 &  0&00000\\
 $n=55,\,56,\,57$ & 0&03680 &  0&00000 & -0&00700 &  0&00000 &  0&00607 &  0&00000 &  0&00735 &  0&00000\\
 $n=58,\,59,\,60$ & 0&03551 &  0&00000 & -0&01305 &  0&00000 & -0&01864 &  0&00000 & -0&01080 &  0&00000\\
\hline\hline
\end{tabular}
\end{center}
\label{tab:dSparamsP}
\end{table*}

\begin{table*}[h!]
\caption{Coefficients of the moir\'e harmonics expansion Eq.\ \eqref{Eq:z_local_P} for the spatial variation of the interlayer distance in $AP$ stacked bilayers with different twist angles. All values are given in units of $10^{-2}\,{\rm \AA}$.}
\begin{center}
\begin{tabular}{P{2.7cm} Q{0.46cm}@{.}R{1.2cm} Q{0.46cm}@{.}R{1.2cm} Q{0.46cm}@{.}R{1.2cm} Q{0.46cm}@{.}R{1.2cm} Q{0.46cm}@{.}R{1.2cm} Q{0.46cm}@{.}R{1.2cm} Q{0.46cm}@{.}R{1.2cm} Q{0.46cm}@{.}R{1.2cm} }
\hline\hline
  \, & \multicolumn{4}{c}{$\theta_{AP}=1.4^\circ$} & \multicolumn{4}{c}{$\theta_{AP}=2.0^\circ$} & \multicolumn{4}{c}{$\theta_{AP}=3.0^\circ$} & \multicolumn{4}{c}{$\theta_{AP}=4.0^\circ$} \\
\, & \multicolumn{2}{c}{$z_n^s$} & \multicolumn{2}{c}{$z_n^a$} & \multicolumn{2}{c}{$z_n^s$} & \multicolumn{2}{c}{$z_n^a$} & \multicolumn{2}{c}{$z_n^s$} & \multicolumn{2}{c}{$z_n^a$} & \multicolumn{2}{c}{$z_n^s$} & \multicolumn{2}{c}{$z_n^a$}\\
 \hline\hline
 $n=0$ &-24&13399 &  0&00000 & -19&05858 &  0&00000 & -11&53592 &  0&00000 & -6&65903 &  0&00000\\
 $n=1,\,2,\,3$ &-5&50791 &  3&24141 & -6&16899 &  4&52594 & -6&74953 &  6&86940 & -7&01133 &  7&94786\\
 $n=4,\,5,\,6$ & 2&82852 & -0&00163 &  2&55399 & -0&00287 &  1&74925 & -0&00282 &  1&10190 &  0&00027\\
 $n=7,\,8,\,9$ & 0&67310 & -2&56605 &  0&32295 & -2&70102 &  0&04658 & -1&84220 & -0&00550 & -1&14069\\
 $n=10,\,11,\,12$ &-0&78417 & -0&91178 & -0&54056 & -0&60187 & -0&13118 & -0&13734 & -0&01039 & -0&01058\\
 $n=13,\,14,\,15$ &-0&80308 &  0&90362 & -0&54585 &  0&59891 & -0&14142 &  0&13902 & -0&02071 &  0&00816\\
 $n=16,\,17,\,18$ & 0&45891 &  0&87199 &  0&41816 &  0&64093 &  0&10759 &  0&18354 &  0&01592 &  0&04758\\
 $n=19,\,20,\,30$ & 0&44231 & -0&00065 &  0&19873 &  0&00296 & -0&00732 &  0&00295 & -0&01067 &  0&00541\\
 $n=31,\,32,\,33$ &-0&02988 & -0&52235 &  0&00011 & -0&21693 &  0&00789 &  0&00567 &  0&00169 &  0&01213\\
 $n=34,\,35,\,36$ &-0&03029 & -0&52334 & -0&00259 & -0&22186 &  0&00403 &  0&00447 & -0&00242 &  0&01619\\
 $n=37,\,38,\,39$ &-0&30441 & -0&10434 & -0&15945 & -0&06188 & -0&00915 & -0&01331 &  0&00372 & -0&00496\\
 $n=40,\,41,\,42$ &-0&11498 & -0&14544 & -0&03342 & -0&03053 &  0&00412 &  0&00503 & -0&00269 &  0&00408\\
 $n=43,\,44,\,45$ &-0&11432 &  0&13318 & -0&02114 &  0&03925 &  0&01359 & -0&00473 &  0&00616 & -0&00671\\
 $n=46,\,47,\,48$ & 0&13680 & -0&15399 &  0&03093 & -0&04459 &  0&00227 &  0&01083 &  0&00313 & -0&00780\\
 $n=49,\,50,\,51$ & 0&13037 &  0&15145 &  0&03238 &  0&01842 & -0&00111 & -0&00694 & -0&00064 & -0&00027\\
 $n=52,\,53,\,54$ & 0&10133 & -0&03621 &  0&03233 & -0&00575 &  0&00822 &  0&00331 &  0&00154 & -0&00116\\
 $n=55,\,56,\,57$ & 0&05078 & -0&00077 & -0&00051 & -0&00329 & -0&00340 & -0&00109 & -0&00098 &  0&00211\\
 $n=58,\,59,\,60$ &-0&01020 &  0&06517 & -0&00164 &  0&00853 &  0&00487 & -0&00159 &  0&00676 &  0&00567\\
 $n=61,\,62,\,63$ & 0&00405 & -0&06227 & -0&00889 & -0&00860 & -0&00023 & -0&00410 & -0&00004 & -0&00092\\
 $n=64,\,65,\,66$ &-0&05478 & -0&01542 & -0&00834 &  0&00509 & -0&00092 &  0&00175 &  0&00065 & -0&00239\\
 $n=67,\,68,\,69$ &-0&06405 &  0&02779 & -0&01315 & -0&00370 &  0&00997 &  0&00561 &  0&00515 &  0&00170\\
\hline\hline
\end{tabular}
\end{center}
\label{tab:dSparamsAP}
\end{table*}

\begin{table*}[h!]
\caption{Coefficients of the moir\'e harmonics expansion Eq.\ \eqref{eq:piezoharmonics} for the piezoelectric potential $\varphi(\rr)$, for $P$ bilayers with different twist angles. The top- and bottom-layer potential coefficients are given by $\varphi_n^{t}=\varphi_n$ and $\varphi_n^{b}=-\varphi_n$. All values are in units of ${\rm mV}$.}
\begin{center}
\begin{tabular}{P{2.7cm} Q{0.46cm}@{.}R{1.2cm} Q{0.46cm}@{.}R{1.2cm} Q{0.46cm}@{.}R{1.2cm} Q{0.46cm}@{.}R{1.2cm} Q{0.46cm}@{.}R{1.2cm} Q{0.46cm}@{.}R{1.2cm} Q{0.46cm}@{.}R{1.2cm} Q{0.46cm}@{.}R{1.2cm} }
\hline\hline
  \, & \multicolumn{4}{c}{$\theta_{P}=1.4^\circ$} & \multicolumn{4}{c}{$\theta_{P}=2.0^\circ$} & \multicolumn{4}{c}{$\theta_{P}=3.0^\circ$} & \multicolumn{4}{c}{$\theta_{P}=4.0^\circ$} \\
\, & \multicolumn{2}{c}{$\mathrm{Re}\,\varphi_n$} & \multicolumn{2}{c}{$\mathrm{Im}\,\varphi_n$} & \multicolumn{2}{c}{$\mathrm{Re}\,\varphi_n$} & \multicolumn{2}{c}{$\mathrm{Im}\,\varphi_n$} & \multicolumn{2}{c}{$\mathrm{Re}\,\varphi_n$} & \multicolumn{2}{c}{$\mathrm{Im}\,\varphi_n$} & \multicolumn{2}{c}{$\mathrm{Re}\,\varphi_n$} & \multicolumn{2}{c}{$\mathrm{Im}\,\varphi_n$}\\
 \hline\hline
 $n=0$ & 0&00000 &  0&00000 &  0&00000 &  0&00000 &  0&00000 &  0&00000 &  0&00000 &  0&00000\\
 $n=1,\,2,\,3$ & 0&00000 & -14&22201 &  0&00000 & -12&33073 &  0&00000 & -7&45156 &  0&00000 & -4&45439\\
 $n=4,\,5,\,6$ & 0&00000 & -0&00115 &  0&00000 &  0&00900 &  0&00000 &  0&00219 &  0&00000 &  0&00051\\
 $n=7,\,8,\,9$ & 0&00000 & -2&59835 &  0&00000 & -1&03104 &  0&00000 & -0&02926 &  0&00000 &  0&11522\\
 $n=10,\,11,\,12$ & 0&00000 &  0&02994 &  0&00000 & -0&06365 &  0&00000 & -0&05794 &  0&00000 & -0&02529\\
 $n=13,\,14,\,15$ & 0&00000 & -0&03119 &  0&00000 &  0&06768 &  0&00000 &  0&05796 &  0&00000 &  0&02520\\
 $n=16,\,17,\,18$ & 0&00000 & -0&52157 &  0&00000 & -0&02203 &  0&00000 &  0&04614 &  0&00000 &  0&01982\\
 $n=19,\,20,\,21$ & 0&00000 & -0&00040 &  0&00000 &  0&00090 &  0&00000 &  0&00002 &  0&00000 &  0&00000\\
 $n=22,\,23,\,24$ & 0&00000 & -0&00543 &  0&00000 &  0&03772 &  0&00000 &  0&00979 &  0&00000 &  0&00000\\
 $n=25,\,26,\,27$ & 0&00000 & -0&00543 &  0&00000 &  0&03788 &  0&00000 &  0&00973 &  0&00000 &  0&00000\\
 $n=28,\,29,\,30$ & 0&00000 & -0&04928 &  0&00000 &  0&02671 &  0&00000 &  0&00597 &  0&00000 &  0&00000\\
\hline\hline
\end{tabular}
\end{center}
\label{tab:piezoharmonicsP}
\end{table*}

\begin{table*}[h!]
\caption{Coefficients of the moir\'e harmonics expansion Eq.\ \eqref{eq:piezoharmonics} for the piezoelectric potential $\varphi(\rr)$, for $AP$ bilayers with different twist angles. The top- and bottom-layer potential coefficients are given by $\varphi_n^{t}=\varphi_n^{b}=\varphi_n$. All values are in units of ${\rm mV}$.}
\begin{center}
\begin{tabular}{P{2.7cm} Q{0.46cm}@{.}R{1.2cm} Q{0.46cm}@{.}R{1.2cm} Q{0.46cm}@{.}R{1.2cm} Q{0.46cm}@{.}R{1.2cm} Q{0.46cm}@{.}R{1.2cm} Q{0.46cm}@{.}R{1.2cm} Q{0.46cm}@{.}R{1.2cm} Q{0.46cm}@{.}R{1.2cm} }
\hline\hline
  \, & \multicolumn{4}{c}{$\theta_{AP}=1.4^\circ$} & \multicolumn{4}{c}{$\theta_{AP}=2.0^\circ$} & \multicolumn{4}{c}{$\theta_{AP}=3.0^\circ$} & \multicolumn{4}{c}{$\theta_{AP}=4.0^\circ$} \\
\, & \multicolumn{2}{c}{$\mathrm{Re}\,\varphi_n$} & \multicolumn{2}{c}{$\mathrm{Im}\,\varphi_n$} & \multicolumn{2}{c}{$\mathrm{Re}\,\varphi_n$} & \multicolumn{2}{c}{$\mathrm{Im}\,\varphi_n$} & \multicolumn{2}{c}{$\mathrm{Re}\,\varphi_n$} & \multicolumn{2}{c}{$\mathrm{Im}\,\varphi_n$} & \multicolumn{2}{c}{$\mathrm{Re}\,\varphi_n$} & \multicolumn{2}{c}{$\mathrm{Im}\,\varphi_n$}\\
 \hline\hline
$n=0$ & 0&00000 &  0&00000 &  0&00000 &  0&00000 &  0&00000 &  0&00000 &  0&00000 &  0&00000\\
$n=1,\,2,\,3$ & -6&61064 &  14&27250 & -6&87717 &  11&26287 & -4&64673 &  6&26576 & -2&89876 &  3&65027\\
$n=4,\,5,\,6$ &  0&76037 & -0&04281 &  0&20305 & -0&00240 &  0&01545 &  0&00166 & -0&00019 &  0&00041\\
 $n=7,\,8,\,9$ & 2&14124 & -1&28374 &  1&30934 & -0&51315 &  0&14263 & -0&09730 & -0&04870 & -0&02681\\
$n=10,\,11,\,12$ & -0&28241 & -0&10708 & -0&01306 & -0&03063 &  0&03092 &  0&02756 &  0&01409 &  0&01381\\
$n=13,\,14,\,15$ & -0&30460 &  0&11894 & -0&01335 &  0&03088 &  0&03070 & -0&02760 &  0&01403 & -0&01380\\
$n=16,\,17,\,18$ & -0&54968 & -0&05502 & -0&16047 &  0&00781 &  0&01366 &  0&02401 &  0&01061 &  0&00995\\
 $n=19,\,20,\,21$ & 0&01943 & -0&00310 & -0&00039 & -0&00202 &  0&00196 & -0&00096 &  0&00162 & -0&00099\\
 $n=22,\,23,\,14$ & 0&09423 & -0&02802 & -0&07428 &  0&08013 & -0&03231 &  0&02338 & -0&01548 &  0&01025\\
 $n=25,\,26,\,27$ & 0&10478 & -0&03522 & -0&01666 & -0&01082 & -0&00502 & -0&00678 & -0&00194 & -0&00352\\
 $n=28,\,29,\,30$ & 0&13256 &  0&06584 &  0&01107 &  0&01535 & -0&00211 &  0&01218 & -0&00192 &  0&00883\\
 $n=31,\,32,\,33$ & 0&01198 &  0&00096 &  0&00000 &  0&00000 &  0&00000 &  0&00000 &  0&00000 &  0&00000\\
 $n=34,\,35,\,36$ & 0&00965 & -0&00181 &  0&00000 &  0&00000 &  0&00000 &  0&00000 &  0&00000 &  0&00000\\
$n=37,\,38,\,39$ & -0&00539 & -0&02158 &  0&00000 &  0&00000 &  0&00000 &  0&00000 &  0&00000 &  0&00000\\
$n=40,\,41,\,42$ & -0&01071 &  0&02171 &  0&00000 &  0&00000 &  0&00000 &  0&00000 &  0&00000 &  0&00000\\
\hline\hline
\end{tabular}
\end{center}
\label{tab:piezoharmonicsAP}
\end{table*}

\begin{table*}[h!]
\caption{Coefficients of the moir\'e harmonics expansion Eq.\ \eqref{eq:ufield} for the in-plane displacement field $\mathbf{u}^t(\rr) - \mathbf{u}^b(\rr)$, for $P$ bilayers with different twist angles. All values are in units of $10^{-2}\,{\rm \AA}$.}
\begin{center}
\begin{tabular}{P{2.7cm} Q{0.46cm}@{.}R{1.2cm} Q{0.46cm}@{.}R{1.2cm} Q{0.46cm}@{.}R{1.2cm} Q{0.46cm}@{.}R{1.2cm} Q{0.46cm}@{.}R{1.2cm} Q{0.46cm}@{.}R{1.2cm} Q{0.46cm}@{.}R{1.2cm} Q{0.46cm}@{.}R{1.2cm} }
\hline\hline
  \, & \multicolumn{4}{c}{$\theta_{P}=1.4^\circ$} & \multicolumn{4}{c}{$\theta_{P}=2.0^\circ$} & \multicolumn{4}{c}{$\theta_{P}=3.0^\circ$} & \multicolumn{4}{c}{$\theta_{P}=4.0^\circ$} \\
\, & \multicolumn{2}{c}{$u_{n,x}$} & \multicolumn{2}{c}{$u_{n,y}$} & \multicolumn{2}{c}{$u_{n,x}$} & \multicolumn{2}{c}{$u_{n,y}$} & \multicolumn{2}{c}{$u_{n,x}$} & \multicolumn{2}{c}{$u_{n,y}$} & \multicolumn{2}{c}{$u_{n,x}$} & \multicolumn{2}{c}{$u_{n,y}$}\\
 \hline\hline
 $n=0$            &  0&00000 &  0&00000 &  0&00000 &  0&00000 &  0&00000 &  0&00000 &  0&00000 &  0&00000\\
 $n=1,\,2,\,3$    & 32&10502 & 18&54762 & 24&52676 & 14&12421 & 13&51045 &  7&79985 &  7&75407 &  4&47794\\
 $n=4,\,5,\,6$    &  0&09762 &  0&00252 &  0&10202 & -0&01842 & -0&74473 & -0&00429 & -0&73187 & -0&00097\\
 $n=7,\,8,\,9$    &  4&76970 &  2&75402 &  1&79569 &  0&51747 &  0&04853 &  0&02943 & -0&19064 & -0&10945\\
 $n=10,\,11,\,12$ & -0&02784 &  0&05540 & -0&11962 & -0&25275 & -0&07374 & -0&19879 & -0&03032 & -0&08454\\
 $n=13,\,14,\,15$ &  0&06456 &  0&00422 & -0&16781 & -0&23527 & -0&13555 & -0&16468 & -0&05791 & -0&06877\\
 $n=16,\,17,\,18$ &  0&90220 &  0&52294 &  0&03769 &  0&02018 & -0&07590 & -0&04364 & -0&03225 & -0&01850\\
 $n=19,\,20,\,21$ &  0&00744 &  0&01132 & -0&06797 & -0&11426 & -0&01627 & -0&02812 & -0&00573 & -0&01457\\
 $n=22,\,23,\,24$ &  0&01042 &  0&00646 & -0&06962 & -0&07217 & -0&01777 & -0&01755 &  0&06136 &  0&03422\\
 $n=25,\,26,\,27$ &  0&01028 &  0&00614 & -0&09693 & -0&02481 & -0&02398 & -0&00656 & -0&00433 &  0&00763\\
 $n=28,\,29,\,30$ &  0&08308 &  0&04820 & -0&04402 & -0&02567 & -0&00973 & -0&00556 &  0&01205 &  0&00590\\
\hline\hline
\end{tabular}
\end{center}
\label{tab:ufieldP}
\end{table*}

\end{document}